\documentclass{article}
\usepackage{amssymb}
\usepackage{amsmath}
\usepackage{amsfonts}

\newtheorem{theorem}{Theorem}

\newtheorem{remark}[theorem]{Remark}

\input{tcilatex}
\begin{document}

\title{\textbf{Integrability vs Supersymmetry:}\\
\textbf{Poisson Structures of the Pohlmeyer Reduction }}
\author{David M. Schmidtt\thanks{%
david.schmidtt@gmail.com} \\
\\
\textit{Instituto de F\'{\i}sica Te\'{o}rica - IFT/UNESP}\\
\textit{Rua Dr. Bento Teobaldo Ferraz, 271, Bloco II}\\
\textit{CEP 01140-070, S\~{a}o Paulo, SP, Brasil.}}
\maketitle

\begin{abstract}
We construct recursively an infinite number of Poisson structures for the
supersymmetric integrable hierarchy governing the Pohlmeyer reduction of
superstring sigma models on the target spaces $AdS_{n}\times S^{n},$ $%
n=2,3,5.$ These Poisson structures are all non-local and not relativistic
except one, which is the canonical Poisson structure of the semi-symmetric
space sine-Gordon model (SSSSG). We verify that the superposition of the
first three Poisson structures corresponds to the canonical Poisson
structure of the reduced sigma model. Using the recursion relations we
construct commuting charges on the reduced sigma model out of those of the
SSSSG model and in the process we explain the integrable origin of the
Zukhovsky map and the twisted inner product used in the sigma model side.
Then, we compute the complete Poisson superalgebra for the conserved
Drinfeld-Sokolov supercharges associated to an exotic kind of extended
non-local rigid 2d supersymmetry recently introduced in the SSSSG context.
The superalgebra has a kink central charge which turns out to be a
generalization to the SSSSG models of the well-known central extensions of
the $N=1$ sine-Gordon and $N=2$ complex sine-Gordon model Poisson
superalgebras computed from 2d superspace. The computation is done in two
different ways concluding the proof of the existence of 2d supersymmetry in
the reduced sigma model phase space under the boost invariant SSSSG Poisson
structure.
\end{abstract}

\tableofcontents

\section{Introduction.}

Since the work of Grigoriev and Tseytlin \cite{Grigo-Tseytlin}, devoted to
the study of the classical Pohlmeyer reduction of the Green-Schwarz
superstring sigma model (GSs$\sigma $) on $AdS_{5}\times S^{5}$, there has
been a relatively intense activity focused on the study of the properties of
a family of 2d integrable field theories models that appear in the reduction
process \cite{Grigo-Tseytlin II}-\cite{Relat magnon S-Matrix}. The
equivalent integrable field theories that are left after reduction of string
and superstring sigma models are known as Symmetric Space Sine-Gordon (SSSG)
and Semi-Symmetric Space Sine-Gordon (SSSSG) models, respectively. There are
two main motivations for studying these models at classical level: the first
one is because the reduced models possesses a manifestly Lorentz invariant
integrable hierarchy structure amenable of quantization and the second one
is because of the possibility of having 2d world-sheet rigid supersymmetry.
At quantum level however, the motivation is even stronger and is essentially
based on the possibility of having an eventual first-principles solution of
the GSs$\sigma $ model based on quantum integrability, at least for the GSs$%
\sigma $ model on $AdS_{5}\times S^{5}$, in which the Pohlmeyer reduction is
expected to survive.

This paper is a continuation of the study of the on-shell 2d supersymmetry
properties of the SSSSG models initiated in \cite{yo03}, based on the
fermionic symmetry flows approach gradually developed in \cite{mKdV susy
flows},\cite{yo01},\cite{yo02},\cite{yo03}. The main results of the present
paper are the following:

\begin{itemize}
\item There is an infinite number of Poisson bi-vectors $\Theta ^{(n)},$ $%
n\in 
\mathbb{Z}
$ on the reduced phase space $\mathcal{P}$ of the GSs$\sigma $ model on the
semi-symmetric space $F/G$ in terms of which the evolution of $\varphi $ can
be written as follows%
\begin{equation}
\frac{\partial \varphi }{\partial t_{a}}=\left\{ \varphi
,H_{z^{-4n}a}\right\} _{-n}=...=\left\{ \varphi ,H_{z^{-4}a}\right\}
_{-1}=\left\{ \varphi ,H_{a}\right\} _{0}=\left\{ \varphi
,H_{z^{4}a}\right\} _{1}=...=\left\{ \varphi ,H_{z^{4n}a}\right\} _{n},
\label{A}
\end{equation}%
where $z$ is the spectral parameter and $a\subset \widehat{\mathfrak{f}}%
^{\perp }\subset \widehat{\mathfrak{f}}$ belongs to an affinization of the
Lie superalgebra $\mathfrak{f.}$

\item The reduced canonical Poisson structure $\Theta _{\sigma }$ of the GSs$%
\sigma $ model can be written as follows%
\begin{equation}
\Theta _{\sigma }=\Theta ^{(-1)}-2\Theta ^{(0)}+\Theta ^{(1)}.  \label{B}
\end{equation}%
This result was already obtained in \cite{Mikha bi-hamiltonian} for the GSs$%
\sigma $ model on $AdS_{5}\times S^{5}$ but here we use different arguments
based on integrability which are also valid for the targets $AdS_{n}\times
S^{n},$ $n=2,3.$

\item The recursion relations (\ref{A}) and the latter expression for $%
\Theta _{\sigma }$, i.e, (\ref{B}) imply the following relation between the
commuting charges of the GSs$\sigma $ models $q_{\sigma }$ and the SSSSG
models $q$%
\begin{equation*}
q_{\sigma }(a)=\dint\nolimits_{-\infty }^{+\infty }dx\left\langle a,%
\mathfrak{d}\right\rangle _{\phi },\text{ \ \ \ \ \ }q(a)=\dint\nolimits_{-%
\infty }^{+\infty }dx\left\langle a,\mathfrak{d}\right\rangle ,\text{ \ \ \
\ \ }\phi (z)=z^{-4}-2+z^{4},
\end{equation*}%
where $\mathfrak{d}$ is a density, $\left\langle X,Y\right\rangle $ and\ $%
\left\langle X,Y\right\rangle _{\phi }$ are inner products on $\widehat{%
\mathfrak{f}}$ and $\left\langle X,Y\right\rangle _{\phi }$ is the twisting
of $\left\langle X,Y\right\rangle $ by the Zukhovsky map $z\rightarrow u(z).$

\item The reduced phase space of the GSs$\sigma $ model on $AdS_{n}\times
S^{n},$ $n=2,3,5$ is supersymmetric under the Poisson structure $\Theta
^{(0)}.$ The mixed Poisson bracket computing the central charge is%
\begin{equation*}
\left\{ q(\epsilon ),\overline{q}(\overline{\epsilon })\right\} =Z,\text{ \
\ \ \ \ }Z=Str\left( \gamma \epsilon \gamma ^{-1}\overline{\epsilon }\right)
\mid _{-\infty }^{+\infty },
\end{equation*}%
where $\gamma $ is the fundamental bosonic field of the SSSSG model. This $Z$
generalizes the central charges of the $N=1$ sine-Gordon and $N=2$ complex
sine-Gordon models.
\end{itemize}

With these results we conclude the proof of the existence of on-shell rigid
2d supersymmetry on the reduced models.

The outline of the paper is as follows. In section 2 we review the basic
properties and results of the integrable hierarchy of the SSSSG models. In
section 3 we construct recursively the Poisson structures focusing mainly in
the first three, which are constructed explicitly. In section 4 we show
several connections between the GSs$\sigma $ models and the SSSSG models,
e.g, the Poisson structures, how to extract the Lax pair of the SSSSG model
from that of the GSs$\sigma $ model, etc. In section 5 we re-study the
supersymmetry flow variations and deduce a set of local transformations
closing on a superalgebra with field dependent parameters. In section 6 we
construct the moment maps associated to the supersymmetry showing, in a
different way in contrast to \cite{yo03}, that they are Hamiltonian flows on
the reduced phase space. In section 7 we compute in two different ways the
mixed Poisson bracket and the kink central charge of the Poisson
supersymmetry algebra of the reduced models. Finally, we make some
concluding remarks. For the sake of completeness and readability we have
tried to be as self-contained as possible.

\section{Essentials of the SSSSG integrable hierarchy.}

In this section we recall some of the definitions and results of the
integrable hierarchy governing the Pohlmeyer reduction of superstring sigma
models that we need in the following. The supersymmetric integrable
hierarchy, the non-local supersymmetry variations and their associated
fermionic conserved charges associated to all the semi-symmetric superspaces
involved in the reduction of\ the GSs$\sigma $ model on $AdS_{n}\times
S^{n}, $ $n=2,3,5$\ was initially introduced in \cite{yo03} and subsequently
rephrased and nicely applied to the $AdS_{5}\times S^{5}$ case in \cite%
{SSSSG AdS5xS5}. However, we will use the notation of \cite{SSSSG AdS5xS5}
for the purpose of notational unification.

Considering a finite dimensional real Lie superalgebra $\mathfrak{f}$
endowed with an order four linear automorphism $\Omega ,$ $\Omega :\mathfrak{%
f\rightarrow f,}$ $\Omega \left( \left[ X,Y\right] \right) =\left[ \Omega
\left( X\right) ,\Omega \left( Y\right) \right] ,$ $\Omega ^{4}=I.$ The
superalgebra $\mathfrak{f}$ then admits a $%
\mathbb{Z}
_{4}$ grade space decomposition satisfying%
\begin{equation}
\mathfrak{f=f}_{0}\mathfrak{\oplus f}_{1}\mathfrak{\oplus f}_{2}\mathfrak{%
\oplus f}_{3},\text{ \ \ \ \ \ }\Omega (\mathfrak{f}_{j})=(i)^{j}\mathfrak{f}%
_{j},\text{ \ \ \ \ \ }\left[ \mathfrak{f}_{i},\mathfrak{f}_{j}\right]
\subset \mathfrak{f}_{(i+j)\func{mod}4}.  \label{Z4 grading}
\end{equation}%
The even subalgebra is $\mathfrak{f}_{even}=\mathfrak{f}_{0}\mathfrak{\oplus
f}_{2}$ while the odd part of $\mathfrak{f}$ is formed by $\mathfrak{f}%
_{odd}=\mathfrak{f}_{1}\mathfrak{\oplus f}_{3}.$

We need to introduce a semisimple element $\Lambda \in \mathfrak{f}_{2}$
inducing the following superalgebra splitting%
\begin{equation*}
\mathfrak{f=}\ker (ad(\Lambda ))\oplus \func{Im}(ad(\Lambda ))\equiv 
\mathfrak{f}^{\perp }\text{ }\mathfrak{\oplus }\text{ }\mathfrak{f}%
^{\parallel },\text{ \ \ \ \ \ }\mathfrak{f}^{\perp }\text{ }\mathfrak{\cap }%
\text{ }\mathfrak{f}^{\parallel }=\oslash ,
\end{equation*}%
and restrict ourselves to the situation in which $\mathfrak{f}$ admits an
extra $%
\mathbb{Z}
_{2}$ gradation $\sigma :$ $\mathfrak{f\rightarrow f,}$ $\sigma \left( \left[
X,Y\right] \right) =\left[ \sigma \left( X\right) ,\sigma \left( Y\right) %
\right] ,$ $\sigma ^{2}=I$ with $\sigma (\mathfrak{f}^{\perp })=\mathfrak{f}%
^{\perp }$ and $\sigma (\mathfrak{f}^{\parallel })=-\mathfrak{f}^{\parallel
},$ implying that $\mathfrak{f}$ is also a symmetric space 
\begin{equation}
\left[ \mathfrak{f}^{\perp },\mathfrak{f}^{\perp }\right] \subset \mathfrak{f%
}^{\perp },\text{ \ \ \ \ \ }\left[ \mathfrak{f}^{\perp },\mathfrak{f}%
^{\parallel }\right] \subset \mathfrak{f}^{\parallel }\text{ , \ \ \ \ \ }%
\left[ \mathfrak{f}^{\parallel },\mathfrak{f}^{\parallel }\right] \subset 
\mathfrak{f}^{\perp }.  \label{kernel-image finite}
\end{equation}

We are mainly interested in the cases $\mathfrak{f=psu(}n,n\mid n\mathfrak{)}
$ for $n=1,2$ and it follows from string theory arguments \cite%
{Grigo-Tseytlin} that it is possible to choose $\Lambda ,$ in an $n\times n$
dimensional supermatrix representation of $\mathfrak{su(}n,n\mid n\mathfrak{)%
}$, as follows 
\begin{equation*}
\Lambda =\frac{i}{2}diag\left( \lambda ,\lambda \right) ,\text{ \ \ \ \ \ }%
\lambda =-diag\left( I_{n},-I_{n}\right) .
\end{equation*}%
This $\Lambda $ satisfies%
\begin{equation*}
\Lambda \mathfrak{f}^{\perp }=\mathfrak{f}^{\perp }\Lambda ,\text{ \ \ \ \ \ 
}\Lambda \mathfrak{f}^{\parallel }=-\mathfrak{f}^{\parallel }\Lambda ,\text{
\ \ \ \ \ }-4\Lambda ^{2}=I_{4n}
\end{equation*}%
and in terms of it the projection operators along $\mathfrak{f}^{\perp }$
and $\mathfrak{f}^{\parallel }$ are given by%
\begin{equation*}
\pi ^{\perp }(\ast )=-\left\{ \Lambda ,\left\{ \Lambda ,\ast \right\}
\right\} ,\text{ \ \ \ \ \ }\pi ^{\parallel }(\ast )=-\left[ \Lambda ,\left[
\Lambda ,\ast \right] \right] .
\end{equation*}

The connection with the Green-Schwarz superstring sigma models on $%
AdS_{n}\times S^{n},$ $n=2,3,5$ involve respectively, the following \cite%
{Grigo-Tseytlin},\cite{Grigo-Tseytlin II},\cite{Zarembo} semi-symmetric
spaces $F/G$ 
\begin{equation*}
\frac{F}{G}:\text{ \ }\frac{PSU(1,1\mid 2)}{SO(1,1)\times SO(2)},\text{ \ \
\ \ \ }\frac{PSU(1,1\mid 2)\times PSU(1,1\mid 2)}{SU(1,1)\times SU(2)},\text{
\ \ \ \ \ }\frac{PSU(2,2\mid 4)}{SO(1,4)\times SO(5)},
\end{equation*}%
where $G=\exp \mathfrak{g,}$ $\mathfrak{g\equiv f}_{0},$ while the
connection with the SSSSG models obtained after Pohlmeyer reduction involve
the special coset spaces $G/H\oplus \left( \mathfrak{f}_{1}^{\parallel }%
\mathfrak{\oplus f}_{3}^{\parallel }\right) $ with%
\begin{equation*}
\frac{G}{H}:\text{ \ }U(1)\times U(1),\text{ \ \ \ \ \ }\frac{SU(1,1)\times
SU(2)}{U(1)\times U(1)},\text{ \ \ \ \ \ }\frac{SO(1,4)\times SO(5)}{%
SU(2)^{\times 4}},
\end{equation*}%
where $H=\exp \mathfrak{h,}$ $\mathfrak{h\equiv f}_{0}^{\perp }.$ These
reduced models exhibits and exotic kind of 2d rigid supersymmetry of the
type $N=(2,2),$ $N=(4,4)$ and $N=(8,8)$ with R-symmetry group $H$ and where
the number of chiral supersymmetries is determined by $N=\dim \mathfrak{f}%
_{1,3}^{\perp },$ see for instance \cite{yo03},\cite{SSSSG AdS5xS5}. We will
soon recall how this supersymmetries appear in our construction after
affinization of $\mathfrak{f}^{\perp }\mathfrak{.}$

The algebraic structure underlying the SSSSG integrable hierarchy, is
defined by the following twisted loop Lie superalgebra%
\begin{equation}
\widehat{\mathfrak{f}}=\dbigoplus\limits_{n\in 
\mathbb{Z}
=-\infty }^{+\infty }\left( z^{4n}\otimes \mathfrak{f}_{0}+z^{4n+1}\otimes 
\mathfrak{f}_{1}+z^{4n+2}\otimes \mathfrak{f}_{2}+z^{4n+3}\otimes \mathfrak{f%
}_{3}\right) ,  \label{superalgebra}
\end{equation}%
which can be rewritten as an integer decomposition%
\begin{equation}
\widehat{\mathfrak{f}}=\widehat{\mathfrak{f}}^{\perp }\mathfrak{\oplus }%
\widehat{\mathfrak{f}}^{\parallel }=\dbigoplus\limits_{r\in 
\mathbb{Z}
=-\infty }^{+\infty }\widehat{\mathfrak{f}}_{r},\text{ \ \ \ \ \ }\left[ Q,%
\widehat{\mathfrak{f}}_{r}\right] =r\widehat{\mathfrak{f}}_{r}
\label{half-integer expansion}
\end{equation}%
in terms of the homogeneous gradation operator $Q_{H}\equiv z\frac{d}{dz}.$
The kernel subalgebra $\widehat{\mathfrak{f}}^{\perp }$ decomposes as%
\footnote{%
The symbol $\ltimes $ denotes central extension.}%
\begin{equation}
\widehat{\mathfrak{f}}^{\perp }=\widehat{\mathfrak{c}}\ltimes \widehat{%
\mathfrak{z}}\mathfrak{,}\text{ \ \ \ \ \ }\widehat{\mathfrak{c}}=\left[ 
\widehat{\mathfrak{f}}^{\perp },\widehat{\mathfrak{f}}^{\perp }\right] ,%
\text{ \ \ \ \ \ }\widehat{\mathfrak{z}}=cent(\widehat{\mathfrak{f}}^{\perp
}),  \label{kernel algebra}
\end{equation}%
where $\widehat{\mathfrak{c}}$ and $\widehat{\mathfrak{z}}$ are the
commutant part and the center of $\mathfrak{\widehat{\mathfrak{f}}^{\perp }}$%
, respectively. The inner product in $\widehat{\mathfrak{f}}$ is to be
defined by\footnote{%
We assume the existence of a supermatrix representation for $\mathfrak{f.}$}%
\begin{equation}
\left\langle X,Y\right\rangle \equiv \doint \frac{dz}{2\pi i}\frac{1}{z}%
Str\left( X(z),Y(z)\right)  \label{simple inner}
\end{equation}%
and selects the term of zero $Q_{H}$ grade, i.e, $z^{0}$. Below, we will
show that in order to describe conserved quantities in the sigma model side
we have to twist this inner product by means of the Zukhovsky map.

The complex variable $z$ enter in the Lax operators as the spectral
parameter and it is important to notice that for the cases of interest we
have two possible superalgebra isomorphisms of $\widehat{\mathfrak{f}}$,
namely%
\begin{equation}
z^{\pm 2n}\text{ }\widehat{\mathfrak{f}}\simeq \widehat{\mathfrak{f}}\text{,
\ \ \ \ \ }z^{\pm 4n}\text{ }\widehat{\mathfrak{f}}\simeq \widehat{\mathfrak{%
f}},\text{ \ \ \ \ \ }n\in 
\mathbb{Z}
.  \label{isomorphisms}
\end{equation}%
The first isomorphism holds in the case of superstrings on $AdS_{3}\times
S^{3}$ because of the identifications $\mathfrak{f}_{0}=\mathfrak{f}_{2}$
and $\mathfrak{f}_{1}=\mathfrak{f}_{3}$ , while the other holds for the
cases of superstrings on $AdS_{2}\times S^{2}$ and $AdS_{5}\times S^{5}$ and
it is because the twisted nature of $\widehat{\mathfrak{f}},$ in contrast to
the KdV hierarchy, that the first Poisson structures of the SSSSG hierarchy
becomes non-local. See, section 3 below.

The phase space of the SSSSG integrable hierarchy and the symmetry flows of
the dynamical system are defined by intersecting the following two
co-adjoint orbits of the dressing groups $\left( \chi ,\gamma ^{-1}%
\widetilde{\chi }\right) ,$ namely $\Xi _{a}(\chi )=\Xi _{a}(\gamma ^{-1}%
\widetilde{\chi }),$ $a\in $ $\widehat{\mathfrak{f}}^{\perp },$ with%
\begin{equation}
\Xi _{a}(\chi )\equiv \mathcal{L}_{a}=\chi \left( \partial _{t_{a}}+a\right)
\chi ^{-1}\in \widehat{\mathfrak{f}}^{\ast },\text{ \ \ \ \ \ }\Xi
_{a}(\gamma ^{-1}\widetilde{\chi })\equiv \gamma ^{-1}\mathcal{L}%
_{a}^{\prime }\gamma =\gamma ^{-1}\widetilde{\chi }\left( \partial
_{t_{a}}+a\right) \widetilde{\chi }^{-1}\gamma \in \widehat{\mathfrak{f}}%
^{\ast },  \label{orbit intersection}
\end{equation}%
where $\gamma \in \exp \mathfrak{g},$ $a\in \widehat{\mathfrak{f}}^{\perp }$
is an element of $Q_{H}$ grade $n\in 
\mathbb{Z}
,$ $t_{a}$ is the time variable associated to $a$ and\footnote{%
The notation $\widehat{\mathfrak{f}}_{<n},\widehat{\mathfrak{f}}_{>n},%
\widehat{\mathfrak{f}}_{\leq n},\widehat{\mathfrak{f}}_{\geq n}$ stands for
an expansion in powers of the spectral parameter $z$ with grade $Q_{H}$ with
values $<n,>n,$ $\leq n$ and $\geq n.$}%
\begin{eqnarray}
\chi &=&\Phi \Omega \overline{u}^{-1}\in \exp \widehat{\mathfrak{f}}_{\leq
0},\text{ \ \ \ \ \ }\Phi (z)=e^{y(z)}\in \exp \widehat{\mathfrak{f}}%
_{<0}^{\parallel },\text{ \ \ \ \ \ }\Omega (z)=e^{\theta (z)}\in \exp 
\widehat{\mathfrak{f}}_{<0}^{\perp },\text{ \ \ \ \ \ }\overline{u}\in \exp 
\mathfrak{f}_{0}^{\perp },  \label{dressing matrices} \\
\widetilde{\chi } &=&\widetilde{\Phi }\widetilde{\Omega }u^{-1}\in \exp 
\widehat{\mathfrak{f}}_{\geq 0},\text{ \ \ \ \ \ }\widetilde{\Phi }(z)=e^{%
\widetilde{y}(z)}\in \exp \widehat{\mathfrak{f}}_{>0}^{\parallel },\text{ \
\ \ \ \ }\widetilde{\Omega }(z)=e^{\widetilde{\theta }(z)}\in \exp \widehat{%
\mathfrak{f}}_{>0}^{\perp },\text{ \ \ \ \ \ }u\in \exp \mathfrak{f}%
_{0}^{\perp },  \notag
\end{eqnarray}%
are the dressing matrices. We identify $\widehat{\mathfrak{f}}\sim \widehat{%
\mathfrak{f}}^{\ast }$ under the inner product (\ref{simple inner}).

The world-sheet light-cone coordinates are associated to the first two
isospectral times $t_{z^{\pm 2}\Lambda }=x^{\pm },$ $z^{\pm 2}\Lambda \in 
\widehat{\mathfrak{z}}$ leading to the following Lax operators\footnote{%
In what follows we use $x^{\pm }=t\pm x$ and $a_{\pm }=\frac{1}{2}(a_{0}\pm
a_{1}).$} 
\begin{equation}
\mathcal{L}_{\pm }=\chi \left( \partial _{\pm }-z^{\pm 2}\Lambda \right)
\chi ^{-1}=\gamma ^{-1}\mathcal{L}_{\pm }^{\prime }\gamma =\gamma ^{-1}%
\widetilde{\chi }\left( \partial _{\pm }-z^{\pm 2}\Lambda \right) \widetilde{%
\chi }^{-1}\gamma .  \label{light-cone lax}
\end{equation}%
The symmetries of the system are introduced through the field variations
induced by the trivial relations $\left[ \mathcal{L}_{a},\mathcal{L}_{\pm }%
\right] =0,$ $a\in \widehat{\mathfrak{c}}$ and are described by the
non-Abelian times $t_{a}.$ As a consequence, the symmetry variations, when $%
u=\overline{u}=I$, form a non-Abelian algebra of flows $\widehat{\mathfrak{S}%
}$ isomorphic to $\widehat{\mathfrak{f}}^{\perp }$, namely $a\in \widehat{%
\mathfrak{f}}^{\perp }\rightarrow \partial _{t_{a}}\equiv \delta _{a}\in 
\widehat{\mathfrak{S}}$ with 
\begin{equation}
\left[ \delta _{a},\delta _{a^{\prime }}\right] (\ast )=\delta _{\left[
a,a^{\prime }\right] }(\ast ).  \label{homomorphism}
\end{equation}

The affine algebra $\widehat{\mathfrak{f}}^{\perp }$ is infinite dimensional
but includes a very special finite dimensional sub-superalgebra \cite{yo03},%
\cite{SSSSG AdS5xS5} $\widehat{\mathfrak{s}}\subset \widehat{\mathfrak{f}}%
^{\perp },$ which is spanned by%
\begin{equation}
\widehat{\mathfrak{s}}=\left( z^{-1}\mathfrak{f}_{3}^{\perp }\text{ }%
\mathfrak{\oplus }\text{ }\mathfrak{h}\text{ }\mathfrak{\oplus }\text{ }z%
\mathfrak{f}_{1}^{\perp }\right) \ltimes 
\mathbb{R}
^{2},\text{ \ \ \ \ \ }%
\mathbb{R}
^{2}=z^{2}\Lambda \oplus z^{-2}\Lambda  \label{s hat}
\end{equation}%
and under (\ref{homomorphism}) turns out to be isomorphic to the following
double central extended superalgebra $\mathfrak{s}\simeq \widehat{\mathfrak{s%
}}$ with\footnote{%
Actually, the supersymmetry algebra of the reduced models is $\mathfrak{s0(}%
1,1\mathfrak{)\rtimes }\left( \mathfrak{a}\ltimes 
\mathbb{R}
^{2}\right) ,$ where $\mathfrak{\rtimes }$ denotes semi-direct sum. We are
dropping the Lorentz group $\mathfrak{s0(}1,1\mathfrak{)}$ in $\mathfrak{s}$
because we are also dropping the grading operator $Q_{H}$ in $\widehat{%
\mathfrak{s}},$ which is its equivalent.} 
\begin{equation}
\mathfrak{s}=\mathfrak{a}\ltimes 
\mathbb{R}
^{2},\text{ \ \ \ \ \ }\mathfrak{a=h\oplus f}_{1}^{\perp }\oplus \mathfrak{f}%
_{3}^{\perp }\text{, \ \ \ \ \ }%
\mathbb{R}
^{2}=\partial _{+}\oplus \partial _{-}.  \label{lie sub-superalgebra}
\end{equation}%
In particular, for $AdS_{n}\times S^{n}$ with $n=2,3,5$ we have,
respectively,%
\begin{equation*}
\mathfrak{a}:\text{ \ }\mathfrak{psu(}1\mid 1\mathfrak{)}^{\oplus 2},\text{
\ \ \ \ \ }\left( \mathfrak{u(}1\mathfrak{)\rtimes psu(}1\mid 1\mathfrak{)}%
^{\oplus 2}\right) ^{\oplus 2}\ltimes \mathfrak{u(}1\mathfrak{)}\text{\ },%
\text{ \ \ \ \ \ }\mathfrak{psu(}2\mid 2\mathfrak{)}^{\oplus 2}\mathfrak{.}
\end{equation*}%
Recently \cite{Hoare-Tseytlin},\cite{Relat magnon S-Matrix} the $q$%
-deformation $\mathfrak{s}\rightarrow \mathfrak{s}_{q},$ with $q=q(k),$ $k$
the level of the WZNW model, have been identified as the\ deformed
supersymmetries of the quantum S-matrix of the Pohlmeyer reduced models.
Needless to say, these superalgebras are of the extended type $(N,N)$ with $%
N=2,4$ and $8$ with R-symmetry groups $\varnothing ,$ $U(1)^{\times 2}$ and $%
SU(2)^{\times 4},$ respectively.

The pre-potentials $u$ and $\overline{u}$ dress the $\partial _{\pm }$
derivatives into the covariant derivatives%
\begin{equation*}
D_{\pm }^{(l)}(\ast )=\partial _{\pm }(\ast )+\left[ A_{\pm }^{(l)},\ast %
\right] ,\text{ \ \ \ \ \ }D_{\pm }^{(r)}(\ast )=\partial _{\pm }(\ast )+%
\left[ A_{\pm }^{(r)},\ast \right] ,
\end{equation*}%
where $A_{\pm }^{(l)}=u^{-1}\partial _{\pm }u,$ $A_{\pm }^{(r)}=\overline{u}%
^{-1}\partial _{\pm }\overline{u}$ and the equations (\ref{light-cone lax})
imply the following form for the Lax operators\footnote{%
The relation with the notation of \cite{Grigo-Tseytlin} is $\psi _{+}=\Psi
_{R},$ $\psi _{-}=\Psi _{L}$ and $\Lambda =-T.$ The mass scale $\mu $ is
taken as 1.}%
\begin{eqnarray}
\mathcal{L}_{+} &=&\partial _{+}+\gamma ^{-1}\partial _{+}\gamma +\gamma
^{-1}A_{+}^{(l)}\gamma +z\psi _{+}-z^{2}\Lambda ,
\label{Lax op with gauge fields} \\
\mathcal{L}_{-} &=&\partial _{-}+A_{-}^{(r)}+z^{-1}\gamma ^{-1}\psi
_{-}\gamma -z^{-2}\gamma ^{-1}\Lambda \gamma  \notag
\end{eqnarray}%
together with the constraints%
\begin{equation}
A_{+}^{(r)}=\left( \gamma ^{-1}\partial _{+}\gamma +\gamma
^{-1}A_{+}^{(l)}\gamma \right) ^{\perp }+2\Lambda \psi _{+}^{2},\text{ \ \ \
\ \ \ }A_{-}^{(l)}=\left( -\partial _{-}\gamma \gamma ^{-1}+\gamma
A_{-}^{(r)}\gamma ^{-1}\right) ^{\perp }+2\Lambda \psi _{-}^{2},
\label{constraints}
\end{equation}%
where we have used the definitions 
\begin{equation*}
\psi _{\pm }=\left[ \Lambda ,y_{\mp 1}\right] ,\text{ \ \ \ \ }\left[
\Lambda ,y_{-2}\right] =\left( \gamma ^{-1}\partial _{+}\gamma +\gamma
^{-1}A_{+}^{(l)}\gamma \right) ^{\parallel },\text{ \ \ \ \ }\left[ \Lambda
,y_{+2}\right] =\left( -\partial _{-}\gamma \gamma ^{-1}+\gamma
A_{-}^{(r)}\gamma ^{-1}\right) ^{\parallel }.
\end{equation*}

The $H_{L}\times H_{R}$ gauge transformations are implemented by 
\begin{eqnarray}
u &\rightarrow &uh_{l}^{-1},\text{ \ \ \ \ \ }\overline{u}\rightarrow \text{%
\ }\overline{u}h_{r}^{-1},\text{ \ \ \ \ \ }\gamma \rightarrow h_{l}\gamma
h_{r}^{-1},\text{ \ \ \ \ \ }\psi _{+}\rightarrow h_{r}\psi _{+}h_{r}^{-1},%
\text{ \ \ \ \ \ }\psi _{-}\rightarrow h_{l}\psi _{-}h_{l}^{-1}
\label{full gauge transf.} \\
A_{+}^{(l)} &\rightarrow &h_{l}A_{+}^{(l)}h_{l}^{-1}-\partial
_{+}h_{l}h_{l}^{-1},\text{ \ \ \ \ \ }A_{-}^{(r)}\rightarrow
h_{r}A_{-}^{(r)}h_{r}^{-1}-\partial _{-}h_{r}h_{r}^{-1}  \notag
\end{eqnarray}%
and the curvature components $F_{+-}=\left[ \mathcal{L}_{+},\mathcal{L}_{-}%
\right] =z^{-1}F_{+-}^{(-1)}+F_{+-}^{(0)}+zF_{+-}^{(1)}$ are given by%
\begin{eqnarray}
F_{+-}^{(0)} &=&-D_{-}^{(r)}\left( \gamma ^{-1}D_{+}^{(l)}\gamma \right) + 
\left[ D_{+}^{(l)},D_{-}^{(r)}\right] +\left[ \Lambda ,\gamma ^{-1}\Lambda
\gamma \right] +\left[ \psi _{+},\gamma ^{-1}\psi _{-}\gamma \right] ,
\label{equations of motion} \\
F_{+-}^{(-1)} &=&-D_{-}^{(r)}\psi _{+}-\left[ \Lambda ,\gamma ^{-1}\psi
_{-}\gamma \right] ,\text{ \ \ \ \ \ }F_{+-}^{(1)}=\gamma ^{-1}\left(
D_{+}^{(l)}\psi _{-}+\left[ \Lambda ,\gamma \psi _{+}\gamma ^{-1}\right]
\right) \gamma .  \notag
\end{eqnarray}%
The equations of motion are defined by $F_{+-}=0.$

An infinite tower of local and non-local, bosonic and fermionic conserved
charges $q_{s}$ of 2d Lorentz spin $s/2$ are hidden in the Lax operators (%
\ref{Lax op with gauge fields}) and to extract them we employ the so-called
Drinfeld-Sokolov (DS) procedure \cite{Generalized DSI}. For simplicity we
will restrict to the on-shell gauge $u=\overline{u}=I$ and introduce the
following coordinates for the phase space $\mathcal{P}$%
\begin{equation}
\mathcal{P}:\text{ \ }\left( Q,\overline{Q}\right) ,\text{ \ \ \ }Q=(q,\psi
),\text{ \ \ \ }\overline{Q}=(\overline{q},\overline{\psi }),\text{ \ \ \ }%
q\equiv \gamma ^{-1}\partial _{+}\gamma ,\text{ \ \ \ }\overline{q}\equiv
-\partial _{-}\gamma \gamma ^{-1},\text{ \ \ \ }\psi \equiv \psi _{+},\text{
\ \ \ }\overline{\psi }\equiv \psi _{-}.  \label{phase coordinates}
\end{equation}

The positive/negative spin $s/2$ charges $q_{-s}/q_{s},$ $s\in 
\mathbb{Z}
^{+}$ are obtained from (\ref{light-cone lax}) and appear in the expansion
of the subtracted monodromy matrix $\mathcal{M}(z)$ around $z=0$ and $%
z=\infty $ as follows \cite{SSSSG AdS5xS5}%
\begin{equation}
\mathcal{M}(z)=\exp \left[ q_{0}+q_{1}z+q_{2}z^{2}+...\right] =\exp \left[
q_{-1}/z+q_{-2}/z^{2}+...\right] .  \label{monodromy}
\end{equation}%
The charges $q_{-s}$ are computed from 
\begin{equation}
\Phi ^{-1}(z)\mathcal{L}_{+}(z)\Phi (z)=\partial _{+}-z^{2}\Lambda +h_{+}(z),%
\text{ \ \ \ \ \ }\Phi ^{-1}(z)\mathcal{L}_{-}(z)\Phi (z)=\partial
_{-}+h_{-}(z),\text{ \ \ \ \ \ }h_{\pm }(z)\in \widehat{\mathfrak{f}}_{\leq
0}^{\perp },  \label{DSI}
\end{equation}%
while the charges $q_{s}$ from 
\begin{equation}
\widetilde{\Phi }^{-1}(z)\mathcal{L}_{-}^{\prime }(z)\widetilde{\Phi }%
(z)=\partial _{-}-z^{-2}\Lambda +\widetilde{h}_{-}(z),\text{ \ \ \ \ \ }%
\widetilde{\Phi }^{-1}(z)\mathcal{L}_{+}^{\prime }(z)\widetilde{\Phi }%
(z)=\partial _{+}+\widetilde{h}_{+}(z),\text{ \ \ \ \ \ }\widetilde{h}_{\pm
}(z)\in \widehat{\mathfrak{f}}_{\geq 0}^{\perp }.  \label{DSII}
\end{equation}

The equations of motion $F_{+-}=0$ imply the following relations%
\begin{equation}
\partial _{+}h_{-}(z)-\partial _{-}h_{+}(z)+\left[ h_{+}(z),h_{-}(z)\right]
=0,\text{ \ \ \ \ \ }\partial _{+}\widetilde{h}_{-}(z)-\partial _{-}%
\widetilde{h}_{+}(z)+\left[ \widetilde{h}_{+}(z),\widetilde{h}_{-}(z)\right]
=0  \label{DS zero curvature}
\end{equation}%
and provide conservations laws for the dynamical system. Due to the fact
that the current components $h(z)$ and $\widetilde{h}(z)$ are related by the
parity transformations $z\rightarrow z^{-1},$ $+\rightarrow -,$ $\gamma
\rightarrow \gamma ^{-1},$ we need to find only one set of these charges.
Another important ingredient of this construction is that we still have the
action of an infinite dimensional group of transformations 
\begin{equation}
\Phi \rightarrow \Phi \eta _{-},\text{ \ \ \ \ \ }\widetilde{\Phi }%
\rightarrow \widetilde{\Phi }\eta _{+},\text{ \ \ \ \ \ }\eta _{\pm }=\exp
\beta _{\pm },\text{ \ \ \ \ \ }\beta _{-},\beta _{+}\in \widehat{\mathfrak{f%
}}_{<0}^{\perp },\widehat{\mathfrak{f}}_{>0}^{\perp },
\label{infinite ambiguity}
\end{equation}%
that does not change the lhs of (\ref{DSI}) and (\ref{DSII}) but changes the
form of the DS currents in the rhs and this is equivalent to a change in the
dressing matrices $\Omega ,\widetilde{\Omega }$ in (\ref{dressing matrices}%
). The change in the currents induced by (\ref{infinite ambiguity}) is 
\begin{equation}
h_{\pm }^{\prime }=\eta _{-}^{-1}h_{\pm }\eta _{-}+\eta _{-}^{-1}\partial
_{\pm }\eta _{-},\text{ \ \ \ \ \ }\widetilde{h}_{\pm }^{\prime }=\eta
_{+}^{-1}\widetilde{h}_{\pm }\eta _{+}+\eta _{+}^{-1}\partial _{\pm }\eta
_{+}.  \label{kernel orbit}
\end{equation}

We are interested in finding the Poisson superalgebra for the conserved
charges associated to the symmetry superalgebra $\mathfrak{s,}$ see (\ref%
{lie sub-superalgebra}) and for this reason we only need to decompose (\ref%
{DS zero curvature}) at $Q_{H}$ grades $0,\pm 1$, $\pm 2$ to find $q_{l/r},$ 
$q_{\pm 1}$ and $q_{\pm 2}.$ The answer is \cite{yo03},\cite{SSSSG AdS5xS5}%
\begin{eqnarray}
q_{r} &=&q^{\perp }+2\Lambda \psi ^{2}=0,\text{ \ \ }q_{-1}=\dint%
\nolimits_{-\infty }^{+\infty }dx\left( \left[ \widetilde{\Lambda }q,\psi %
\right] -\left( \gamma ^{-1}\overline{\psi }\gamma \right) ^{\perp }\right) ,%
\text{ \ \ \ }Str\left( \Lambda \text{,}q_{-2}\right)
=\dint\nolimits_{-\infty }^{+\infty }dx\left( T_{++}+T_{-+}\right) ,
\label{DS charges} \\
q_{l} &=&\overline{q}^{\perp }+2\Lambda \overline{\psi }^{2}=0,\text{ \ \ }%
q_{+1}=\dint\nolimits_{-\infty }^{+\infty }dx\left( \left[ \widetilde{%
\Lambda }\overline{q},\overline{\psi }\right] -\left( \gamma \psi \gamma
^{-1}\right) ^{\perp }\right) ,\text{ \ \ \ }Str\left( \Lambda \text{,}%
q_{+2}\right) =\dint\nolimits_{-\infty }^{+\infty }dx\left(
T_{--}+T_{+-}\right) ,  \notag
\end{eqnarray}%
where we have used the notation $\widetilde{\Lambda }=ad\Lambda $ and the
definitions (\ref{phase coordinates}). The explicit form for the components $%
T_{\mu \nu }$ will be written below. One important comment concerning the
first equations $q_{l/r}$ is in order. At grade zero, the equations (\ref{DS
zero curvature}) are $\partial _{-}q_{r}=\partial _{+}q_{l}=0,$ implying
that the quantities $q_{r}=q_{r}(x^{+})$ and $q_{l}=q_{l}(x^{-})$ are
chiral, but these are precisely the constraints (\ref{constraints}) in the
gauge\footnote{%
In this gauge the symmetry $H_{L}\times H_{R}$ is reduced from (\ref{full
gauge transf.}) to the chiral Kac-Moody symmetry of the fermionic extension
of perturbed WZNW model.} $u=\overline{u}=I.$ Thus, they vanish and they are
not the true gauge charges. However, this constraints have important
consequences not only in dictating the explicit form of $q_{\pm 1},q_{\pm 2}$
but also in the geometric interpretation of $q_{\pm 1},q_{\pm 2}$ as moment
maps, see section 6 below. The true gauge charge is the $q_{0}$ appearing in
(\ref{monodromy}) and it cannot be found by the DS procedure, it is of
kink-type and is given \cite{SSSSG AdS5xS5} by $\exp q_{0}=\gamma (\infty
)^{-1}\gamma (-\infty ).$

In what follows we split the elements $a\in \widehat{\mathfrak{f}}^{\perp }$
into two groups, $b\in \widehat{\mathfrak{f}}_{\geq 0}^{\perp }$ and $%
\overline{b}\in \widehat{\mathfrak{f}}_{\leq 0}^{\perp }$ in order to
separate the two infinite sets of symmetry flows. The equations (\ref%
{light-cone lax}) allows us to write (\ref{Lax op with gauge fields}) in the
following form%
\begin{equation}
\mathcal{L}_{+}=\partial _{+}-A(z^{2}\Lambda )_{\geq 0},\text{ \ \ \ \ \ }%
\mathcal{L}_{-}=\partial _{-}-\gamma ^{-1}\widetilde{A}(z^{-2}\Lambda
)_{<0}\gamma ,\text{ \ \ \ \ \ }A(b)\equiv \Phi b\Phi ^{-1},\text{ \ \ \ \ \ 
}\widetilde{A}(\overline{b})\equiv \widetilde{\Phi }\overline{b}\widetilde{%
\Phi }^{-1},  \label{A and A tilda}
\end{equation}%
which are manifestly invariant under the action of (\ref{infinite ambiguity}%
). We can also show that

\begin{equation}
\mathcal{L}_{b}=\partial _{t_{b}}+\left( \chi b\chi ^{-1}\right) _{\geq 0},%
\text{ \ \ \ \ }\mathcal{L}_{\overline{b}}=\gamma ^{-1}\left( \partial _{t_{%
\overline{b}}}+\left( \widetilde{\chi }\overline{b}\widetilde{\chi }%
^{-1}\right) _{\leq 0}\right) \gamma .  \label{NA Lax}
\end{equation}

To end this section we recall how to compute the differentials for the
current functionals $h(z)$ and $\widetilde{h}(z)$ on the co-adjoint orbits $%
\mathcal{L}_{\pm }$ and $\mathcal{L}_{\pm }^{\prime }.$ For simplicity, we
consider $h_{+}$ only$.$

Now that we have introduced the space-time coordinates $x^{\pm },$ define
the following integrated inner product%
\begin{equation}
\left( X,Y\right) ^{\pm }\equiv \dint_{-\infty }^{+\infty }dx^{\pm
}\left\langle X,Y\right\rangle ,  \label{integrated inner product}
\end{equation}%
which we will used extensively in what follows\footnote{%
When it is clear from the context, we will drop the $\pm $ signs in the $%
\left( X,Y\right) ^{\pm }$ integrations.}.

Consider the Hamiltonian $H_{b}\equiv \left( b,h_{+}(z)\right) $ associated
to the flow $\partial _{t_{b}},$ for some element $b\in \widehat{\mathfrak{f}%
}^{\perp }$ of positive $Q_{H}$ grade$.$ The differentials are defined
through the usual relation

\begin{equation*}
\frac{d}{d\varepsilon }H_{b}\left[ \mathcal{L}_{+}+\varepsilon r_{+}\right]
\mid _{\varepsilon =0}\equiv \left( d_{Q}H_{b},r_{+}\right) ,\text{ \ \ \ \ }%
r_{+}=cte\in \widehat{\mathfrak{f}}_{\geq 0},\text{ \ \ \ \ \ \ }%
d_{Q}H_{b}\in \widehat{\mathfrak{f}}\text{ mod }\widehat{\mathfrak{f}}_{>0}.
\end{equation*}%
Explicitly, we have that

\begin{equation}
\frac{d}{d\varepsilon }H_{b}\left[ \mathcal{L}_{+}+\varepsilon r_{+}\right]
\mid _{\varepsilon =0}=\left( b,\frac{d}{d\varepsilon }\mathfrak{L}%
_{+}^{\varepsilon }\right) \mid _{\varepsilon =0}=\left( \Phi b\Phi
^{-1},r_{+}\right) +\left( b,\left[ \mathfrak{L}_{+}\mathfrak{,}T_{y}\right]
\right) ,  \label{differential}
\end{equation}%
where $\mathfrak{L}_{+}^{\varepsilon }\equiv \Phi ^{-1}(\varepsilon )%
\mathcal{L}_{+}(\varepsilon )\Phi (\varepsilon ),$ $\mathfrak{L}_{+}\equiv 
\mathfrak{L}_{+}^{\varepsilon =0},$ $T_{y}\equiv \Phi ^{-1}\widehat{y}%
(z)\Phi $ and $\widehat{y}(z)\equiv \frac{d}{d\varepsilon }y(\varepsilon
)\mid _{\varepsilon =0}.$ When the second term in the rhs of (\ref%
{differential}) vanishes, which is valid for $b\in \mathfrak{z}$ and $%
b=z\epsilon $, $\epsilon \in \mathfrak{f}_{1}^{\perp }$ , the differential
of $H_{b}$ is given by%
\begin{equation*}
d_{Q}H_{b}=A(b)_{\leq 0}=d_{q}H_{b}+z^{-1}d_{\psi }H_{b},
\end{equation*}%
where we have defined%
\begin{equation}
d_{q}H_{b}=\frac{\delta H_{b}}{\delta q}\equiv A(b)_{0},\text{ \ \ \ \ \ }%
d_{\psi }H_{b}=\frac{\delta H_{b}}{\delta \psi }\equiv zA(b)_{-1}.
\label{positive differentials}
\end{equation}%
Similar results holds also for $h_{-}$ and $\widetilde{h}_{\pm }$ and will
be written later. As shown above, the Lax operators $\mathcal{L}_{\pm }$ are
invariant under the action of (\ref{infinite ambiguity}) and its effect on
the differentials $d_{Q}H_{b}$ is simply a conjugation $b\rightarrow \eta
_{-}b\eta _{-}^{-1}.$

\section{Recursion relations and the SSSSG Poisson Structures.}

In this section we show how construct recursively an infinite number of
Poisson structures for the SSSSG integrable hierarchy\footnote{%
In \cite{yo03}, this hierarchy was named extended homogeneous integrable
hierarchy. A better name would be mKdV/SSSSG or SSSSG for short.}. It turns
out that due to the twisted nature of the superalgebra (\ref{superalgebra}),
all Poisson structures are non-local except one which is precisely the
canonical structure associated to the fermionic extension of the WZNW model
having (\ref{equations of motion}) as equations of motion (in the gauge $u=%
\overline{u}=I$). For simplicity, we will consider only the positive
symmetry flows $\partial _{t_{b}}\mathcal{L}_{+}$. The analysis for any
other combination follows exactly the same lines as a consequence of (\ref{A
and A tilda}) and (\ref{NA Lax}).

Taking $b\in $ $\widehat{\mathfrak{f}}_{\geq 0}^{\perp }$ and noting that
the spectral parameter $z$ has $Q_{H}$ grade +1, we have from (\ref{A and A
tilda}) the following recursion relations%
\begin{equation}
A\left( b\right) _{n-1}=z^{-1}A(zb)_{n},\text{ \ \ \ \ \ }A\left( b\right)
_{n-2}=z^{-2}A(z^{2}b)_{n},\text{ \ \ \ \ \ }A\left( b\right)
_{n-4}=z^{-4}A(z^{4}b)_{n},  \label{recursion}
\end{equation}%
where we have used the first relation twice and fourth times in order to get
the second and third relations. When the underlying affine algebra $\widehat{%
\mathfrak{f}}$ is untwisted we have that $z\widehat{\mathfrak{f}}\simeq 
\widehat{\mathfrak{f}}$ and $b,$ $zb\in \widehat{\mathfrak{f}}.$ In this
case the first relation leads to the well known local first Poisson
structure of the KdV hierarchy and for this reason will not be considered
here any further. For the details of its construction in the bosonic limit,
the reader is refered to \cite{generalized DS II}. The second relation is to
be used when $\mathfrak{f}_{0}=\mathfrak{f}_{2}$ and $\mathfrak{f}_{1}=%
\mathfrak{f}_{3}$ implying that $z^{2}\widehat{\mathfrak{f}}\simeq \widehat{%
\mathfrak{f}},$ corresponding to the GS superstring in $AdS_{3}\times S^{3}.$
The third relation is to be used when $z^{4}\widehat{\mathfrak{f}}\simeq 
\widehat{\mathfrak{f}},$ corresponding to the GS superstring in $%
AdS_{n}\times S^{n}$ with $n=2,5.$

We now proceed to construct the non-local Poisson structures associated to
the twisted superalgebra $\widehat{\mathfrak{f}},$ i.e (\ref{superalgebra}).

\subsection{Second Poisson structure.}

The second structure is canonical, it is the most natural and it is the same
in all cases, then we consider it first. Taking an element $b\in \mathfrak{z}
$ of positive $Q_{H}$ grade and dressing the trivial relation $\left[
b,\partial _{+}-z^{2}\Lambda \right] =0,$ we have the following
compatibility relation%
\begin{equation}
\left[ \mathcal{L}_{+},A(b)\right] =0  \label{compatibility}
\end{equation}%
allowing to represent the flow $\partial _{t_{b}}\mathcal{L}_{+},$ using (%
\ref{NA Lax}), in two equivalent forms

\begin{equation*}
\frac{\partial \mathcal{L}_{+}}{\partial t_{b}}=-\left[ A(b)_{\geq 0},%
\mathcal{L}_{+}\right] =\left[ A(b)_{<0},\mathcal{L}_{+}\right] .
\end{equation*}

The first form is equivalent to%
\begin{equation}
\frac{\partial q}{\partial t_{b}}=D_{+}d_{q}H_{b},\text{ \ \ \ \ \ }z\frac{%
\partial \psi }{\partial t_{b}}=D_{+}A(b)_{1}+z\widetilde{\psi }d_{q}H_{b},
\label{second form}
\end{equation}%
where we have denoted $\widetilde{\psi }=ad\psi $ and defined $D_{+}(\ast )=%
\left[ \partial _{+}+q,\ast \right] .$ The recursion relation induced by (%
\ref{compatibility}) is 
\begin{equation}
D_{+}A(b)_{n}=-z\widetilde{\psi }A(b)_{n-1}+z^{2}\widetilde{\Lambda }%
A(b)_{n-2}  \label{local recursion}
\end{equation}%
and can be written in two different but equivalent ways%
\begin{eqnarray}
A(b)_{n} &=&zuA(b)_{n-1}+z^{2}vA(b)_{n-2},\text{ \ \ \ \ \ }u\equiv
-D_{+}^{-1}\widetilde{\psi },\text{ \ \ \ \ \ }v\equiv D_{+}^{-1}\widetilde{%
\Lambda },  \label{standard recursion} \\
A(b)_{n} &=&z^{-1}wA(b)_{n+1}+z^{-2}yA(b)_{n+2},\text{ \ }w\equiv \widetilde{%
\Lambda }^{-1}\widetilde{\psi },\text{ \ \ \ \ \ \ }y\equiv \widetilde{%
\Lambda }^{-1}D_{+}.  \notag
\end{eqnarray}

Using (\ref{local recursion}) and (\ref{positive differentials}) with $n=1,$
we have 
\begin{equation*}
D_{+}A(b)_{1}=-z\widetilde{\psi }d_{q}H_{b}+z\widetilde{\Lambda }d_{\psi
}H_{b}
\end{equation*}%
and inserting this result in the second equation in (\ref{second form}) we
have\footnote{%
In what follows we will denote row and column vectors by the same letter.}%
\begin{equation}
\frac{\partial Q}{\partial t_{b}}=\Theta ^{(0)}d_{Q}H_{b},\text{ \ \ \ \ \ }%
\Theta ^{(0)}\equiv \left( 
\begin{array}{cc}
D_{+} & 0 \\ 
0 & \widetilde{\Lambda }%
\end{array}%
\right) ,\text{ \ \ \ \ \ }d_{Q}H_{b}=\left( 
\begin{array}{c}
d_{q}H_{b} \\ 
d_{\psi }H_{b}%
\end{array}%
\right)  \label{teta 0}
\end{equation}%
where $Q$ was defined in (\ref{phase coordinates}). The flow of any
functional $\varphi (Q)$ is given by%
\begin{equation*}
\frac{\partial \varphi (Q)}{\partial t_{b}}=\left( d_{q}\varphi ,\frac{%
\partial q}{\partial t_{b}}\right) +\left( d_{\psi }\varphi ,\frac{\partial
\psi }{\partial t_{b}}\right) =\left( d_{q}\varphi ,D_{+}d_{q}H_{b}\right)
+\left( d_{\psi }\varphi ,\text{\ }\widetilde{\Lambda }d_{\psi }H_{b}\right)
\end{equation*}%
and motivates the following definition for the second Poisson bracket%
\begin{equation}
\left\{ \varphi ,\psi \right\} _{2}\equiv \left( d_{Q}\varphi ,\text{\ }%
\Theta ^{(0)}d_{Q}\psi \right) =\dint\nolimits_{-\infty }^{+\infty
}dx^{+}Str\left( \frac{\delta \varphi }{\delta Q},\Theta ^{(0)}\frac{\delta
\psi }{\delta Q}\right)  \label{second bracket}
\end{equation}%
in terms of which%
\begin{equation*}
\frac{\partial \varphi }{\partial t_{b}}=\left\{ \varphi ,H_{b}\right\} _{2}.
\end{equation*}

This is the boost-invariant Poisson structure already considered in \cite%
{yo03}. Note that in the whole derivation we did not used the relations (\ref%
{recursion}) at all, hence the second bracket (\ref{second bracket}) is the
same for the two possible situations of interest, cf (\ref{isomorphisms}).

To compute the Poisson superalgebra corresponding to $\mathfrak{s}$ in (\ref%
{lie sub-superalgebra}) we will need the light-cone components of the
conserved currents $h(z)$, $\widetilde{h}(z)$ associated to the elements in $%
\mathfrak{s.}$ They are $b=z\epsilon ,$ $b=z^{2}\Lambda ,$ with $\epsilon
\in \mathfrak{f}_{1}^{\perp }$ and $\overline{b}=z^{-1}\overline{\epsilon },$
$\overline{b}=z^{-2}\Lambda $ with $\overline{\epsilon }\in \mathfrak{f}%
_{3}^{\perp }.$ The DS current components take the form

\begin{equation}
j_{+}(b)\equiv \left\langle A(b),\mathcal{L}_{+}\right\rangle ,\text{ \ \ \
\ }j_{-}(b)\equiv -\left\langle A(b),\mathcal{L}_{-}\right\rangle ,\text{ \
\ \ \ }\overline{j}_{+}(\overline{b})\equiv -\left\langle \widetilde{A}(%
\overline{b}),\mathcal{L}_{+}^{\prime }\right\rangle ,\text{\ \ \ \ \ }%
\overline{j}_{-}(\overline{b})\equiv \left\langle \widetilde{A}(\overline{b}%
),\mathcal{L}_{-}^{\prime }\right\rangle  \label{DS currents}
\end{equation}%
and their associated differentials\footnote{%
For the differentials we have only two terms in the expansions, $(\ast
)_{\leq 0}=(\ast )_{0}+(\ast )_{-1}$ and $(\ast )_{\geq 0}=(\ast )_{0}+(\ast
)_{+1}.$} are%
\begin{eqnarray}
d_{+}j_{+}(b) &=&A(b)_{\leq 0},\text{ \ \ \ \ \ \ \ \ \ }d_{+}^{\prime
}j_{+}(b)=\gamma d_{+}j_{+}(b)\gamma ^{-1},  \label{differentials} \\
\text{\ }d_{-}j_{-}(b) &=&-A(b)_{\geq 0},\text{ \ \ \ \ \ \ \ }d_{-}^{\prime
}j_{-}(b)=\gamma d_{-}j_{-}(b)\gamma ^{-1},  \notag \\
d_{+}^{\prime }\overline{j}_{+}(\overline{b}) &=&-\widetilde{A}(\overline{b}%
)_{\leq 0},\text{ \ \ \ \ \ }d_{+}\overline{j}_{+}(\overline{b})=\gamma
^{-1}d_{+}^{\prime }\overline{j}_{+}(\overline{b})\gamma ,  \notag \\
d_{-}^{\prime }\overline{j}_{-}(\overline{b}) &=&\widetilde{A}(\overline{b}%
)_{\geq 0},\text{ \ \ \ \ \ \ \ \ }d_{-}\overline{j}_{-}(\overline{b}%
)=\gamma ^{-1}d_{-}^{\prime }\overline{j}_{-}(\overline{b})\gamma .  \notag
\end{eqnarray}%
The notation $d_{\pm }$ and $d_{\pm }^{\prime }$ helps to keep track the
domains of definitions in which the current components are defined.

The light-cone second brackets in the Kostant-Kirillov form are \cite{yo03}%
\begin{eqnarray}
\left\{ \varphi ,\psi \right\} _{2}(\mathcal{L}_{+}) &=&-\left( \mathcal{L}%
_{+},\left[ d_{+}\varphi ,d_{+}\psi \right] _{R_{-}}\right) ,\text{ \ \ \ \
\ \ \ \ \ \ \ \ }\left\{ \varphi ,\psi \right\} _{2}(\mathcal{L}_{-})\text{ }%
=\text{ }-\left( \mathcal{L}_{-},\gamma ^{-1}\left[ d_{-}^{\prime }\varphi
,d_{-}^{\prime }\psi \right] _{R_{+}}\gamma \right) ,
\label{second brackets} \\
\left\{ \varphi ,\psi \right\} _{2}(\mathcal{L}_{+}^{\prime }) &=&-\left( 
\mathcal{L}_{+}^{\prime },\gamma \left[ d_{+}\varphi ,d_{+}\psi \right]
_{R_{-}}\gamma ^{-1}\right) ,\text{ \ \ \ \ \ }\left\{ \varphi ,\psi
\right\} _{2}(\mathcal{L}_{-}^{\prime })\text{ }=\text{ }-\left( \mathcal{L}%
_{-}^{\prime },\left[ d_{-}^{\prime }\varphi ,d_{-}^{\prime }\psi \right]
_{R_{+}}\right) ,  \notag
\end{eqnarray}%
where $R_{-}\equiv \left( \mathcal{\pi }_{0}\mathcal{-\pi }_{<0}\right) /2$
and $R_{+}\equiv \left( \mathcal{\pi }_{0}\mathcal{-\pi }_{>0}\right) /2$
are the ususal $R$-matrices defined in terms of projectors $\pi $. The
Poisson bracket on the spatial orbit $\mathcal{L}_{x}\mathcal{=L}_{+}-%
\mathcal{L}_{-}$ is defined by

\begin{equation}
\left\{ \varphi ,\psi \right\} _{2}(\mathcal{L}_{x})=\left\{ \varphi ,\psi
\right\} _{2}(\mathcal{L}_{+})-\left\{ \varphi ,\psi \right\} _{2}(\mathcal{L%
}_{-})  \label{spatial orbit}
\end{equation}%
and a similar definition holds for $\left\{ \varphi ,\psi \right\} _{2}(%
\mathcal{L}_{x}^{\prime }).$ All these brackets have to be restricted to the
level sets\footnote{%
These conditions are automatically satisfied by the soliton solutions, see 
\cite{SSSSG AdS5xS5}.}$q_{l/r}=0$ consequence of (\ref{DS charges}) 
\begin{equation}
q^{\perp }+2\Lambda \psi ^{2}=0,\text{ \ \ \ \ }\overline{q}^{\perp
}+2\Lambda \overline{\psi }^{2}=0.\text{\ }  \label{level sets}
\end{equation}

\subsection{First Poisson structures.}

The evolution equations (\ref{second form}) can be written in two different
but completely equivalent ways. By using the two equations of (\ref{standard
recursion}) in (\ref{second form}), we have%
\begin{equation}
\frac{\partial Q}{\partial t_{b}}=\left( 
\begin{array}{c}
D_{+}u\left( zA(b)_{-1}\right) +D_{+}v\left( z^{2}A(b)_{-2}\right) \\ 
\widetilde{\Lambda }u\left( z^{2}A(b)_{-2}\right) +\widetilde{\Lambda }%
v\left( z^{3}A(b)_{-3}\right)%
\end{array}%
\right) _{\Theta ^{(1)}}=\left( 
\begin{array}{c}
D_{+}w\left( z^{-1}A(b)_{1}\right) +D_{+}y\left( z^{-2}A(b)_{2}\right) \\ 
\widetilde{\Lambda }w\left( A(b)_{0}\right) +\widetilde{\Lambda }y\left(
z^{-1}A(b)_{1}\right)%
\end{array}%
\right) _{\Theta ^{(-1)}}.  \label{first form}
\end{equation}%
Now we need to rewrite these expressions in terms of differential forms and
to do it this time we need to take into account the recursion relations (\ref%
{recursion}), thus the first Poisson structures are sensitive to the degree
of twisting in (\ref{isomorphisms}) and in turn this is reflected in the
degree of non-locality of the Poisson brackets.

Starting with the second relation in (\ref{recursion}) which is relevant in
the $AdS_{3}\times S^{3}$ case, we get the following expressions for $\Theta
^{(1)}$%
\begin{equation*}
A(b)_{-2}=z^{-2}d_{q}H_{z^{2}b},\text{ \ \ \ \ \ }A(b)_{-3}=z^{-3}d_{\psi
}H_{z^{2}b},\text{ \ \ \ \ \ }zA(b)_{-1}=ud_{q}H_{z^{2}b}+vd_{\psi
}H_{z^{2}b}
\end{equation*}%
and for $\Theta ^{(-1)}$%
\begin{equation*}
A(b)_{1}=zd_{\psi }H_{z^{-2}b},\text{ \ \ \ \ \ }%
A(b)_{2}=z^{2}d_{q}H_{z^{-2}b},\text{ \ \ \ \ \ }%
A(b)_{0}=yd_{q}H_{z^{-2}b}+wd_{\psi }H_{z^{-2}b}.
\end{equation*}

Inserting these expressions in (\ref{first form}), the first forms become%
\begin{equation}
\frac{\partial Q}{\partial t_{b}}=\Theta ^{(\pm 1)}d_{Q}H_{z^{\pm 2}b},\text{
\ \ \ \ \ }\Theta ^{(1)}\equiv \left( 
\begin{array}{cc}
D_{+}\left( u^{2}+v\right) & D_{+}uv \\ 
\widetilde{\Lambda }u & \widetilde{\Lambda }v%
\end{array}%
\right) ,\text{ \ \ \ \ \ }\Theta ^{(-1)}\equiv \left( 
\begin{array}{cc}
D_{+}y & D_{+}w \\ 
\widetilde{\Lambda }wy & \widetilde{\Lambda }\left( w^{2}+y\right)%
\end{array}%
\right)  \label{first for AdS3xS3}
\end{equation}%
and we can explicitly verify that the three Poisson structures $\Theta
^{(\pm 1)},\Theta ^{(0)}$ satisfy the relation%
\begin{equation*}
\Theta ^{(-1)}=\Theta ^{(0)}\Theta ^{(1)-1}\Theta ^{(0)}
\end{equation*}%
where%
\begin{equation*}
\Theta ^{(1)-1}=\left( 
\begin{array}{cc}
y & wy \\ 
wy & \left( w^{2}y+y^{2}\right)%
\end{array}%
\right) D_{+}^{-1}
\end{equation*}%
and this means that they are all compatible.

Now we consider the third relation in (\ref{recursion}) which is relevant to
the cases $AdS_{n}\times S^{n},$ $n=2,5.$ We get the following expressions
for $\Theta ^{(1)}$

\begin{eqnarray*}
A(b)_{-4} &=&z^{-4}d_{q}H_{z^{4}b},\text{ \ \ \ \ \ }A(b)_{-5}=z^{-5}d_{\psi
}H_{z^{4}b}, \\
zA(b)_{-1} &=&\left( u^{3}+uv+vu\right) d_{q}H_{z^{4}b}+\left(
u^{2}v+v^{2}\right) d_{\psi }H_{z^{4}b}, \\
z^{2}A(b)_{-2} &=&\left( u^{2}+v\right) d_{q}H_{z^{4}b}+uvd_{\psi
}H_{z^{4}b}, \\
z^{3}A(b)_{-3} &=&ud_{q}H_{z^{4}b}+vd_{\psi }H_{z^{4}b}.
\end{eqnarray*}%
and for $\Theta ^{(-1)}$%
\begin{eqnarray*}
A(b)_{3} &=&z^{3}d_{\psi }H_{z^{-4}b},\text{ \ \ \ \ \ }%
A(b)_{4}=z^{4}d_{q}H_{z^{-4}b}, \\
A(b)_{0} &=&\left( y^{2}+w^{2}y\right) d_{q}H_{z^{-4}b}+\left(
w^{3}+wy+yw\right) d_{\psi }H_{z^{-4}b}, \\
\text{\ }z^{-1}A(b)_{1} &=&wyd_{q}H_{z^{-4}b}+\left( w^{2}+y\right) d_{\psi
}H_{z^{-4}b}, \\
\text{\ }z^{-2}A(b)_{2} &=&yd_{q}H_{z^{-4}b}+wd_{\psi }H_{z^{-4}b}.
\end{eqnarray*}

Inserting this expression in (\ref{first form}) above we have%
\begin{eqnarray}
\frac{\partial Q}{\partial t_{b}} &=&\Theta ^{(\pm 1)}d_{Q}H_{z^{\pm 4}b},\
\ \   \notag \\
\Theta ^{(1)} &\equiv &\left( 
\begin{array}{cc}
D_{+}\left( u^{4}+u^{2}v+uvu+vu^{2}+v^{2}\right) & D_{+}\left(
u^{3}v+uv^{2}+vuv\right) \\ 
\widetilde{\Lambda }\left( u^{3}+uv+vu\right) & \widetilde{\Lambda }\left(
u^{2}v+v^{2}\right)%
\end{array}%
\right) ,\ \ \   \label{teta 1} \\
\Theta ^{(-1)} &\equiv &\left( 
\begin{array}{cc}
D_{+}\left( w^{2}y+y^{2}\right) & D_{+}\left( w^{3}+wy+yw\right) \\ 
\widetilde{\Lambda }\left( wy^{2}+w^{3}y+ywy\right) & \widetilde{\Lambda }%
\left( w^{4}+w^{2}y+wyw+yw^{2}+y^{2}\right)%
\end{array}%
\right)  \label{teta -1}
\end{eqnarray}%
and in a similar way we can verify the compatibility relation%
\begin{equation*}
\Theta ^{(-1)}=\Theta ^{(0)}\Theta ^{(1)-1}\Theta ^{(0)}
\end{equation*}%
with 
\begin{equation*}
\Theta ^{(1)-1}=\left( 
\begin{array}{cc}
w^{2}y+y^{2} & w^{3}y+wy^{2}+ywy \\ 
wy^{2}+w^{3}y+ywy & w^{4}y+w^{2}y^{2}+\left( wy\right) ^{2}+yw^{2}y+y^{3}%
\end{array}%
\right) D_{+}^{-1}.
\end{equation*}

The flow of any functional $\varphi (Q)$ is given by%
\begin{equation*}
\frac{\partial \varphi (Q)}{\partial t_{b}}=\left( d_{Q}\varphi ,\frac{%
\partial Q}{\partial t_{b}}\right) =\left( d_{Q}\varphi ,\Theta ^{(\pm
1)}d_{Q}H_{z^{\pm 4}b}\right)
\end{equation*}%
and motivates the following definition for the first Poisson brackets%
\begin{equation*}
\left\{ \varphi ,\psi \right\} _{\pm 1}\equiv \left( d_{Q}\varphi ,\text{\ }%
\Theta ^{(\pm 1)}d_{Q}\psi \right) =\dint\nolimits_{-\infty }^{+\infty
}dx^{+}Str\left( \frac{\delta \varphi }{\delta Q},\Theta ^{(\pm 1)}\frac{%
\delta \psi }{\delta Q}\right)
\end{equation*}%
in terms of which%
\begin{equation*}
\frac{\partial \varphi }{\partial t_{b}}=\left\{ \varphi ,H_{z^{\pm
4}b}\right\} _{\pm 1}.
\end{equation*}%
Then, we have shown that%
\begin{equation}
\left\{ \varphi ,H_{b}\right\} _{2}=\left\{ \varphi ,H_{z^{\pm 4}b}\right\}
_{\pm 1}.  \label{Poisson recursion}
\end{equation}

In this way we can construct recursively an infinite family of non-local
Poisson structures for the SSSSG integrable hierarchy governing the
Pohlmeyer reduced models%
\begin{equation*}
\frac{\partial Q}{\partial t_{b}}=\Theta ^{(\pm n)}d_{Q}H_{z^{\pm 4n}b},\ \
\ \ \ \ n\in 
\mathbb{Z}
^{+}.
\end{equation*}%
The computation for the higher Poisson bi-vectors $\Theta ^{(\pm n)}$, $%
n\geq 2$ becomes rather cumbersome. However, in the bosonic limit we can
write $\Theta ^{(\pm n)}$ in closed form%
\begin{equation*}
\Theta ^{(n)}=\left( \widetilde{\Lambda }D_{+}^{-1}\widetilde{\Lambda }%
D_{+}^{-1}\right) ^{n}\Theta ^{(0)},\text{ \ \ \ \ \ }\Theta ^{(0)}=D_{+},%
\text{ \ \ \ \ \ }n\in 
\mathbb{Z}
.
\end{equation*}

The degree of non-locality $n$ and the behavior of the brackets under
Lorentz boosts $x^{\pm }\rightarrow $ $\lambda ^{\pm 1}x^{\pm }$ are
correlated and the only boost invariant bracket is $\Theta ^{(0)}.$ It is
very important to recall that all Poisson structures have to be restricted,
in field space, to the slices $q_{l/r}=0.$

As we will see in the next section, the special combination%
\begin{equation}
\Theta _{\sigma }=\Theta ^{(-1)}-2\Theta ^{(0)}+\Theta ^{(+1)}
\label{announce}
\end{equation}%
corresponds to the canonical Poisson structure of the GS superstring $\sigma 
$-model after gauge fixing all the local symmetries. From this we conclude
that the Pohlmeyer reduced model described by $\mathcal{L}_{\pm }$ and
supplemented by the constraints $q_{l/r}=0$ carry all the classical
information of the $\sigma $-model that generated it.

\section{Connection with the GS superstring sigma model.}

In this section we rewrite some known results of the GS superstring $\sigma $%
-model and show how they fit in the SSSSG integrable hierarchy approach. In
particular, the relation between the canonical Poisson structure of the $%
\sigma $-model and the Poisson structures constructed from the recursion
operators, the relation between commuting charges in terms of the Zukhovsky
map and the relation between Lax pair representations.

\subsection{Relation between Poisson structures.}

Due to the relevance of the result of \cite{Mikha bi-hamiltonian} in
relation to ours, here we briefly review it. The goal is to find the
explicit form of the canonical Poisson bi-vector $\Theta _{\sigma }$ when
restricted to the symplectic leaves left after fixing all the local gauge
symmetries of the $\sigma $-model.

The GSs$\sigma $ model is defined by the following action functional 
\begin{equation}
S_{GS}=\frac{1}{2}\dint_{\Sigma }Str\left( J^{(2)}\wedge \ast J^{(2)}+\kappa
J^{(1)}\wedge J^{(3)}\right) ,\text{ \ \ \ \ \ }J\equiv f^{-1}df,
\label{GS action}
\end{equation}%
where $\Sigma $ denotes the string world-sheet. \ In the conformal gauge%
\footnote{%
The conventions used in conformal gauge are $\epsilon ^{+-}=-1,$ $\eta
_{+-}=1,$ $\kappa =-1,$ $\left( \ast J\right) _{\alpha }=J_{\rho }\epsilon
^{\rho \lambda }\eta _{\lambda \alpha }.$ The covariant derivative is $%
D\equiv d+ad_{J}.$%
\par
{}} and in the partially fixed kappa symmetry gauge \cite{Grigo-Tseytlin} $%
J_{+}^{(3)}=J_{-}^{(1)}=0$, the canonical symplectic form of (\ref{GS action}%
) constructed in \cite{Mikha bi-hamiltonian} takes the form $\Omega _{\sigma
}=\Omega _{\sigma }^{+}-\Omega _{\sigma }^{-},$ with\footnote{%
As used before, we write $(X,Y)^{\pm }\equiv \dint dx^{\pm }Str(X,Y)$ and
omit the $\pm $ indices when there is not ambiguity.} 
\begin{equation}
\Omega _{\sigma }^{+}=\frac{1}{2}\left( \left( f^{-1}\delta f\right)
^{(2)},\wedge D_{+}^{(0)}\left( f^{-1}\delta f\right)
^{(2)}+2J_{+}^{(2)}\left( f^{-1}\delta f\right) ^{(1)},\wedge \left(
f^{-1}\delta f\right) ^{(1)}\right) ,  \label{combination}
\end{equation}%
and $\Omega _{\sigma }^{-}=\Omega _{\sigma }^{+}(+\leftrightarrow
-,1\leftrightarrow 3).$

The idea is to find the inverse $\Omega _{\sigma }^{-1}=\Theta _{\sigma }$
by using Hamiltonian vectors fields. Consider a left-invariant vector field $%
X_{\xi }=f\xi $ in which $f\in F=\exp \mathfrak{f,}$ $\mathfrak{f}$ is
defined by (\ref{Z4 grading}) and $\xi $ is the vector field at the
supergroup identity. This vector field satisfy $\left( f^{-1}\delta f\right)
\left( X_{\xi }\right) =\xi $ and as consequence we have that

\begin{equation}
\Omega _{\sigma }^{+}\left( X_{\eta },X_{\xi }\right) =\frac{1}{2}\left(
\eta ^{(2)}\overleftrightarrow{,D_{+}^{(0)}}\xi ^{(2)}-\eta ^{(1)},%
\overleftrightarrow{a_{2+}}\xi ^{(1)}\right) ,  \label{contraction}
\end{equation}%
where $X\overleftrightarrow{\mathcal{O}}Y\equiv X\left( \mathcal{O}Y\right)
-\left( \mathcal{O}Y\right) X$ and $a_{i+}=ad_{J_{+}^{(i)}}.$ Considering $%
X_{\eta },X_{\xi }$ as the Hamiltonian vector fields generated by the
functionals $F,G,$ we have%
\begin{equation*}
\left\{ F,G\right\} _{\sigma }=\Omega _{\sigma }^{+}\left( X_{\eta },X_{\xi
}\right) =X_{\eta }\left( G\right) =\delta G\left( X_{\eta }\right) =-X_{\xi
}\left( F\right) =-\delta F\left( X_{\xi }\right) .
\end{equation*}

To find the components $\eta $ of the Hamiltonian vector field $X_{\eta },$
we have to solve the relation%
\begin{equation}
\Omega _{\sigma }\left( X_{\eta },X_{\xi }\right) =-\delta F\left( X_{\xi
}\right) ,\text{ \ \ \ \ \ }\forall \xi .  \label{Hamiltonian vector field}
\end{equation}%
There is a freedom in choosing the component $\xi ^{(0)}$ and in the
following we will take it as zero. Using the relation $\delta J_{\mu
}=D_{\mu }\left( f^{-1}\delta f\right) $ we get%
\begin{equation*}
\delta J_{+}^{(3)}\left( X_{\xi }\right) =0=D_{+}^{(0)}\xi ^{(3)}+a_{1+}\xi
^{(2)}+a_{2+}\xi ^{(1)}
\end{equation*}%
because we need to maintain the gauge $J_{+}^{(3)}=0\ $and this allows to
determine one of the components, in this case we take%
\begin{equation}
\xi ^{(3)}=-D_{+}^{(0)-1}\left( a_{1+}\xi ^{(2)}+a_{2+}\xi ^{(1)}\right) .
\label{third}
\end{equation}

By replacing this component on the other contractions we have%
\begin{eqnarray}
\delta J_{+}^{(2)}\left( X_{\xi }\right) &=&a_{1+}\xi ^{(1)}+D_{+}^{(0)}\xi
^{(2)},  \label{current form} \\
\delta J_{+}^{(1)}\left( X_{\xi }\right) &=&\left(
D_{+}^{(0)}-a_{2+}D_{+}^{(0)-1}a_{2+}\right) \xi
^{(1)}-a_{2+}D_{+}^{(0)-1}a_{1+}\xi ^{(2)},  \notag \\
\delta J_{+}^{(0)}\left( X_{\xi }\right) &=&-a_{1+}D_{+}^{(0)-1}a_{2+}\xi
^{(1)}+\left( a_{2+}-a_{1+}D_{+}^{(0)-1}a_{1+}\right) \xi ^{(2)}.  \notag
\end{eqnarray}%
With this in mind we can compute the rhs of (\ref{Hamiltonian vector field})
by using (\ref{current form}) and the obvious relation%
\begin{equation*}
\delta F\left( X_{\xi }\right) =\left( \frac{\delta F}{\delta J_{+}^{(a)}}%
,\delta J_{+}^{(a)}\left( X_{\xi }\right) \right) .
\end{equation*}%
The answer is%
\begin{eqnarray*}
\left( \frac{\delta F}{\delta J_{+}^{(0)}},\delta J_{+}^{(0)}\left( X_{\xi
}\right) \right) &=&\left( a_{2+}D_{+}^{(0)-1}a_{1+}\frac{\delta F}{\delta
J_{+}^{(0)}},\xi ^{(1)}+\left( -a_{2+}+a_{1+}D_{+}^{(0)-1}a_{1+}\right) 
\frac{\delta F}{\delta J_{+}^{(0)}},\xi ^{(2)}\right) , \\
\left( \frac{\delta F}{\delta J_{+}^{(1)}},\delta J_{+}^{(1)}\left( X_{\xi
}\right) \right) &=&\left( \left(
-D_{+}^{(0)}+a_{2+}D_{+}^{(0)-1}a_{2+}\right) \frac{\delta F}{\delta
J_{+}^{(1)}},\xi ^{(1)}+a_{1+}D_{+}^{(0)-1}a_{2+}\frac{\delta F}{\delta
J_{+}^{(1)}},\xi ^{(2)}\right) , \\
\left( \frac{\delta F}{\delta J_{+}^{(2)}},\delta J_{+}^{(2)}\left( X_{\xi
}\right) \right) &=&\left( -a_{1+}\frac{\delta F}{\delta J_{+}^{(2)}},\xi
^{(1)}-D_{+}^{(0)}\frac{\delta F}{\delta J_{+}^{(2)}},\xi ^{(2)}\right) .
\end{eqnarray*}

The lhs of (\ref{Hamiltonian vector field}) was already computed in (\ref%
{contraction}) and it is given by%
\begin{equation*}
\Omega _{\sigma }\left( X_{\eta },X_{\xi }\right) =\left( a_{2+}\eta
^{(1)},\xi ^{(1)}-D_{+}^{(0)}\eta ^{(2)},\xi ^{(2)}\right) .
\end{equation*}%
By equating both sides, we determine all the components of the vector field $%
X_{\eta }$, $\eta ^{(3)}$ is determined by (\ref{third}),%
\begin{eqnarray}
\eta ^{(1)} &=&-D_{+}^{(0)-1}a_{1+}\frac{\delta F}{\delta J_{+}^{(0)}}%
+\left( a_{2+}^{-1}D_{+}^{(0)}-D_{+}^{(0)-1}a_{2+}\right) \frac{\delta F}{%
\delta J_{+}^{(1)}}+a_{2+}^{-1}a_{1+}\frac{\delta F}{\delta J_{+}^{(2)}},
\label{components} \\
\eta ^{(2)} &=&-\left( D_{+}^{(0)-1}a_{2+}-\left( D_{+}^{(0)-1}a_{1+}\right)
^{2}\right) \frac{\delta F}{\delta J_{+}^{(0)}}%
+D_{+}^{(0)-1}a_{1+}D_{+}^{(0)-1}a_{2+}\frac{\delta F}{\delta J_{+}^{(1)}}-%
\frac{\delta F}{\delta J_{+}^{(2)}}  \notag
\end{eqnarray}%
and this is enough to find the inverse of the symplectic form.

Once the Hamiltonian vector field $X_{\eta }$ of $F$ is found, we easily
obtain the Poisson brackets of the functional $F$ with the currents $J$ by
taking $G=J_{+}^{(b)}$, i.e, 
\begin{equation*}
\left\{ F,J_{+}^{(b)}\right\} _{\sigma }=\delta J_{+}^{(b)}\left( X_{\eta
}\right) .
\end{equation*}%
The components of the Poisson bi-vector $\Theta _{\sigma }=\Omega _{\sigma
}^{-1}$are, by definition, given by the Poisson brackets of the phase space
coordinates $J$ among themselves%
\begin{equation*}
\left\{ J_{+}^{(a)},J_{+}^{(b)}\right\} _{\sigma }\equiv \left[ \Theta
_{\sigma }\right] _{ab}.
\end{equation*}%
Replacing the components (\ref{components}) and specializing to the case $%
F=J_{+}^{(a)},$ we obtain%
\begin{eqnarray*}
\left\{ J_{+}^{(2)},J_{+}^{(2)}\right\} _{\sigma }
&=&D_{+}^{(0)}-a_{1+}a_{2+}^{-1}a_{1+},\text{ \ \ \ \ \ }\left\{
J_{+}^{(2)},J_{+}^{(1)}\right\} =-a_{1+}a_{2+}^{-1}D_{+}^{(0)},\text{ \ \ \
\ \ }\left\{ J_{+}^{(2)},J_{+}^{(0)}\right\} =a_{2+}, \\
\left\{ J_{+}^{(1)},J_{+}^{(0)}\right\} _{\sigma }
&=&a_{1+}-a_{2+}D_{+}^{(0)-1}a_{1+}D_{+}^{(0)-1}a_{2+}+a_{2+}\left(
D_{+}^{(0)-1}a_{1+}\right) ^{3}-\left( a_{2+}D_{+}^{(0)-1}\right) ^{2}a_{1+},
\\
\left\{ J_{+}^{(1)},J_{+}^{(1)}\right\} _{\sigma } &=&-\left(
D_{+}^{(0)}-a_{2+}D_{+}^{(0)-1}a_{2+}\right) a_{2+}^{-1}\left(
D_{+}^{(0)}-a_{2+}D_{+}^{(0)-1}a_{2+}\right) +a_{2+}\left(
D_{+}^{(0)-1}a_{1+}\right) ^{2}D_{+}^{(0)-1}a_{2+}, \\
\left\{ J_{+}^{(0)},J_{+}^{(0)}\right\} _{\sigma }
&=&a_{2+}D_{+}^{(0)-1}a_{2+}-\left( a_{1+}D_{+}^{(0)-1}\right)
^{2}a_{2+}-a_{2+}\left( D_{+}^{(0)-1}a_{1+}\right) ^{2}- \\
&&-a_{1+}D_{+}^{(0)-1}a_{2+}D_{+}^{(0)-1}a_{1+}+a_{1+}\left(
D_{+}^{(0)-1}a_{1+}\right) ^{3}.
\end{eqnarray*}%
This brackets can be split according to their behavior under Lorentz boost 
\cite{Mikha bi-hamiltonian}%
\begin{equation}
\left\{ J_{+}^{(2)},J_{+}^{(2)}\right\} _{\sigma
}^{[2]}=D_{+}^{(0)}-a_{1+}a_{2+}^{-1}a_{1+},\text{ \ \ \ \ \ }\left\{
J_{+}^{(2)},J_{+}^{(1)}\right\} _{\sigma
}^{[2]}=-a_{1+}a_{2+}^{-1}D_{+}^{(0)},\text{ \ \ \ \ \ }\left\{
J_{+}^{(1)},J_{+}^{(1)}\right\} _{\sigma
}^{[2]}=-D_{+}^{(0)}a_{2+}^{-1}D_{+}^{(0)}  \label{1}
\end{equation}%
and%
\begin{equation}
\left\{ J_{+}^{(2)},J_{+}^{(0)}\right\} _{\sigma }^{[0]}=a_{2+},\text{ \ \ \
\ \ }\left\{ J_{+}^{(1)},J_{+}^{(0)}\right\} _{\sigma }^{[0]}=a_{1+},\text{
\ \ \ \ \ }\left\{ J_{+}^{(1)},J_{+}^{(1)}\right\} _{\sigma }^{[0]}=2a_{2+}
\label{2}
\end{equation}%
and%
\begin{eqnarray}
\left\{ J_{+}^{(1)},J_{+}^{(0)}\right\} _{\sigma }^{[-2]}
&=&-a_{2+}D_{+}^{(0)-1}a_{1+}D_{+}^{(0)-1}a_{2+}+a_{2+}\left(
D_{+}^{(0)-1}a_{1+}\right) ^{3}-\left( a_{2+}D_{+}^{(0)-1}\right) ^{2}a_{1+},
\label{3} \\
\left\{ J_{+}^{(1)},J_{+}^{(1)}\right\} _{\sigma }^{[-2]} &=&-a_{2+}\left(
D_{+}^{(0)-1}a_{2+}\right) ^{2}+a_{2+}\left( D_{+}^{(0)-1}a_{1+}\right)
^{2}D_{+}^{(0)-1}a_{2+},  \notag \\
\left\{ J_{+}^{(0)},J_{+}^{(0)}\right\} _{\sigma }^{[-2]}
&=&a_{2+}D_{+}^{(0)-1}a_{2+}-\left( a_{1+}D_{+}^{(0)-1}\right)
^{2}a_{2+}+a_{2+}\left( D_{+}^{(0)-1}a_{1+}\right) ^{2}-  \notag \\
&&-a_{1+}D_{+}^{(0)-1}a_{2+}D_{+}^{(0)-1}a_{1+}+a_{1+}\left(
D_{+}^{(0)-1}a_{1+}\right) ^{3}.  \notag
\end{eqnarray}%
The functionals $\left\{ \ast ,\ast \right\} $ $_{\sigma }^{[\lambda ]},$ $%
\lambda =-2,0,2$ define three mutually compatible Poisson brackets providing
the bi-Hamiltonian structure of the $\sigma $-model in terms of which the
Poisson bracket decomposes as follows%
\begin{equation}
\left\{ F,G\right\} _{\sigma }=\left\{ F,G\right\} ^{[-2]}+\left\{
F,G\right\} ^{[0]}+\left\{ F,G\right\} ^{[2]}.  \label{mikha bracket}
\end{equation}

In order to make contact with the results of section 3 we need to write the
bracket $\left\{ F,G\right\} _{\sigma }$ in terms of the currents $%
J_{+}^{(0)},J_{+}^{(1)}$ only, which are the fundamental variables we are
using for the Pohlmeyer reduced models, see (\ref{phase coordinates}). All
these brackets make sense only on gauge invariant functionals $F,$ i.e $%
F\left( J+\delta _{\xi ^{(0)}}J\right) =F(J)$ and this condition imply that%
\begin{equation}
D_{+}^{(0)}\frac{\delta F}{\delta J_{+}^{(0)}}+a_{1+}\frac{\delta F}{\delta
J_{+}^{(1)}}+a_{2+}\frac{\delta F}{\delta J_{+}^{(2)}}=0\rightarrow \frac{%
\delta F}{\delta J_{+}^{(2)}}=-a_{2+}^{-1}\left( D_{+}^{(0)}\frac{\delta F}{%
\delta J_{+}^{(0)}}+a_{1+}\frac{\delta F}{\delta J_{+}^{(1)}}\right) ,
\label{gauge invariance}
\end{equation}%
allowing to eliminate the functional derivative $\frac{\delta F}{\delta
J_{+}^{(2)}}$ in all the expressions$.$ The last bracket (\ref{3}) is
already in the desired form and we can write it as 
\begin{eqnarray*}
\left\{ F,G\right\} ^{[-2]} &=&\left( \frac{\delta F}{\delta J_{+}^{(a)}}%
,\Theta _{ab}^{[-2]}\frac{\delta F}{\delta J_{+}^{(b)}}\right) ,\text{ \ \ \
\ \ }U\equiv -D_{+}^{(0)-1}a_{1+},\text{ \ \ \ \ \ }V\equiv
-D_{+}^{(0)-1}a_{2+}, \\
\Theta ^{\lbrack -2]} &=&\left( 
\begin{array}{cc}
D_{+}^{(0)}\left( U^{4}+U^{2}V+UVU+VU^{2}+V^{2}\right) & D_{+}^{(0)}\left(
U^{3}V+UV^{2}+VUV\right) \\ 
-a_{2+}\left( U^{3}+UV+VU\right) & -a_{2+}\left( U^{2}V+V^{2}\right)%
\end{array}%
\right) ,
\end{eqnarray*}%
where $a,b=0,1$. Using (\ref{gauge invariance}) in (\ref{2}) we easily get%
\begin{equation*}
\left\{ F,G\right\} ^{[0]}=\left( \frac{\delta F}{\delta J_{+}^{(a)}},\Theta
_{ab}^{[0]}\frac{\delta F}{\delta J_{+}^{(b)}}\right) ,\text{ \ \ \ \ \ }%
\Theta ^{\lbrack 0]}=-2\left( 
\begin{array}{cc}
D_{+}^{(0)} & 0 \\ 
0 & -a_{2+}%
\end{array}%
\right)
\end{equation*}%
and in (\ref{1}) we get%
\begin{eqnarray*}
\left\{ F,G\right\} ^{[2]} &=&\left( \frac{\delta F}{\delta J_{+}^{(a)}}%
,\Theta _{ab}^{[2]}\frac{\delta F}{\delta J_{+}^{(b)}}\right) ,\text{ \ \ \
\ \ }W\equiv -a_{2+}^{-1}a_{1+},\text{ \ \ \ \ \ }Y\equiv
-a_{2+}^{-1}D_{+}^{(0)}, \\
\Theta ^{\lbrack 2]} &=&\left( 
\begin{array}{cc}
D_{+}^{(0)}\left( W^{2}Y+Y^{2}\right) & D_{+}^{(0)}\left( W^{3}+WY+YW\right)
\\ 
-a_{2+}\left( WY^{2}+W^{3}Y+YWY\right) & -a_{2+}\left(
W^{2}+W^{2}Y+WYW+YW^{2}+Y^{2}\right)%
\end{array}%
\right) .
\end{eqnarray*}

The full restriction of $\Omega _{\sigma }$ to the reduced phase space is
accomplished by the currents $J_{+}^{(2)}$ satisfying the Virasoro
constraints by the currents $J_{+}^{(1)}$ satisfying the condition $%
J_{+}^{(1)}\in \func{Im}$ $a_{2+}$, fixing the residual kappa symmetry. Both
conditions are satisfied by the first set of Pohlmeyer variables 
\begin{equation*}
J_{+}^{(0)}=q,\text{ \ \ \ \ \ }J_{+}^{(1)}=\psi ,\text{ \ \ \ \ \ }%
J_{+}^{(2)}=-\Lambda ,\text{ \ \ \ \ \ }D_{+}^{(0)}=D_{+}
\end{equation*}%
and this imply that $U=u,$ $V=v,$ $W=w,$ $Y=y$, where $u,v,w,$ and $y$ were
defined in section 3. Comparing with the previous results (\ref{teta 0}), (%
\ref{teta 1}) and (\ref{teta -1}) we conclude that 
\begin{eqnarray*}
\Theta ^{\lbrack -2]} &=&\Theta ^{(1)},\text{ \ \ \ \ \ }\Theta ^{\lbrack
0]}=-2\theta ^{(0)},\text{ \ \ \ \ \ }\Theta ^{\lbrack 2]}=\Theta ^{(-1)}, \\
\Theta _{\sigma } &=&\Theta ^{\lbrack -2]}+\text{\ }\Theta ^{\lbrack
0]}+\Theta ^{\lbrack 2]}=\Theta ^{(1)}-2\Theta ^{(0)}+\Theta ^{(1)}
\end{eqnarray*}%
as announce in (\ref{announce}). To find $\Theta _{-}^{\sigma }$ we simply
replace $\left( +\leftrightarrow -\right) ,(1\leftrightarrow 3)$ in the
results above. In this case the second set of Pohlmeyer variables is%
\begin{equation}
J_{-}^{(0)}=0,\text{ \ \ \ \ \ }J_{-}^{(3)}=\gamma ^{-1}\overline{\psi }%
\gamma ,\text{ \ \ \ \ \ }J_{-}^{(2)}=-\gamma ^{-1}\Lambda \gamma ,\text{ \
\ \ \ \ }D_{-}^{(0)}=\partial _{-}.  \label{second set}
\end{equation}%
Notice that there is a subtlety in the replacement of (\ref{second set}) to
get $\Theta _{-}^{\sigma }.$ The current $J_{-}^{(0)}$ is zero on the (-)
sector of the light-cone and in principle the functionals $\frac{\delta F}{%
\delta J_{-}^{(0)}}$ are not well defined. In order for these results to
make sense, we interpret the reduced sigma model light cone phase space as
being parametrized by $(Q,\overline{Q})$ with $Q$ on $\mathcal{L}_{+}$ and $%
\overline{Q}$ on $\mathcal{L}_{-}^{\prime }$, see the definitions (\ref%
{phase coordinates}), with the Poisson brackets defined in section 3.

What is remarkable in the computation of \cite{Mikha bi-hamiltonian}
reproduced here, is that the degree of non-locality of the first Poisson
structures is determined entirely by the gauge fixing of all the local
symmetries of the GS superstring $\sigma $-model (\ref{GS action}), while in
the SSSSG model the degree of non-locality is determined entirely by the
twisted nature of the superalgebra $\widehat{\mathfrak{f}}$ through the
recursion relation determined by $z^{\pm 4n}\widehat{\mathfrak{f}}\simeq 
\widehat{\mathfrak{f}}$ and traced back to the $%
\mathbb{Z}
_{4}$ grading of the semi-symmetric space in which the string propagates.
The reduced model although manifestly relativistic carries all the
non-relativistic information of the former sigma model it comes from. As
mentioned at the end of the last section, the Poisson structures $\Theta
^{\lbrack 0]},\Theta ^{\lbrack \pm 2]}$ have to restricted to the level
manifolds $q_{l/r}=0.$

\subsection{Relation between commuting charges.}

In this section we show how to construct $\sigma $- model commuting charges
out of the SSSSG commuting charges. In the process we also show the
integrable origin of the Zukhovski variable and the need for twisting the
inner product in the sigma model side, in the sense of \cite{vicedo}.

In the (+) sector of the light-cone phase space, the reduced $\sigma $-model
and the SSSSG model are parametrized by the same coordinates $Q=(q,\psi ),$
the only difference being their Poisson structures. Then, it is natural to
try to relate the symmetry flows on both sides. For simplicity we analyze
one sector only. We want to construct a sigma model current component $%
I_{+}(b)$ from the SSSSG model component $j_{+}(b),$ both generating the
same flow on functionals of the phase space $\mathcal{L}_{+}$. In other
words, we want to solve the following relation

\begin{equation*}
\frac{\partial \varphi (\mathcal{L}_{+})}{\partial t_{b}}=\left\{ \varphi
,j_{+}(b)\right\} _{2}(\mathcal{L}_{+})=\left\{ \varphi ,I_{+}(b)\right\}
_{\sigma }(\mathcal{L}_{+}),\text{ \ \ \ \ \ }b\in \mathfrak{z,}
\end{equation*}%
which is equivalent to

\begin{equation}
\Theta ^{(0)}d_{Q}j_{+}(b)=\Theta _{\sigma }d_{Q}I_{+}(b).  \label{problem}
\end{equation}%
To find $I_{+}(b)$ we make the following superposition ansatz%
\begin{equation}
d_{Q}I_{+}(b)=d_{Q}j_{+}(\phi (z)b),\text{ \ \ \ \ \ }\phi (z)\equiv
\dsum\nolimits_{n\in 
\mathbb{Z}
}c_{n}z^{4n},  \label{ansatz}
\end{equation}%
because we need $\phi (z)b\in \widehat{\mathfrak{f}}$ to be in the
superalgebra $\widehat{\mathfrak{f}},$ cf (\ref{isomorphisms}). The equation
(\ref{problem}) can be written as follows 
\begin{equation*}
\Theta ^{(0)}d_{Q}j_{+}(b)=\Theta _{\sigma }d_{Q}j_{+}(\phi (z)b)=\Theta
^{(0)}d_{Q}j_{+}(\phi (z)(z^{4}-2+z^{-4})b),
\end{equation*}%
where we have used the relation (\ref{Poisson recursion}) connecting the
first set of isospectral flows of the SSSSG hierarchy and this determines $%
\phi (z)$ unambiguously to be\footnote{%
In the reduction of the $\sigma $-model in $AdS_{3}\times S^{3},$ we have $%
\phi (z)=\left( z^{2}-2+z^{-2}\right) .$ This decrease in the grade of
non-locality, cf (\ref{first for AdS3xS3}), in contrast to the $%
AdS_{n}\times S^{n},$ $n=2,5$ situation can be understood to be a
consequence of the group structure of the target space of the $\sigma $%
-model.} 
\begin{equation}
\phi (z)=\frac{1}{(z^{4}-2+z^{-4})}=\frac{1}{\left( z^{2}-z^{-2}\right) ^{2}}%
=\frac{z^{\pm 4}}{(1-z^{\pm 4})^{2}}=\dsum\nolimits_{n=0}^{\infty }nz^{\pm
4n}.  \label{fundamental poly}
\end{equation}%
In the last term we have expanded around $z=0$ and $z=+\infty $ and this
fixes all the coefficients $c_{n}$ to be 
\begin{eqnarray*}
\text{around }z &=&0,\text{ \ \ \ \ \ \ \ }c_{n}=0\text{ for }n\leq 0\text{
\ \ \ \ \ and \ \ \ }c_{n}=n\text{ \ \ for }n>0, \\
\text{around }z &=&+\infty ,\text{ \ \ \ \ }c_{n}=0\text{ for }n\geq 0\text{
\ \ \ \ \ and \ \ \ }c_{n}=-n\text{ \ \ for }n<0.
\end{eqnarray*}

For this result to make sense in the superalgebra $\widehat{\mathfrak{f}},$
we have to consider $\phi (z)$ as a power series in $z^{\pm 4n}$ around $z=0$
for $b$ of positive $Q_{H}$ grade with associated $j_{+}(b)$ and around $%
z=+\infty $ for $\overline{b}$ of negative $Q_{H}$ grade with associated $%
\overline{j}_{+}(\overline{b})$ and this is because the positive and
negative flows are not connected by the recursion relations$.$ Thus, the
sigma model current component $I_{+}(b)$ is expressed as an infinite linear
combination of SSSSG components $j_{+}(b)$ over $b$. What we have is a
linear combination of the $j_{+}(b)%
{\acute{}}%
s$ attached to the south pole on the Riemann sphere parametrized by the
spectral parameter $z$. The same happens for $\overline{I}_{+}(\overline{b})$
and $\overline{j}_{+}(\overline{b})$ at the north pole. Then, we have%
\begin{equation}
I_{+}(b)=j_{+}(\phi (z)b),\text{ \ \ \ \ \ }\overline{I}_{+}(\overline{b})=%
\overline{j}_{+}(\phi (z)\overline{b}).  \label{positive recur}
\end{equation}

To include the (-) sector of the light-cone formulation we use (\ref{A and A
tilda}) and (\ref{NA Lax}) to prove the equivalent recursion relations%
\footnote{%
Using $\theta _{\pm }^{(0)}$ we can rewrite the equations of motion (\ref%
{equations of motion}), in the gauge $u=\overline{u}=I,$ in an elegant
compact form%
\begin{equation*}
\partial _{-}Q=-\theta _{+}^{(0)}d_{Q}\overline{j}_{+}(z^{-2}\Lambda ),\text{
\ \ \ \ \ }\partial _{+}\overline{Q}=-\gamma \theta _{-}^{(0)}d_{\overline{Q}%
}j_{-}(z^{2}\Lambda )\gamma ^{-1}.
\end{equation*}%
}

\begin{eqnarray*}
\Theta _{-}^{(0)}d_{\overline{Q}}j_{-}(b) &=&\Theta _{-}^{(-1)}d_{\overline{Q%
}}j_{-}(z^{-4}b)=\Theta _{-}^{(1)}d_{\overline{Q}}j_{-}(z^{4}b), \\
\Theta _{-}^{(0)}d_{\overline{Q}}\overline{j}_{-}(\overline{b}) &=&\Theta
_{-}^{(-1)}d_{\overline{Q}}\overline{j}_{-}(z^{-4}\overline{b})=\Theta
_{-}^{(1)}d_{\overline{Q}}\overline{j}_{-}(z^{4}\overline{b}),
\end{eqnarray*}%
which can alternatively be obtained by parity. Then, for the negative
current components we obtain%
\begin{equation}
I_{-}(b)=j_{-}(\phi (z)b),\text{ \ \ \ \ \ }\overline{I}_{-}(\overline{b})=%
\overline{j}_{-}(\phi (z)\overline{b}).  \label{negative recur}
\end{equation}

From (\ref{DS charges}) we define $q(z^{2}\Lambda )\equiv Str(\Lambda
,q_{-2})$ and $\overline{q}(z^{-2}\Lambda )\equiv Str(\Lambda ,q_{-2}),$
which clearly generates flows along the $\left( x^{+},x^{-}\right) $
directions in the flat world-sheet$.$ From these results we see that the $%
\sigma $-model charges generating the same flows on $\mathcal{L}_{x}$
induced by $q(z^{2}\Lambda ),$ $\overline{q}(z^{-2}\Lambda )$ are

\begin{eqnarray}
q_{\sigma }(z^{2}\Lambda ) &=&\dint\nolimits_{-\infty }^{+\infty }dx\left(
I_{+}(z^{2}\Lambda )+I_{-}(z^{2}\Lambda )\right) =\dint\nolimits_{-\infty
}^{+\infty }dx\left( j_{+}(\phi (z)z^{2}\Lambda )+j_{-}(\phi (z)z^{2}\Lambda
)\right) ,  \label{recursion relation} \\
\overline{q}_{\sigma }(z^{-2}\Lambda ) &=&\dint\nolimits_{-\infty }^{+\infty
}dx\left( \overline{I}_{+}(z^{-2}\Lambda )+\overline{I}_{-}(z^{-2}\Lambda
)\right) =\dint\nolimits_{-\infty }^{+\infty }dx\left( \overline{j}_{+}(\phi
(z)z^{-2}\Lambda )+\overline{j}_{-}(\phi (z)z^{-2}\Lambda )\right) .  \notag
\end{eqnarray}%
This is a nice result because we can associate conserved charges generating $%
\left( x^{+},x^{-}\right) $ translations despite of the fact that the
Virasoro constraints have been already imposed $Str(J_{\pm }^{(2)},J_{\pm
}^{(2)})=0.$

A closer look to the result (\ref{recursion relation}) suggests that the
inner product on sigma model have to be twisted, in the sense of \cite%
{vicedo}. What we have is the following: from (\ref{positive recur}), (\ref%
{negative recur}) we notice, considering $j_{+}(b)$ only$,$ that 
\begin{equation*}
j_{+}(b)=\dint\nolimits_{-\infty }^{+\infty }dx^{+}\left[ \doint \frac{dz}{%
2\pi i}\frac{1}{z}Str\left( \Phi b\Phi ^{-1},\mathcal{L}_{+}\right) \right] ,%
\text{ \ \ \ \ \ }I_{+}(b)=\dint\nolimits_{-\infty }^{+\infty }dx^{+}\left[
\doint \frac{dz}{2\pi i}\frac{\phi (z)}{z}Str\left( \Phi b\Phi ^{-1},%
\mathcal{L}_{+}\right) \right] .
\end{equation*}%
From this we naturally identify two different inner products that can be
defined on the superalgebra $\widehat{\mathfrak{f}},$ cf (\ref{superalgebra})%
$.$ The simplest one for the SSSSG model already defined in (\ref{simple
inner}) and a twisted one for the $\sigma -$model and defined by, see \cite%
{vicedo},%
\begin{equation}
\left\langle X,Y\right\rangle _{\phi }\equiv \doint \frac{du}{2\pi i}%
Str\left( X(z),Y(z)\right) ,  \label{twisted inner}
\end{equation}%
where $du=\frac{dz}{z}\phi (z),$ $u=\frac{1}{16}Z$ and $Z=2\frac{1+z^{4}}{%
1-z^{4}}$ is the Zukhovsky variable and we see how the map SSSSG Model$%
\rightarrow \sigma $-Model is implemented by the change of variables $%
z\rightarrow u=u(z).$ Using these products we get, in compact form 
\begin{equation}
j_{+}(b)=\left( A(b),\mathcal{L}_{+}\right) ,\text{ \ \ \ \ \ }%
I_{+}(b)=\left( A(b),\mathcal{L}_{+}\right) _{\phi },
\label{sigma/SSSSG current relation}
\end{equation}%
where we have taken (\ref{integrated inner product}) into account. For $%
\overline{j}_{+}(\overline{b})$ and $\overline{I}_{-}(\overline{b})$ we use
the parity transformations. Notice that by writing the evolution equations
on the $\sigma $-model in terms of $\Theta ^{(0)}$ instead of $\Theta
_{\sigma }$ all the non-localities of the Poisson brackets are removed. This
also explains the integrable origin of the Zukhovsky variable $Z$.

Finally, we have shown that%
\begin{eqnarray*}
\left\{ q(b),q(b^{\prime })\right\} _{2}(\mathcal{L}_{x}) &=&0\rightarrow
\left\{ q_{\sigma }(b),q_{\sigma }(b^{\prime })\right\} _{\sigma }(\mathcal{L%
}_{x})=0, \\
\left\{ \overline{q}(\overline{b}),\overline{q}(\overline{b}^{\prime
})\right\} _{2}(\mathcal{L}_{x}^{\prime }) &=&0\rightarrow \left\{ \overline{%
q}_{\sigma }(\overline{b}),\overline{q}_{\sigma }(\overline{b}^{\prime
})\right\} _{\sigma }(\mathcal{L}_{x}^{\prime })=0,
\end{eqnarray*}%
for $b,b^{\prime }\in \mathfrak{z}$ with positive $Q_{H}$ grade and $%
\overline{b},\overline{b}^{\prime }\in \mathfrak{z}$ with negative $Q_{H}$
grade.

One comment is in order, the relation between the charges seems to be valid
only for the elements $b,$ $\overline{b}\in \mathfrak{z,}$ because for the
differentials of the currents $j(b),$ $\overline{j}(\overline{b}),$ the
second term in the rhs of (\ref{differential}) is absent. In this case the
ansatz (\ref{ansatz}) works fine. It is important to see under what
conditions we can apply these results to the elements $b,$ $\overline{b}\in 
\widehat{\mathfrak{c}}$ because this would allow to define 2d fermionic
symmetry flows on the sigma model by maping the fermionic conserved charges
of the SSSSG model.

\begin{remark}
It would be interested to study the relation between the reduction of the
Kostant-Kirillov bracket on the $\sigma $-model using the twisted inner
product $(\ast ,\ast )_{\phi }$ defined in \cite{vicedo}, and the non-local
Poisson bracket $\left\{ \ast ,\ast \right\} _{\sigma }$ constructed in \cite%
{Mikha bi-hamiltonian}, i.e (\ref{mikha bracket}).
\end{remark}

\subsection{Relation between Lax representations.}

In this section we show the relation between the Lax representation of the $%
\sigma $-model and the SSSSG model. We follow \cite{BPR} and use the
light-cone frame formulation of \cite{Grigo-Tseytlin}. The idea is to recall
the construction of the Lax pair for the $\sigma $-model in an arbitrary
world-sheet $\Sigma $ and then apply the Pohlmeyer reduction to recover the
Lax operators defined in (\ref{Lax op with gauge fields}).

The Lagrangian of the GS $\sigma $-model action (\ref{GS action}) is%
\footnote{%
We use $J^{(2)}\wedge \ast J^{(2)}=\gamma ^{\mu \nu }J_{\mu }^{(2)}J_{\nu
}^{(2)}dx^{0}\wedge dx^{1},$ $\gamma ^{\mu \nu }$\ =$\sqrt{\left\vert
h\right\vert }h^{\mu \nu }$ and the light-cone conventions introduced in the
section 4.1.}%
\begin{equation*}
L_{GS}=\frac{1}{2}Str\left( \gamma ^{\mu \nu }J_{\mu }^{(2)}J_{\nu
}^{(2)}+\kappa \epsilon ^{\mu \nu }J_{\mu }^{(1)}J_{\nu }^{(3)}\right)
dx^{0}\wedge dx^{1}.
\end{equation*}%
Introduce the light-cone frame zweibein $e$ in order to write $h^{\mu \nu
}=e_{\alpha }^{\mu }e_{\beta }^{\nu }\eta ^{\alpha \beta },$ $h_{\mu \nu
}=e_{\mu }^{\alpha }e_{\nu }^{\beta }\eta _{\alpha \beta },$ $J_{\mu
}=e_{\mu }^{\alpha }J_{\alpha },$ $J_{\alpha }=e_{\alpha }^{\mu }J_{\mu },$
where $\alpha ,\beta =+,-$ are the tangent space light-cone indices. The
Lagrangian takes the form 
\begin{equation}
L_{GS}=Str\left( J_{+}^{(2)}J_{-}^{(2)}+\frac{\kappa }{2}\left(
J_{+}^{(1)}J_{-}^{(3)}-J_{-}^{(1)}J_{+}^{(3)}\right) \right) e^{+}\wedge
e^{-},\text{ \ \ \ \ \ }e^{+}\wedge e^{-}=\frac{dx^{0}\wedge dx^{1}}{\det
e_{\alpha }^{\mu }}  \label{frame GS action}
\end{equation}%
and the equations of motion can be written as follows%
\begin{equation}
\delta _{J}L_{GS}=-Str\left( \left( f^{-1}\delta f\right) ,D_{\alpha
}\Lambda ^{\alpha }\right) ,\text{ \ \ \ \ \ }\Lambda ^{\alpha }\equiv \eta
^{\alpha \beta }J_{\beta }^{(2)}-\frac{1}{2}\kappa \epsilon ^{\alpha \beta
}\left( J_{\beta }^{(1)}-J_{\beta }^{(3)}\right) ,  \label{eqs of motion}
\end{equation}%
where we have suppressed the term\ $e^{+}\wedge e^{-}.$ The equations of
motion for the frame field $e$ imply the Virasoro constrains $Str(J_{\pm
}^{(2)},J_{\pm }^{(2)})=0$.

To introduce the Lax connection for the sigma model we combine the
Maurer-Cartan (MC) and the Euler-Lagrange (EL) equation into a single flat
connection. To make an ansatz for the Lax connection it is useful to recast
everything in terms of differential forms and see what are the independent
current components to be used in the ansatz.

The EL equations of motion (\ref{eqs of motion}) can be written, in the form%
\begin{eqnarray*}
d\ast J^{(2)}+J^{(0)}\wedge \ast J^{(2)}+\ast J^{(2)}\wedge J^{(0)}+\kappa
\left( J^{(1)}\wedge J^{(1)}-J^{(3)}\wedge J^{(3)}\right) &=&0, \\
J^{(1)}\wedge \ast J^{(2)}+\ast J^{(2)}\wedge J^{(1)}+\kappa \left(
J^{(1)}\wedge J^{(2)}+J^{(2)}\wedge J^{(1)}\right) &=&0, \\
J^{(3)}\wedge \ast J^{(2)}+\ast J^{(2)}\wedge J^{(3)}-\kappa \left(
J^{(3)}\wedge J^{(2)}+J^{(2)}\wedge J^{(3)}\right) &=&0,
\end{eqnarray*}%
where we have used the the MC equations $dJ+J\wedge J=0$ on the fermionic
part and the condition $\kappa ^{2}=1$. This is a good representation in the
sense that the world-sheet metric $h$ appears only in the term $\ast J^{(2)}$
through the Hodge star $\ast $ as in the Lagrangian $L_{GS}.$ We see that
the only forms involved are $J^{(0)},J^{(2)},\ast J^{(2)},J^{(1)}$,$%
J^{(3)},\ $thus the ansantz is 
\begin{equation}
L=l_{0}J^{(0)}+l_{1}J^{(2)}+l_{2}\ast J^{(2)}+l_{3}J^{(1)}+l_{4}J^{(3)},%
\text{ \ \ \ \ \ }l_{i}\in 
\mathbb{C}
.  \label{Lax kappa}
\end{equation}

For the curvature we have, after finding $l_{0}=1,$ that%
\begin{eqnarray*}
dL+L\wedge L &=&c_{1}J^{(1)}\wedge J^{(1)}+c_{2}J^{(2)}\wedge
J^{(2)}+c_{3}J^{(3)}\wedge J^{(3)}+c_{4}\left( J^{(1)}\wedge
J^{(2)}+J^{(2)}\wedge J^{(1)}\right) + \\
&&+c_{5}\left( J^{(1)}\wedge J^{(3)}+J^{(3)}\wedge J^{(1)}\right)
+c_{6}\left( J^{(2)}\wedge J^{(3)}+J^{(3)}\wedge J^{(2)}\right) ,
\end{eqnarray*}%
where%
\begin{eqnarray*}
c_{1} &=&l_{3}^{2}-l_{1}-\kappa l_{2},\text{ \ \ \ \ \ }%
c_{2}=l_{1}^{2}-l_{2}^{2}-1,\text{ \ \ \ \ \ }c_{3}=l_{4}^{2}-l_{1}+\kappa
l_{2}, \\
c_{4} &=&l_{1}l_{3}-\kappa l_{2}l_{3}-l_{4},\text{ \ \ \ }c_{5}=l_{3}l_{4}-1,%
\text{ \ \ \ \ \ }c_{6}=l_{1}l_{4}+\kappa l_{2}l_{4}-l_{3}.
\end{eqnarray*}%
This connection is flat when%
\begin{equation*}
l_{0}=1,\text{ \ \ \ \ \ }l_{1}\equiv w,\text{ \ \ \ \ \ }l_{2}=s_{2}\sqrt{%
w^{2}-1},\text{ \ \ \ \ \ }l_{3}=s_{3}\sqrt{w+\kappa \text{\ }l_{2}},\text{
\ \ \ \ \ }l_{4}=s_{4}\sqrt{w-\kappa \text{\ }l_{2}},
\end{equation*}%
where $s_{2},s_{3},s_{4}$ can be $\pm 1$ in any combination and $w$ is the
only independent complex parameter. The equations for $c_{4},c_{5},c_{6}$
are redundant and trivially satisfied.

In order to make contact with the superalgebra structure of the SSSSG
integrable hierarchy, we consider the following complex curve in $%
\mathbb{C}
^{2}$%
\begin{equation*}
w^{2}\equiv f(z),\text{ \ \ \ \ \ }f(z)=1+\frac{1}{4}\phi (z)^{-1},\text{ \
\ \ \ \ }\phi (z)^{-1}=\left( z^{2}-z^{-2}\right) ^{2},
\end{equation*}%
with $\phi (z)$ defined in (\ref{fundamental poly}). The relation between
the the spectral parameters $w,z$ of the $\sigma $-model and the SSSSG model
is defined through the curve above inducing the map $l_{i}(w)\rightarrow
l_{i}(z),$ with 
\begin{eqnarray*}
l_{0} &=&1,\text{ \ \ \ \ }l_{1}=\text{\ }w=s_{1}\frac{1}{2}\left(
z^{2}+z^{-1}\right) ,\text{ \ \ \ \ \ }l_{2}=s_{2}\frac{1}{2}\left(
z^{2}-z^{-2}\right) , \\
l_{3} &=&s_{3}\sqrt{\left( \frac{s_{1}+\kappa s_{2}}{2}\right) z^{2}+\left( 
\frac{s_{1}-\kappa s_{2}}{2}\right) z^{-2}},\text{ \ \ \ \ \ }l_{4}=s_{4}%
\sqrt{\left( \frac{s_{1}-\kappa s_{2}}{2}\right) z^{2}+\left( \frac{%
s_{1}+\kappa s_{2}}{2}\right) z^{-2}},
\end{eqnarray*}%
where $s_{2}=\pm 1.$ To recover the Lax operators (\ref{Lax op with gauge
fields}) consider the particular solution $s_{1}=-s_{2}=s_{3}=s_{4}=-\kappa
=1\ $implying%
\begin{equation*}
l_{1}-l_{2}=z^{2},\text{ \ \ \ \ \ }l_{1}+l_{2}=z^{-2},\text{ \ \ \ \ \ }%
l_{3}=z,\text{ \ \ \ \ \ }l_{4}=z^{-1}
\end{equation*}%
and the following form for the $\sigma $-model Lax connection%
\begin{equation}
L_{+}=J_{+}^{(0)}+zJ_{+}^{(1)}+z^{2}J_{+}^{(2)}+z^{-1}J_{+}^{(3)},\text{ \ \
\ \ \ }L_{-}=J_{-}^{(0)}+z^{-1}J_{-}^{(3)}+z^{-2}J_{-}^{(2)}+zJ_{-}^{(1)},
\label{Lax no kappa}
\end{equation}%
which is clearly valued in the twisted affine superalgebra $\widehat{%
\mathfrak{f}}$ defined in (\ref{superalgebra})$.$ It becomes the SSSSG Lax
connection after going to conformal gauge and fixing all the local
symmetries by means the Pohlmeyer reduction (recall that $%
J_{+}^{(3)}=J_{-}^{(1)}=0$)%
\begin{eqnarray}
J_{+}^{(0)} &=&\gamma ^{-1}\partial _{+}\gamma +\gamma
^{-1}A_{+}^{(l)}\gamma ,\text{ \ \ \ \ \ }J_{+}^{(1)}=\psi _{+},\text{ \ \ \
\ \ }J_{+}^{(2)}=-\Lambda ,\text{ \ \ \ \ \ }A_{+}^{(l)}=u^{-1}\partial
_{+}u,  \label{full Pohlmeyer} \\
J_{-}^{(0)} &=&A_{-}^{(r)},\text{ \ \ \ \ \ }J_{-}^{(3)}=\gamma ^{-1}\psi
_{-}\gamma ,\text{ \ \ \ \ \ }J_{-}^{(2)}=-\gamma ^{-1}\Lambda \gamma ,\text{
\ \ \ \ \ }A_{-}^{(r)}=\overline{u}^{-1}\partial _{-}\overline{u}.  \notag
\end{eqnarray}

The key point to show the existence of extended 2d supersymmetry in the
phase space of the reduced GS $\sigma $-model is to exploit the integrable
properties of the SSSSG integrable hierarchy, namely (\ref{orbit
intersection}) and rewrite the Lax operators in the following two equivalent
forms%
\begin{eqnarray}
\mathcal{L}_{\pm } &=&\chi \overline{u}^{-1}\left( \partial _{\pm }-z^{\pm
2}\Lambda \right) \overline{u}\chi ^{-1}=\gamma ^{-1}\widetilde{\chi }%
u^{-1}\left( \partial _{\pm }-z^{\pm 2}\Lambda \right) u\widetilde{\chi }%
^{-1}\gamma ,  \label{equivalent representations} \\
\mathcal{L}_{\epsilon } &=&\chi \overline{u}^{-1}\left( \delta _{\epsilon
}+z\epsilon \right) \overline{u}\chi ^{-1}=\gamma ^{-1}\widetilde{\chi }%
u^{-1}\left( \delta _{\epsilon }+z\epsilon \right) u\widetilde{\chi }%
^{-1}\gamma ,  \notag \\
\mathcal{L}_{\overline{\epsilon }} &=&\chi \overline{u}^{-1}\left( \delta _{%
\overline{\epsilon }}+z^{-1}\overline{\epsilon }\right) \overline{u}\chi
^{-1}=\gamma ^{-1}\widetilde{\chi }u^{-1}\left( \delta _{\overline{\epsilon }%
}+z^{-1}\overline{\epsilon }\right) u\widetilde{\chi }^{-1}\gamma ,  \notag
\end{eqnarray}%
supplemented by the constraints (\ref{constraints}). The operators $\mathcal{%
L}_{\epsilon }$ and $\mathcal{L}_{\overline{\epsilon }}$ are responsible for
the supersymmetry transformations.

In the next section we construct two possible sets of supersymmetry
transformations that, by construction, preserve the equations of motion $%
F_{+-}=0.$

\begin{remark}
In (\ref{Lax no kappa}) we have the action of the kappa symmetry
transformations and in (\ref{equivalent representations}) we have the action
of the 2d rigid supersymmetry transformations. The number of supersymmetries
and the number of kappa symmetries is the same and perhaps there is a
relation between them in the passage from (\ref{Lax no kappa}) to (\ref%
{equivalent representations}). However, we have not succeeded in showing
such a connection.
\end{remark}

\section{General supersymmetry variations.}

We now proceed to find the supersymmetry variations for all the fields $%
\gamma ,u,\overline{u},\psi _{\pm }$. This is an important result and for
this reason we do this in some detail.

Consider the $Q_{H}$ grade $\pm $1 SUSY flows associated to $b=z\epsilon ,$ $%
\epsilon \in \mathfrak{f}_{1}^{\perp }$ and $\overline{b}=z^{-1}\overline{%
\epsilon },$ $\overline{\epsilon }\in \mathfrak{f}_{3}^{\perp }$. From the
second expression of (\ref{equivalent representations}) we have two
equivalent representations for $\mathcal{L}_{\epsilon }$ (recall that $%
\mathcal{L}_{\epsilon }$ is valued on $Q_{H}$ grades 0 and 1 only)%
\begin{eqnarray*}
\mathcal{L}_{\epsilon } &=&\delta _{\epsilon }+\overline{u}^{-1}\delta
_{\epsilon }\overline{u}+\left[ y_{-1}+\theta _{-1},\epsilon _{\overline{u}}%
\right] +z\epsilon _{\overline{u}}, \\
\mathcal{L}_{\epsilon } &=&\delta _{\epsilon }+\gamma ^{-1}\delta _{\epsilon
}\gamma +\gamma ^{-1}u^{-1}\delta _{\epsilon }u\gamma -z\gamma ^{-1}\left(
\delta _{\epsilon }y_{+1}+\delta _{\epsilon }\theta _{+1}\right) \gamma
+z\gamma ^{-1}\left[ y_{+1}+\theta _{+1},u^{-1}\delta _{\epsilon }u\right]
\gamma +z\gamma ^{-1}\epsilon _{u}\gamma ,
\end{eqnarray*}%
where we have used the definitions (\ref{dressing matrices}) and defined $%
\epsilon _{\overline{u}}\equiv \overline{u}^{-1}\epsilon \overline{u},$ \ $%
\epsilon _{u}\equiv u^{-1}\epsilon u$, $\theta _{\pm 1}\in \widehat{%
\mathfrak{f}}^{\perp }$ $.$ By equating we have%
\begin{eqnarray*}
\gamma ^{-1}\delta _{\epsilon }\gamma &=&\left[ y_{-1}+\theta _{-1},\epsilon
_{\overline{u}}\right] +\overline{u}^{-1}\delta _{\epsilon }\overline{u}%
-\gamma ^{-1}u^{-1}\delta _{\epsilon }u\gamma , \\
\delta _{\epsilon }y_{+1} &=&-\left( \gamma \epsilon _{\overline{u}}\gamma
^{-1}\right) ^{\parallel }+\left[ y_{+1},u^{-1}\delta _{\epsilon }u\right] ,
\\
\delta _{\epsilon }\theta _{+1} &=&-\left( \gamma \epsilon _{\overline{u}%
}\gamma ^{-1}\right) ^{\perp }+\left[ \theta _{+1},u^{-1}\delta _{\epsilon }u%
\right] +\epsilon _{u},
\end{eqnarray*}%
where in the second and third lines we have splitted with respect to $%
\mathfrak{f}^{\parallel }$ and $\mathfrak{f}^{\perp }.$ The second line is
equivalent to%
\begin{equation*}
\delta _{\epsilon }\psi _{-}=-\left[ \Lambda ,\gamma \epsilon _{\overline{u}%
}\gamma ^{-1}\right] +\left[ \psi _{-},u^{-1}\delta _{\epsilon }u\right] .
\end{equation*}

To simplify, we introduce the notation%
\begin{equation*}
\mathcal{L}_{\epsilon }=\delta _{\epsilon }+w+z\epsilon _{\overline{u}},%
\text{ \ \ \ \ \ }w\equiv \overline{u}^{-1}\delta _{\epsilon }\overline{u}+%
\left[ y_{-1}+\theta _{-1},\epsilon _{\overline{u}}\right]
\end{equation*}%
and compute the variations coming from $\left[ \mathcal{L}_{\epsilon },%
\mathcal{L}_{+}\right] =0.$ This relation decomposes along $Q_{H}$ grades
0,1 and 2. The $Q_{H}$ grade 0 and 2 equation are satisfied while the $Q_{H}$
grade 1 equation splits along $\mathfrak{f}^{\parallel }$ and $\mathfrak{f}%
^{\perp }$ with the kernel part also satisfied and we are left with%
\begin{equation*}
\delta _{\epsilon }\psi _{+}=\left[ \left( \gamma ^{-1}D_{+}^{(l)}\gamma
\right) ^{\parallel },\epsilon _{\overline{u}}\right] -\left[ w^{\perp
},\psi _{+}\right] .
\end{equation*}%
Now we compute $\left[ \mathcal{L}_{\epsilon },\mathcal{L}_{-}\right] =0.$
This splits along $Q_{H}$ grades -2,-1,0 and 1. The $Q_{H}$ grade -1,-2 and
1 equations are all satisfied as well as the $\mathfrak{f}^{\parallel }$
part of the $Q_{H}$ grade 0 equation and we are left with%
\begin{equation*}
\delta _{\epsilon }A_{-}^{(r)}=D_{-}^{(r)}w^{\perp }+\left[ \left( \gamma
^{-1}\psi _{-}\gamma \right) ^{\perp },\epsilon _{\overline{u}}\right] .
\end{equation*}

Putting all together we have a raw expression for the SUSY variations 
\begin{eqnarray}
\gamma ^{-1}\delta _{\epsilon }\gamma &=&-\left[ \Lambda ,\left[ \psi
_{+},\epsilon _{\overline{u}}\right] \right] +w^{\perp }-\gamma
^{-1}u^{-1}\delta _{\epsilon }u\gamma ,  \label{general susy var} \\
\delta _{\epsilon }\psi _{+} &=&\left[ \left( \gamma ^{-1}D_{+}^{(L)}\gamma
\right) ^{\parallel },\epsilon _{\overline{u}}\right] -\left[ w^{\perp
},\psi _{+}\right] ,  \notag \\
\delta _{\epsilon }\psi _{-} &=&-\left[ \Lambda ,\gamma \epsilon _{\overline{%
u}}\gamma ^{-1}\right] +\left[ \psi _{-},u^{-1}\delta _{\epsilon }u\right] ,
\notag \\
\delta _{\epsilon }A_{-}^{(R)} &=&D_{-}^{(R)}w^{\perp }+\left[ \left( \gamma
^{-1}\psi _{-}\gamma \right) ^{\perp },\epsilon _{\overline{u}}\right] , 
\notag
\end{eqnarray}%
where $w^{\perp }=\overline{u}^{-1}\delta _{\epsilon }\overline{u}+\left[
\theta _{-1},\epsilon _{\overline{u}}\right] .$ The $\delta _{\overline{%
\epsilon }}$ SUSY variations are found by parity. We see that there is a
large amount of freedom in these expressions and this can be exploited
according to our needs. There are two important special cases to be
considered below.

\subsection{Local supersymmetry variations.}

These set of supersymmetry variations was originally introduced by hand in 
\cite{Goykhman} to be symmetries of a variant of the fermionic extension of
the perturbed gauge WZNW model associated to the equations of motion (\ref%
{equations of motion}) in the vector gauge $u=\overline{u}.$ Here we deduce
this kind of supersymmetry and show that it is a consequence of the
integrable hierarchy structure governing the SSSSG models.

The very form of (\ref{general susy var}) suggest the following choice%
\begin{equation*}
w^{\perp }=0,\text{ \ \ \ \ \ }u^{-1}\delta _{\epsilon }u=0
\end{equation*}%
and this imply that 
\begin{equation*}
\overline{u}^{-1}\delta _{\epsilon }\overline{u}=-\left[ \theta
_{-1},\epsilon _{\overline{u}}\right] ,\text{ \ \ \ \ }\ \delta _{\epsilon
}A_{+}^{(l)}=0.
\end{equation*}%
The consistency between the last equation of (\ref{general susy var}) and
the $\overline{u}^{-1}\delta _{\epsilon }\overline{u}$ variation right above
(use the identity $\delta _{\epsilon }A_{\pm }^{(r)}=D_{\pm }^{(r)}\left( 
\overline{u}^{-1}\delta _{\epsilon }\overline{u}\right) $) allow to
determine $\theta _{-1}$ for this case and we do not need its explicit form.
From the variation $\delta _{\epsilon }A_{-}^{(r)}$ we have 
\begin{equation*}
\ \delta _{\epsilon }\overline{u}\overline{u}^{-1}=\partial _{-}^{-1}\left\{ 
\overline{u}\left[ \left( \gamma ^{-1}\psi _{-}\gamma \right) ^{\perp
},\epsilon _{\overline{u}}\right] \overline{u}^{-1}\right\} .
\end{equation*}

The full set of SUSY variations is then given by%
\begin{eqnarray}
\gamma ^{-1}\delta _{\epsilon }\gamma &=&-\left[ \Lambda ,\left[ \psi
_{+},\epsilon _{\overline{u}}\right] \right] ,\text{ \ \ \ \ \ }\delta
_{\epsilon }\psi _{+}=\left[ \left( \gamma ^{-1}D_{+}^{(l)}\gamma \right)
^{\parallel },\epsilon _{\overline{u}}\right] ,\text{ \ \ \ \ \ }\delta
_{\epsilon }\psi _{-}=-\left[ \Lambda ,\gamma \epsilon _{\overline{u}}\gamma
^{-1}\right] ,  \label{off-shell susy} \\
\ \delta _{\epsilon }A_{+}^{(l)} &=&0,\text{ \ \ \ \ \ }\delta _{\epsilon
}A_{-}^{(r)}=\left[ \left( \gamma ^{-1}\psi _{-}\gamma \right) ^{\perp
},\epsilon _{\overline{u}}\right] ,\text{ \ \ \ \ \ \ }  \notag \\
u^{-1}\delta _{\epsilon }u &=&0,\text{ \ \ \ \ \ }\ \delta _{\epsilon }%
\overline{u}\overline{u}^{-1}=\partial _{-}^{-1}\left\{ \overline{u}\left[
\left( \gamma ^{-1}\psi _{-}\gamma \right) ^{\perp },\epsilon _{\overline{u}}%
\right] \overline{u}^{-1}\right\} ,  \notag
\end{eqnarray}%
and%
\begin{eqnarray}
\delta _{\overline{\epsilon }}\gamma \gamma ^{-1} &=&\left[ \Lambda ,\left[
\psi _{-},\overline{\epsilon }_{u}\right] \right] ,\text{ \ \ \ \ \ }\delta
_{\overline{\epsilon }}\psi _{-}=\left[ \left( \gamma D_{-}^{(r)}\gamma
^{-1}\right) ^{\parallel },\overline{\epsilon }_{u}\right] ,\text{ \ \ \ \ \ 
}\delta _{\overline{\epsilon }}\psi _{+}=-\left[ \Lambda ,\gamma ^{-1}%
\overline{\epsilon }_{u}\gamma \right] ,  \label{off-shell susy II} \\
\ \delta _{\overline{\epsilon }}A_{-}^{(r)} &=&0,\text{ \ \ \ \ }\delta _{%
\overline{\epsilon }}A_{+}^{(l)}=\left[ \left( \gamma \psi _{+}\gamma
^{-1}\right) ^{\perp },\overline{\epsilon }_{u}\right] ,\text{ \ \ \ \ \ \ }
\notag \\
\overline{u}^{-1}\delta _{\overline{\epsilon }}\overline{u} &=&0,\text{ \ \
\ \ \ }\ \delta _{\overline{\epsilon }}uu^{-1}=\partial _{+}^{-1}\left\{ u%
\left[ \left( \gamma \psi _{+}\gamma ^{-1}\right) ^{\perp },\overline{%
\epsilon }_{u}\right] u^{-1}\right\} .  \notag
\end{eqnarray}%
Notice that our construction shows that the natural gauge objects are $u$
and $\overline{u}$ and also explains the integrable origin of these local
symmetries.

Now, we want to see if (\ref{homomorphism}) holds. Unfortunately, the price
paid for introducing locality is that the SUSY algebra becomes field
dependent and highly non-trivial\footnote{%
In the rest of the paper we assume that $\left[ \epsilon ,\epsilon ^{\prime }%
\right] =2\epsilon \cdot \epsilon ^{\prime }\Lambda $ and $\left[ \overline{%
\epsilon },\overline{\epsilon }^{\prime }\right] =2\overline{\epsilon }\cdot 
\overline{\epsilon }^{\prime }\Lambda ,$ which is valid for the
superalgebras of interest.}. For the $\delta _{\epsilon }$ variations it is
given by\footnote{%
The $\mathfrak{f}^{\perp }$\ valued term $\widetilde{Q}$ is given by $%
\widetilde{Q}=Q-2\epsilon \cdot \epsilon ^{\prime }\left(
A_{+}^{(l)}-A_{+}^{(r)}\right) ,$ $Q=\mathcal{O}-2\epsilon \cdot \epsilon
^{\prime }(2\Lambda \psi _{+}^{2}),$ \ $\mathcal{O=}\left[ \left[ \widetilde{%
\Lambda }\psi _{+},\epsilon _{\overline{u}}\right] ,\left[ \widetilde{%
\Lambda }\psi _{+},\epsilon _{\overline{u}}^{\prime }\right] \right] .$}%
\begin{eqnarray*}
\left[ \delta _{\epsilon },\delta _{\epsilon ^{\prime }}\right] \gamma
&=&-2\epsilon \cdot \epsilon ^{\prime }\partial _{+}\gamma +\delta
_{\epsilon _{l/r}}\gamma ,\text{ \ \ \ }\left[ \delta _{\epsilon },\delta
_{\epsilon ^{\prime }}\right] \psi _{+}=-2\epsilon \cdot \epsilon ^{\prime
}\partial _{+}\psi _{+}+\delta _{\epsilon _{r}}\psi _{+}, \\
\left[ \delta _{\epsilon },\delta _{\epsilon ^{\prime }}\right] \psi _{-}
&=&-2\epsilon \cdot \epsilon ^{\prime }\partial _{+}\psi _{-}+\delta
_{\epsilon _{l}}\psi _{-},\text{ \ \ \ \ }\left[ \delta _{\epsilon },\delta
_{\epsilon ^{\prime }}\right] A_{-}^{(r)}=-2\epsilon \cdot \epsilon ^{\prime
}\partial _{+}A_{-}^{(r)}+\delta _{\epsilon _{r}}A_{-}^{(r)} \\
\left[ \delta _{\epsilon },\delta _{\epsilon ^{\prime }}\right] A_{+}^{(l)}
&=&-2\epsilon \cdot \epsilon ^{\prime }\partial _{+}A_{+}^{(l)}+\delta
_{\epsilon _{l}}A_{+}^{(l)}=0,\text{ \ \ \ \ \ }\epsilon _{l}\equiv -\text{\ 
}2\epsilon \cdot \epsilon ^{\prime }A_{+}^{(l)},\text{ \ \ \ \ }\epsilon
_{r}\equiv -\left( 2\epsilon \cdot \epsilon ^{\prime }A_{+}^{(l)}+\widetilde{%
Q}\right) ,\text{ \ \ \ \ \ }
\end{eqnarray*}%
where we have used the infinitesimal version of (\ref{full gauge transf.})
given by (take $h_{l/r}=\exp \epsilon _{l/r})$ 
\begin{eqnarray*}
\delta _{\epsilon _{l/r}}\gamma &=&\epsilon _{l}\gamma -\gamma \epsilon _{r},%
\text{ \ \ \ \ \ }\delta _{\epsilon _{r}}\psi _{+}=\left[ \epsilon _{r},\psi
_{+}\right] ,\text{ \ \ \ \ \ }\delta _{\epsilon _{l}}\psi _{-}=\left[
\epsilon _{l},\psi _{-}\right] , \\
\delta _{\epsilon _{l}}A_{+}^{(l)} &=&-D_{+}^{(l)}\epsilon _{l},\text{ \ \ \
\ \ }\delta _{\epsilon _{r}}A_{-}^{(r)}=-D_{-}^{(r)}\epsilon _{r}.
\end{eqnarray*}%
For the $\delta _{\overline{\epsilon }}$ variations it is given by\footnote{%
The $\mathfrak{f}^{\perp }$\ valued term $\widetilde{Q}^{\prime }$ is given
by $\widetilde{Q}^{\prime }=Q^{\prime }-2\overline{\epsilon }\cdot \overline{%
\epsilon }^{\prime }\left( A_{-}^{(r)}-A_{-}^{(l)}\right) ,$ \ $Q^{\prime }=%
\mathcal{O}^{\prime }-2\overline{\epsilon }\cdot \overline{\epsilon }%
^{\prime }(2\Lambda \psi _{-}^{2}),$ \ $\mathcal{O}^{\prime }\mathcal{=}%
\left[ \left[ \widetilde{\Lambda }\psi _{-},\overline{\epsilon }_{u}\right] ,%
\left[ \widetilde{\Lambda }\psi _{-},\overline{\epsilon }_{u}^{\prime }%
\right] \right] .$}%
\begin{eqnarray*}
\left[ \delta _{\overline{\epsilon }},\delta _{\overline{\epsilon }^{\prime
}}\right] \gamma &=&-2\overline{\epsilon }\cdot \overline{\epsilon }^{\prime
}\partial _{-}\gamma +\delta _{\epsilon _{l/r}}\gamma ,\text{ \ \ \ }\left[
\delta _{\overline{\epsilon }},\delta _{\overline{\epsilon }^{\prime }}%
\right] \psi _{-}=-2\overline{\epsilon }\cdot \overline{\epsilon }^{\prime
}\partial _{-}\psi _{-}+\delta _{\epsilon _{l}}\psi _{-}, \\
\left[ \delta _{\overline{\epsilon }},\delta _{\overline{\epsilon }^{\prime
}}\right] \psi _{+} &=&-2\overline{\epsilon }\cdot \overline{\epsilon }%
^{\prime }\partial _{-}\psi _{+}+\delta _{\epsilon _{r}}\psi _{+},\text{ \ \
\ \ }\left[ \delta _{\overline{\epsilon }},\delta _{\overline{\epsilon }%
^{\prime }}\right] A_{+}^{(l)}=-2\overline{\epsilon }\cdot \overline{%
\epsilon }^{\prime }\partial _{-}A_{+}^{(l)}+\delta _{\epsilon
_{l}}A_{+}^{(l)}, \\
\left[ \delta _{\overline{\epsilon }},\delta _{\overline{\epsilon }^{\prime
}}\right] A_{-}^{(r)} &=&-2\overline{\epsilon }\cdot \overline{\epsilon }%
^{\prime }\partial _{-}A_{-}^{(r)}+\delta _{\epsilon _{r}}A_{-}^{(r)}=0,%
\text{ \ \ \ \ \ }\epsilon _{r}\equiv -2\overline{\epsilon }\cdot \overline{%
\epsilon }^{\prime }A_{-}^{(r)},\text{ \ \ \ \ }\epsilon _{l}\equiv -\left( 2%
\overline{\epsilon }\cdot \overline{\epsilon }^{\prime }A_{-}^{(r)}+%
\widetilde{Q}^{\prime }\right) ,\text{ \ \ \ \ \ }
\end{eqnarray*}%
while for the mixed variations it is%
\begin{eqnarray*}
\left[ \delta _{\epsilon },\delta _{\overline{\epsilon }}\right] \gamma
&=&\epsilon _{l}\gamma -\gamma \epsilon _{r},\text{ \ \ \ \ \ }\left[ \delta
_{\epsilon },\delta _{\overline{\epsilon }}\right] \psi _{+}=\delta
_{\epsilon _{r}}\psi _{+},\text{ \ \ \ \ \ }\left[ \delta _{\epsilon
},\delta _{\overline{\epsilon }}\right] \psi _{-}=\delta _{\epsilon
_{l}}\psi _{-}, \\
\left[ \delta _{\epsilon },\delta _{\overline{\epsilon }}\right] A_{+}^{(l)}
&=&-D_{+}^{(l)}\epsilon _{l},\text{ \ \ \ \ \ }\left[ \delta _{\epsilon
},\delta _{\overline{\epsilon }}\right] A_{-}^{(r)}=-D_{-}^{(r)}\epsilon
_{r}, \\
\epsilon _{r} &\equiv &-\left[ \epsilon _{\overline{u}},\left( \gamma ^{-1}%
\overline{\epsilon }_{u}\gamma \right) ^{\perp }\right] ,\text{ \ \ \ \ \ }%
\epsilon _{l}\equiv -\left[ \left( \gamma \epsilon _{\overline{u}}\gamma
^{-1}\right) ^{\perp },\overline{\epsilon }_{u}\right] .
\end{eqnarray*}

Due to the fact that these superalgebra of flows is field dependent and does
not satisfy (\ref{homomorphism}), we will not consider it any further, at
least in this paper.

\subsection{Non-local supersymmetry variations.}

These rather simpler transformations were recently introduced in \cite{yo03}
as the on-shell supersymmetry variations associated to the phase space of
the Pohlmeyer reduced GSs$\sigma $ models on $AdS_{n}\times S^{n},$ $n=2,3,5$%
, namely the SSSSG models. See also \cite{SSSSG AdS5xS5} to see the same
results and for further details on soliton solutions and quantization.

These transformations are found in the on-shell gauge $u=\overline{u}=I.$
From (\ref{general susy var}) we have%
\begin{eqnarray}
\gamma ^{-1}\delta _{\epsilon }\gamma &=&-\left[ \widetilde{\Lambda }\psi
_{+},\epsilon \right] +w^{\perp },\text{ \ \ \ \ \ }\delta _{\epsilon }\psi
_{+}=\left[ \left( \gamma ^{-1}\partial _{+}\gamma \right) ^{\parallel
},\epsilon \right] -\left[ w^{\perp },\psi _{+}\right] ,
\label{on-shell susy} \\
\delta _{\epsilon }\psi _{-} &=&-\left[ \Lambda ,\gamma \epsilon \gamma ^{-1}%
\right] ,\text{ \ \ \ \ \ }w^{\perp }=\left[ \theta _{-1},\epsilon \right] ,%
\text{ \ \ \ \ \ }\theta _{-1}=-\partial _{-}^{-1}\left( \gamma ^{-1}\psi
_{-}\gamma \right) ^{\perp },  \notag \\
\delta _{\epsilon }\theta _{+1} &=&-\left( \gamma \epsilon \gamma
^{-1}\right) ^{\perp }+\epsilon  \notag
\end{eqnarray}%
and%
\begin{eqnarray}
\delta _{\overline{\epsilon }}\gamma \gamma ^{-1} &=&\left[ \widetilde{%
\Lambda }\psi _{-},\overline{\epsilon }\right] -\overline{w}^{\perp },\text{
\ \ \ \ }\delta _{\overline{\epsilon }}\psi _{-}=-\left[ \left( \partial
_{-}\gamma \gamma ^{-1}\right) ^{\parallel },\overline{\epsilon }\right] -%
\left[ \overline{w}^{\perp },\psi _{-}\right]  \label{on-shell susy II} \\
\delta _{\overline{\epsilon }}\psi _{+} &=&-\left[ \Lambda ,\gamma ^{-1}%
\overline{\epsilon }\gamma \right] ,\text{ \ \ \ \ \ }\overline{w}^{\perp }=%
\left[ \theta _{+1},\overline{\epsilon }\right] ,\text{ \ \ \ \ \ }\theta
_{+1}=-\partial _{+}^{-1}\left( \gamma \psi _{+}\gamma ^{-1}\right) ^{\perp
},  \notag \\
\delta _{\overline{\epsilon }}\theta _{-1} &=&-\left( \gamma ^{-1}\overline{%
\epsilon }\gamma \right) ^{\perp }+\overline{\epsilon },  \notag
\end{eqnarray}%
where $\theta _{-1}$, $\theta _{+1}$ were determined by the condition $%
\delta _{\epsilon }A_{-}^{(r)}=0$, $\delta _{\overline{\epsilon }%
}A_{+}^{(l)}=0,$ respectively$.$ These last conditions means that the
supersymmetry variations (\ref{on-shell susy}),(\ref{on-shell susy II})
preserve the constraints (\ref{level sets}), hence they are symmetries of
the SSSSG phase space.

Using these variations we can confirm (\ref{homomorphism}) by explicit
action on the fields\footnote{%
This is the full algebra, not the free limit approximation as was considered
in \cite{yo03}.} $\gamma ,\psi _{\pm }$%
\begin{eqnarray}
\left[ \delta _{\epsilon },\delta _{\epsilon ^{\prime }}\right] (\ast )
&=&2\epsilon \cdot \epsilon ^{\prime }\partial _{+}(\ast ),\text{ \ \ \ \ \ }%
\left[ \delta _{\overline{\epsilon }},\delta _{\overline{\epsilon }^{\prime
}}\right] (\ast )=2\overline{\epsilon }\cdot \overline{\epsilon }^{\prime
}\partial _{-}(\ast )  \label{flow superalgebra} \\
\left[ \delta _{\epsilon },\delta _{\overline{\epsilon }}\right] (\ast )
&=&\delta _{h}\left( \ast \right) ,\text{ \ \ \ \ \ }\delta _{h}\left( \ast
\right) =\left[ h,\ast \right] ,\text{ \ \ \ \ \ }h\equiv -\left[ \epsilon ,%
\overline{\epsilon }\right] .  \notag
\end{eqnarray}%
which is nothing but the algebra of $\mathfrak{s}$ written above in (\ref%
{lie sub-superalgebra})$.$ From this we see that the role of the non-local
terms $\theta _{\pm 1}$ in the variations is to maintain the isomorphism (%
\ref{homomorphism}), i.e, $\mathfrak{s}\simeq \widehat{\mathfrak{s}}$. The
key step in the proof of the mixed commutator, which is the most involved,
are the relations 
\begin{equation*}
\delta _{\epsilon }\overline{w}^{\perp }+\left[ \left( \gamma \epsilon
\gamma ^{-1}\right) ,\overline{\epsilon }\right] =\left[ \epsilon ,\overline{%
\epsilon }\right] ,\text{ \ \ \ \ \ }\delta _{\overline{\epsilon }}w^{\perp
}-\left[ \epsilon ,\left( \gamma ^{-1}\overline{\epsilon }\gamma \right)
^{\perp }\right] =-\left[ \epsilon ,\overline{\epsilon }\right] .
\end{equation*}

\section{Moment maps for the action of the superalgebra $\mathfrak{s}$.}

In this section we construct the moment maps for the actions on the phase
space of the symmetry flows of the sub-superalgebra $\mathfrak{s}$. This
leads to another identification of the reduced phase of the SSSSG model and
also explains the geometric origin of the non-Abelian R-symmetry group that
rotates the DS supercharges.

Consider the symplectic 2-form of the fermionic extension of the perturbed
WZNW model. It is given by the inverse of the Poisson bi-vector $\Theta
^{(0)}$ defined in (\ref{teta 0}), i.e, $\Omega _{WZNW}=\left( \Omega
_{+}-\Omega _{-}\right) ,$ with%
\begin{eqnarray}
-\Omega _{+} &=&\frac{1}{2}\left( \delta \gamma \gamma ^{-1}\wedge \partial
_{+}\left( \delta \gamma \gamma ^{-1}\right) \right) +\frac{1}{2}\left(
\delta \psi \wedge \widetilde{\Lambda }\delta \psi \right) ,
\label{SSSSG symplectic} \\
-\Omega _{-} &=&\frac{1}{2}\left( \gamma ^{-1}\delta \gamma \wedge \partial
_{-}\left( \gamma ^{-1}\delta \gamma \right) \right) +\frac{1}{2}\left(
\delta \overline{\psi }\wedge \widetilde{\Lambda }\delta \overline{\psi }%
\right) ,  \notag
\end{eqnarray}%
where we have re-scaled with the factor 1/2. We will need the following
identities 
\begin{equation*}
\delta q=D_{+}\left( \gamma ^{-1}\delta \gamma \right) ,\text{ \ \ \ }\delta 
\overline{q}=-D_{-}\left( \delta \gamma \gamma ^{-1}\right) ,\text{ \ \ \ }%
\gamma \delta q\gamma ^{-1}=\partial _{+}\left( \delta \gamma \gamma
^{-1}\right) ,\text{ \ \ \ }\gamma ^{-1}\delta \overline{q}\gamma =-\partial
_{-}\left( \gamma ^{-1}\delta \gamma \right) ,
\end{equation*}%
where we have defined $D_{-}(\ast )=\left[ \partial _{-}+\overline{q},\ast %
\right] .$

This symplectic form is invariant under the action of the chiral $%
H_{l}\times H_{r}$ gauge transformations%
\begin{equation}
\widetilde{\gamma }=h_{l}\gamma h_{r},\text{ \ \ \ \ \ }\widetilde{\psi }%
=h_{r}^{-1}\psi h_{r},\text{ \ \ \ \ }\widetilde{\text{\ }\overline{\psi }}%
=h_{l}\overline{\psi }h_{l}^{-1}\text{ \ \ \ \ \ }h_{l}\equiv h_{l}(x^{-}),%
\text{ \ \ \ \ \ }h_{r}\equiv h_{r}(x^{+})  \label{KM symmetry}
\end{equation}%
and to compute the moments $\mu _{l/r}$ for these actions, we use the
following definitions%
\begin{equation}
\delta H_{\epsilon }(\ast )=-\Omega \left( X_{\epsilon },\ast \right) ,\text{
\ \ \ \ \ }H_{\epsilon }=\dint\nolimits_{-\infty }^{+\infty }dxStr\left( \mu
,\epsilon \right) ,  \label{moment definition}
\end{equation}%
where $X_{\epsilon }$ is the vector field in the phase space induced by the
action of the symmetry $\epsilon ,$ i.e $X_{\epsilon }=\delta _{\epsilon
}\phi ^{i}.$ In our case we have $X_{\epsilon _{l/r}}=\left( \delta
_{\epsilon _{l/r}}\gamma ,\delta _{\epsilon _{l/r}}\psi ,\delta _{\epsilon
_{l/r}}\overline{\psi }\right) $ with%
\begin{equation*}
X_{\epsilon _{r}}=\left( \gamma \epsilon _{r},-\left[ \epsilon _{r},\psi %
\right] ,0\right) ,\text{ \ \ \ \ \ \ }X_{\epsilon _{l}}=\left( \epsilon
_{l}\gamma ,0,\left[ \epsilon _{l},\overline{\psi }\right] \right) ,
\end{equation*}%
where we have taken $h_{l/r}=\exp \epsilon _{l/r}$ in (\ref{KM symmetry})$.$
Thus, we need to compute the contractions 
\begin{equation*}
-\Omega _{WZNW}(X_{\epsilon _{r}},\ast )=-\Omega _{+}(X_{\epsilon _{r}},\ast
),\text{ \ \ \ \ \ }-\Omega _{WZNW}(X_{\epsilon _{l}},\ast )=\Omega
_{-}(X_{\epsilon _{l}},\ast ).
\end{equation*}

Inserting $X_{\epsilon _{r}}$ in the first equation of (\ref{SSSSG
symplectic}) yields\footnote{%
Recall that for left-right invariant vector fields $X_{l/r}$ we have $\delta
\gamma (X_{l})=\gamma X_{l}(e),$ $\delta \gamma (X_{r})=X_{r}(e)\gamma .$
Identifying $X_{l}=\gamma \epsilon _{r}$, $X_{r}=\epsilon _{l}\gamma $ and $%
X_{l}(e)=\epsilon _{r},$ $X_{r}(e)=\epsilon _{l}$ we obtain $\delta \gamma
(X)=\epsilon _{l}\gamma +\gamma \epsilon _{r}.$}%
\begin{equation*}
-\Omega _{+}(X_{\epsilon _{r}},\ast )=\left( \gamma \epsilon _{r}\gamma
^{-1},\partial _{+}\left( \delta \gamma \gamma ^{-1}\right) \right) -\left( 
\left[ \epsilon _{r},\psi \right] ,\widetilde{\Lambda }\delta \psi \right) .
\end{equation*}%
Now, using $\delta q=\gamma ^{-1}\partial _{+}\left( \delta \gamma \gamma
^{-1}\right) \gamma $ and $\delta (2\Lambda \psi \psi )=-\left[ \psi
,2\Lambda \delta \psi \right] $ we have%
\begin{equation*}
-\Omega _{+}(X_{\epsilon _{r}},\ast )=\delta \dint\nolimits_{-\infty
}^{+\infty }dxStr\left( q+2\Lambda \psi ^{2},\epsilon _{r}\right)
\end{equation*}%
and from this we identify, modulo $\delta $-exact terms, the moment for the
chiral right gauge action%
\begin{equation*}
\mu _{r}=q^{\perp }+2\Lambda \psi ^{2}.
\end{equation*}%
In a similar way we obtain%
\begin{equation*}
\Omega _{-}(X_{\epsilon _{l}},\ast )=\delta \dint\nolimits_{-\infty
}^{+\infty }dxStr\left( \overline{q}+2\Lambda \overline{\psi }^{2},-\epsilon
_{l}\right) \rightarrow \mu _{l}=\left( \overline{q}^{\perp }+2\Lambda 
\overline{\psi }\overline{\psi }\right) .
\end{equation*}

The moments $\mu _{l/r}$ are precisely the constraints (\ref{constraints})
naturally imposed by the dressing formalism of the SSSSG model (\ref{orbit
intersection}). From this we see that the SSSSG model corresponds to the
Hamiltonian reduction of the perturbed WZNW model by the chiral symmetry (%
\ref{KM symmetry}), i.e, $\mu _{l/r}=0,$ see \cite{Mikha II}$.$ Hence the
importance of restricting the Poisson structures to these level sets$.$ They
also affect the form of the DS charges as shown in (\ref{DS charges}). By
contracting again we easily show that%
\begin{equation*}
\Omega _{WZNW}(X_{\epsilon _{l}},X_{\epsilon _{l}^{\prime
}})=\dint\nolimits_{-\infty }^{+\infty }dxStr\left( \mu _{l},\left[ \epsilon
_{l},\epsilon _{l}^{\prime }\right] \right) ,\text{ \ \ }\Omega
_{WZNW}(X_{\epsilon _{r}},X_{\epsilon _{r}^{\prime
}})=\dint\nolimits_{-\infty }^{+\infty }dxStr\left( \mu _{r},\left[ \epsilon
_{r},\epsilon _{r}^{\prime }\right] \right)
\end{equation*}%
with the mixed contraction vanishing.

Now, we proceed to compute the moments for the supersymmetries (\ref%
{on-shell susy}), (\ref{on-shell susy II}) which is the novel part here. We
assume that the invariance of the symplectic form holds as a consequence of
the invariance of the Lagrangian \cite{yo03}. The advantage of this
computation is that it shows that the non-local SUSY variations are
Hamiltonian flows because they obey (\ref{moment definition}).

Let us write again the $\delta _{\epsilon }$ field variations (\ref{on-shell
susy}) in a more compact form%
\begin{eqnarray*}
\delta _{\epsilon }\gamma &=&\gamma \widehat{\epsilon },\text{ \ \ \ \ \ }%
\widehat{\epsilon }\equiv -\left[ \widetilde{\Lambda }\psi ,\epsilon \right]
+w^{\perp }, \\
\delta _{\epsilon }\psi &=&\left[ q^{\parallel },\epsilon \right] -\left[
w^{\perp },\psi \right] ,\text{ \ \ \ \ \ }\delta _{\epsilon }\overline{\psi 
}=-\left[ \Lambda ,\gamma \epsilon \gamma ^{-1}\right] .
\end{eqnarray*}%
Using the $\psi $ fermion equation of motion (\ref{equations of motion}) and
the explicit form of $w^{\perp }$ we can show that $\partial _{-}$\ $%
\widehat{\epsilon }=\left[ \epsilon ,\gamma ^{-1}\overline{\psi }\gamma %
\right] ,$ which will be used below. What we need to compute is the
following contraction%
\begin{equation}
-\Omega _{WZNW}(X_{\epsilon },\ast )=-\Omega _{+}(X_{\epsilon },\ast
)+\Omega _{-}(X_{\epsilon },\ast ).  \label{susy moment +1}
\end{equation}

Denoting by $X_{\epsilon }=\left( \delta _{\epsilon }\gamma ,\delta
_{\epsilon }\psi ,\delta _{\epsilon }\overline{\psi }\right) $ and using $%
\delta \gamma (X_{\epsilon })=\gamma \widehat{\epsilon },$ $\delta \psi
(X_{\epsilon })=\left[ q^{\parallel },\epsilon \right] -\left[ w^{\perp
},\psi \right] ,$ $\delta \overline{\psi }(X_{\epsilon })=-\left[ \Lambda
,\gamma \epsilon \gamma ^{-1}\right] $ we get%
\begin{eqnarray}
-\Omega _{+}(X_{\epsilon },\ast ) &=&\left( \gamma \text{\ }\widehat{%
\epsilon }\gamma ^{-1},\partial _{+}\left( \delta \gamma \gamma ^{-1}\right)
\right) +\left( \left[ q^{\parallel },\epsilon \right] -\left[ w^{\perp
},\psi \right] ,\widetilde{\Lambda }\delta \psi \right)  \label{omega +} \\
&=&\left( \widehat{\epsilon },\delta q\right) -\left( \epsilon ,\left[ q,%
\widetilde{\Lambda }\delta \psi \right] \right) -\left( w^{\perp },\left[
\psi ,\widetilde{\Lambda }\delta \psi \right] \right)  \notag \\
&=&\delta \left( \epsilon ,\left[ \widetilde{\Lambda }\psi ,q\right] \right)
+\left( w^{\perp },\delta \left( q+2\Lambda \psi \psi \right) \right)  \notag
\\
&=&\delta \left( \epsilon ,\left[ \widetilde{\Lambda }\psi ,q\right] \right)
+\left( \delta \mu _{R},w^{\perp }\right) .  \notag
\end{eqnarray}%
In a similar way, we have%
\begin{eqnarray}
-\Omega _{-}(X_{\epsilon },\ast ) &=&\left( \widehat{\epsilon },\partial
_{-}\left( \gamma ^{-1}\delta \gamma \right) \right) -\left( \left[ \Lambda
,\gamma \epsilon \gamma ^{-1}\right] ,\widetilde{\Lambda }\delta \overline{%
\psi }\right)  \label{omega -} \\
&=&-\left( \partial _{-}\widehat{\epsilon },\gamma ^{-1}\delta \gamma
\right) -\left( \left[ \Lambda ,\gamma \epsilon \gamma ^{-1}\right] ,%
\widetilde{\Lambda }\delta \overline{\psi }\right)  \notag \\
&=&-\left( \epsilon ,\gamma ^{-1}\delta \overline{\psi }\gamma +\left[
\gamma ^{-1}\overline{\psi }\gamma ,\gamma ^{-1}\delta \gamma \right] \right)
\notag \\
&=&-\delta \left( \epsilon ,\gamma ^{-1}\overline{\psi }\gamma \right) . 
\notag
\end{eqnarray}

Now, at fixed time ($dx^{\pm }=\pm dx)$ we find that%
\begin{equation*}
-\Omega _{+}(X_{\epsilon },\ast )=\delta \dint\nolimits_{-\infty }^{+\infty
}dxStr\left( \epsilon ,\left[ \widetilde{\Lambda }\psi ,q\right] \right)
+\dint\nolimits_{-\infty }^{+\infty }dxStr\left( \delta \mu _{R},w^{\perp
}\right) ,\text{ \ \ \ \ \ }\Omega _{-}(X_{\epsilon },\ast )=-\delta
\dint\nolimits_{-\infty }^{+\infty }dxStr\left( \epsilon ,\gamma ^{-1}%
\overline{\psi }\gamma \right) .
\end{equation*}%
Using (\ref{susy moment +1}) and repeating the computations for (\ref%
{on-shell susy II}) we obtain, modulo $\delta $-exact terms, the SUSY
moments, i.e, the supercurrent densities of (\ref{DS charges}), for the $%
\delta _{\epsilon }$ and $\delta _{\overline{\epsilon }}$ susy actions%
\begin{equation}
\mu =\left[ \widetilde{\Lambda }q,\psi \right] -\left( \gamma ^{-1}\overline{%
\psi }\gamma \right) ^{\perp },\text{ \ \ \ \ \ }\overline{\mu }=\left[ 
\widetilde{\Lambda }\overline{q},\overline{\psi }\right] -\left( \gamma \psi
\gamma ^{-1}\right) ^{\perp }.  \label{susy moments}
\end{equation}

Note that we can define moments for this rigid supersymmetry only when $\mu
_{l/r}=0,$ which is precisely when we restrict ourselves to the reduced
phase space in consistency with the Drinfeld-Sokolov procedure. Note also
that the non local terms in the SUSY variations are crucial for the result.
As above, we perform a second contraction to find the action of the gauge
group on $\mu $ and\ $\overline{\mu }$ 
\begin{eqnarray}
\Omega _{WZNW}(X_{\epsilon _{r}},X_{\epsilon }) &=&\dint\nolimits_{-\infty
}^{+\infty }dxStr\left( \mu ,\left[ \epsilon _{r},\epsilon \right] \right)
+\dint\nolimits_{-\infty }^{+\infty }dxStr\left( \mu _{r},\left[ \epsilon
_{r},w^{\perp }\right] \right) ,  \label{rotation} \\
\Omega _{WZNW}(X_{\epsilon _{l}},X_{\overline{\epsilon }})
&=&\dint\nolimits_{-\infty }^{+\infty }dxStr\left( \overline{\mu },\left[
-\epsilon _{l},\overline{\epsilon }\right] \right) +\dint\nolimits_{-\infty
}^{+\infty }dxStr\left( \mu _{l},\left[ -\epsilon _{l},\overline{w}^{\perp }%
\right] \right)  \notag
\end{eqnarray}%
with the other mixed brackets vanishing.

Recall that besides (\ref{KM symmetry}), we have the action of the residual
group $H$ of global gauge transformations 
\begin{equation*}
\widetilde{\gamma }=u\gamma v^{-1},\text{ \ \ \ \ \ }\widetilde{\psi }=v\psi
v^{-1},\text{ \ \ \ \ }\widetilde{\text{\ }\overline{\psi }}=u\overline{\psi 
}u^{-1}
\end{equation*}%
and under this action we can check explicitly that the moments obey the
usual equivariance property $\mu (g\cdot \phi )=\left( Ad^{\ast }g\right)
\mu (\phi )=g\mu (\phi )g^{-1}.$ Taking $g\cdot \phi =\widetilde{\phi }$ and 
$\phi ^{i}=(\gamma ,\psi ,\overline{\psi })$ we have%
\begin{eqnarray*}
\mu _{r}(\widetilde{\phi }) &=&\left( Ad^{\ast }v\right) \mu _{r}(\phi ),%
\text{ \ \ \ \ \ }\mu (\widetilde{\phi })\text{ }=\text{ }\left( Ad^{\ast
}v\right) \mu (\phi ), \\
\mu _{l}(\widetilde{\phi }) &=&\left( Ad^{\ast }u\right) \mu _{l}(\phi ),%
\text{ \ \ \ \ \ \ }\overline{\mu }(\widetilde{\phi })\text{ }=\text{ }%
\left( Ad^{\ast }u\right) \overline{\mu }(\phi ),
\end{eqnarray*}%
cf (\ref{rotation}) with $\epsilon _{r}=\epsilon _{r}=cte.$ The supercharges
transform under the global gauge group $H$ or R-symmetry group. Note that
the residual gauge transformations become the stabilizer of the moments $\mu
_{l/r}$ only when they vanish $\mu _{l/r}=0,$ becoming a symmetry of the
symplectic quotient $\mathcal{P}$, defined right below, and with coordinates
(\ref{phase coordinates}). The non-Abelian R-symmetry action is just the
equivariance of the susy moments and the SSSSG symplectic form on the
reduced phase space $\mathcal{P}$ are now%
\begin{equation}
\Omega _{SSSSG}=\Omega _{WZNW}\mid _{\mu _{l/r}=0},\text{ \ \ \ \ \ }%
\mathcal{P=}\widehat{\mu }^{-1}(0)//H,  \label{reduced symplectic form}
\end{equation}%
where $\widehat{\mu }\equiv (\mu _{l},\mu _{r}).$ Clearly, this way of
writing $\mathcal{P}$ is equivalent to the co-adjoint orbit formulation of
section 2.

The moments for the light-cone translations $x^{\pm }\rightarrow x^{\pm
}+\epsilon ^{\pm }\ $can be extracted directly from (\ref{DS charges}), by
writing $Str\left( \Lambda \text{,}q_{\pm 2}\right) =\dint\nolimits_{-\infty
}^{+\infty }dxStr\left( \mu _{\pm },\Lambda ^{\pm }\right) ,$ $\Lambda ^{\pm
}\equiv \epsilon ^{\pm }\Lambda .$ Their explicit form can be easily found
from the $T_{\mu \nu }$ components written in below in section 7.1.

To finish, we write $\mathfrak{s}$ and its associated moment maps $\mu :$ $%
\mathcal{P\rightarrow }\mathfrak{s}^{\ast }:$%
\begin{equation}
\mathfrak{s=}\left( \mathfrak{h\oplus f}_{1}^{\perp }\oplus \mathfrak{f}%
_{3}^{\perp }\right) \ltimes 
\mathbb{R}
^{2},\text{ \ \ \ \ \ }\mu _{l/r}\in \mathfrak{h}^{\ast },\text{ \ \ \ \ \ }%
\mu \in (\mathfrak{f}_{1}^{\perp })^{\ast },\text{ \ \ \ \ \ }\overline{\mu }%
\in (\mathfrak{f}_{3}^{\perp })^{\ast },\text{ \ \ \ \ \ }\mu _{\pm }\in
\left( 
\mathbb{R}
^{2}\right) ^{\ast }.  \label{moments for s}
\end{equation}

\section{The 2d supersymmetry algebra with kink central charges.}

We will find, under the second Poisson structure (\ref{second bracket}), the
full\footnote{%
In \cite{yo03} only half of the SUSY algebra were found, here we compute the
whole algebra.} Poisson superalgebra satisfied by the supercharges $q(b)$
and $\overline{q}(\overline{b})$ with $b=z\epsilon ,z^{2}\Lambda $ and $%
\overline{b}=z^{-1}\overline{\epsilon },z^{-2}\Lambda .$ There are no
charges associated to $k\in \mathfrak{h}$ because of (\ref{DS charges}) and (%
\ref{level sets}), hence all the brackets are computed on (\ref{reduced
symplectic form}) with $k\in \mathfrak{h}$ generating the R-symmetry group.
The generators $(b,\overline{b},k)$ span the sub-superalgebra $\widehat{%
\mathfrak{s}}$ of (\ref{s hat}) and the Poisson superalgebra is that of the
symmetries of $\mathfrak{s}$ of (\ref{lie sub-superalgebra}). As one would
expect from (\ref{flow superalgebra}) and (\ref{monodromy}), the boundary
central charge $Z$ of the mixed bracket $\left\{ q(z\epsilon ),\overline{q}%
(z^{-1}\overline{\epsilon })\right\} _{2}\sim Z$ should be related to the
kink electric charge $q_{0}$ of the global gauge symmetry $H$ but, as we
shall see, $Z$ and $q_{0}$ are not related at all and in turn, $Z$ turns out
to be a generalization of the well known surface terms of the $N=1$ and $N=2$
sine-Gordon and complex sine-Gordon model superalgebras \cite{Witten-Olive}
and \cite{Napo-Sciuto}, respectively.

\subsection{The superalgebra using the Poisson bi-vector $\Theta _{SSSSG}.$}

From (\ref{DS charges}), define $q(z\epsilon )\equiv Str(\epsilon ,q_{-1})$, 
$\overline{q}(z^{-1}\overline{\epsilon })\equiv Str(\overline{\epsilon }%
,q_{+1})$ and recall $q(z^{2}\Lambda )=Str(\Lambda ,q_{-2})$ and $\overline{q%
}(z^{-2}\Lambda )=Str(\Lambda ,q_{-2}).$ Using (\ref{DS currents}) we have%
\begin{eqnarray*}
q(z\epsilon ) &=&\dint\nolimits_{-\infty }^{+\infty }dx\left(
j_{+}(z\epsilon )+j_{-}(z\epsilon )\right) ,\text{ \ \ \ \ \ \ \ \ \ \ \ \ \ 
}q(z^{2}\Lambda )=\dint\nolimits_{-\infty }^{+\infty }dx\left(
j_{+}(z^{2}\Lambda )+j_{-}(z^{2}\Lambda )\right) , \\
\text{ }\overline{q}(z^{-1}\overline{\epsilon }) &=&\dint\nolimits_{-\infty
}^{+\infty }dx\left( \overline{j}_{+}(z^{-1}\overline{\epsilon }%
)+j_{-}(z^{-1}\overline{\epsilon })\right) ,\text{ \ \ \ \ \ }\overline{q}%
(z^{-2}\Lambda )=\dint\nolimits_{-\infty }^{+\infty }dx\left( \overline{j}%
_{+}(z^{-2}\Lambda )+\overline{j}_{-}(z^{-2}\Lambda )\right) ,
\end{eqnarray*}%
where $T_{++}\equiv j_{+}(z^{2}\Lambda ),$ $T_{-+}\equiv j_{-}(z^{2}\Lambda
),$ $T_{+-}\equiv \overline{j}_{+}(z^{-2}\Lambda )$ and $T_{--}\equiv 
\overline{j}_{-}(z^{-2}\Lambda ).$ We use the brackets (\ref{second brackets}%
) and (\ref{spatial orbit}) in the computation.

One important comment is in order here. It is not difficult to show that the
brackets $\left\{ q(z\epsilon ),q(z\epsilon ^{\prime })\right\} _{2}$ and $%
\left\{ \overline{q}(z^{-1}\overline{\epsilon }),\overline{q}(z^{-1}%
\overline{\epsilon }^{\prime })\right\} _{2}$ are insensitive to the action
of (\ref{infinite ambiguity}), hence they take the same value along the
orbits (\ref{kernel orbit}) and the computation can be done by using the
local diffentials (\ref{differentials}). The subtlety appears with the mixed
bracket $\left\{ q(z\epsilon ),\overline{q}(z^{-1}\overline{\epsilon }%
)\right\} _{2}$, which is not invariant. In this case we perform a
compensating gauge transformation $\eta _{\pm }$ with $\beta _{\pm 1}=\theta
_{\pm 1},$ where $\theta _{\pm 1}$ are defined by (\ref{on-shell susy}),(\ref%
{on-shell susy II}).

Let us consider first the bracket of $q(z\epsilon )$ with itself. This
bracket was already computed in \cite{yo03}, then we pose only the items
necessary for compute it. From (\ref{differentials}) we find the
supercurrent differentials%
\begin{eqnarray*}
d_{q}\text{\ }j_{+}(z\epsilon ) &=&\left[ y_{-1},\epsilon \right] ,\text{ \
\ \ \ \ }d_{\psi }j_{+}(z\epsilon )=\left[ y_{-2},\epsilon \right] +\frac{1}{%
2}\left[ y_{-1},\left[ y_{-1},\epsilon \right] \right] , \\
d_{\overline{q}}\text{\ }j_{-}(z\epsilon ) &=&-\left[ y_{-1},\epsilon \right]
,\text{ \ \ \ \ \ }d_{\overline{\psi }}j_{-}(z\epsilon )=-\epsilon .
\end{eqnarray*}%
Using them, we have%
\begin{eqnarray*}
\left\{ \left( j_{+}(z\epsilon )\right) ,\left( j_{+}(z\epsilon ^{\prime
})\right) \right\} _{2}(\mathcal{L}_{+}) &=&-\left( D_{+},\left[ d_{q}\text{%
\ }j_{+}(z\epsilon ),d_{q}\text{\ }j_{+}(z\epsilon ^{\prime })\right]
\right) -\left( \Lambda ,\left[ d_{\psi }j_{+}(z\epsilon ),d_{\psi
}j_{+}(z\epsilon ^{\prime })\right] \right) , \\
\left( D_{+},\left[ d_{q}\text{\ }j_{+}(z\epsilon ),d_{q}\text{\ }%
j_{+}(z\epsilon ^{\prime })\right] \right) &=&-2\epsilon \cdot \epsilon
^{\prime }\left( \Lambda \psi \partial _{+}\psi +\left( q^{\perp }\right)
^{2}\right) , \\
\left( \Lambda ,\left[ d_{\psi }j_{+}(z\epsilon ),d_{\psi }j_{+}(z\epsilon
^{\prime })\right] \right) &=&-2\epsilon \cdot \epsilon ^{\prime }\left( 
\frac{1}{2}\left( q^{\parallel }\right) ^{2}\right) ,
\end{eqnarray*}%
where we have used $\left[ \epsilon ,\epsilon ^{\prime }\right] =2\epsilon
\cdot \epsilon ^{\prime }\Lambda .$ Then, we get%
\begin{equation*}
\left\{ \left( j_{+}(z\epsilon )\right) ,\left( j_{+}(z\epsilon ^{\prime
})\right) \right\} _{2}(\mathcal{L}_{+})=-2\epsilon \cdot \epsilon ^{\prime
}\dint\nolimits_{-\infty }^{+\infty }dx^{+}T_{++},\text{ \ \ \ \ \ }%
T_{++}=-Str\left( \frac{1}{2}\left( q^{\parallel }\right) ^{2}+\left(
q^{\perp }\right) ^{2}+\Lambda \psi \partial _{+}\psi \right) .
\end{equation*}

In a similar way we have%
\begin{eqnarray*}
\left\{ \left( j_{-}(z\epsilon )\right) ,\left( j_{-}(z\epsilon ^{\prime
})\right) \right\} _{2}(\mathcal{L}_{-}) &=&-\left( \partial _{-},\left[ d_{%
\overline{q}}\text{\ }j_{-}(z\epsilon ),d_{\overline{q}}\text{\ }%
j_{-}(z\epsilon ^{\prime })\right] \right) -\left( \gamma ^{-1}\Lambda
\gamma ,\left[ d_{\overline{\psi }}j_{-}(z\epsilon ),d_{\overline{\psi }%
}j_{-}(z\epsilon ^{\prime })\right] \right) , \\
\left( \partial _{-},\left[ d_{\overline{q}}\text{\ }j_{-}(z\epsilon ),d_{%
\overline{q}}\text{\ }j_{-}(z\epsilon ^{\prime })\right] \right)
&=&2\epsilon \cdot \epsilon ^{\prime }\left( \frac{1}{2}\gamma ^{-1}%
\overline{\psi }\gamma \psi \right) , \\
\left( \gamma ^{-1}\Lambda \gamma ,\left[ d_{\overline{\psi }%
}j_{-}(z\epsilon ),d_{\overline{\psi }}j_{-}(z\epsilon ^{\prime })\right]
\right) &=&2\epsilon \cdot \epsilon ^{\prime }\left( \gamma ^{-1}\Lambda
\gamma \Lambda \right)
\end{eqnarray*}%
and 
\begin{equation*}
\left\{ \left( j_{-}(z\epsilon )\right) ,\left( j_{-}(z\epsilon ^{\prime
})\right) \right\} _{2}(\mathcal{L}_{-})=-2\epsilon \cdot \epsilon ^{\prime
}\dint\nolimits_{-\infty }^{+\infty }dx^{-}T_{-+},\text{ \ \ \ \ \ }%
T_{-+}=Str\left( \gamma ^{-1}\Lambda \gamma \Lambda +\frac{1}{2}\gamma ^{-1}%
\overline{\psi }\gamma \psi \right) .
\end{equation*}

At fixed time ($dx^{\pm }\rightarrow \pm dx$) we have%
\begin{eqnarray*}
\left\{ q(z\epsilon ),q(z\epsilon ^{\prime })\right\} _{2}(\mathcal{L}_{x})
&=&\left\{ \left( j_{+}(z\epsilon )\right) ,\left( j_{+}(z\epsilon ^{\prime
})\right) \right\} _{2}(\mathcal{L}_{+})-\left\{ \left( j_{-}(z\epsilon
)\right) ,\left( j_{-}(z\epsilon ^{\prime })\right) \right\} _{2}(\mathcal{L}%
_{-}) \\
&=&-2\epsilon \cdot \epsilon ^{\prime }\dint_{-\infty }^{+\infty }dx\left(
T_{++}+T_{-+}\right) =-q\left( z^{2}\left[ \epsilon ,\epsilon ^{\prime }%
\right] \right) ,
\end{eqnarray*}%
and we have shown that 
\begin{equation}
\left\{ q(z\epsilon ),q(z\epsilon ^{\prime })\right\} _{2}(\mathcal{P}%
)=-q\left( z^{2}\left[ \epsilon ,\epsilon ^{\prime }\right] \right) \text{.}
\label{++ super}
\end{equation}

By parity we have%
\begin{equation}
\left\{ \overline{q}(z^{-1}\overline{\epsilon }),\overline{q}(z^{-1}%
\overline{\epsilon }^{\prime })\right\} _{2}(\mathcal{L}^{\prime })=-2%
\overline{\epsilon }\cdot \overline{\epsilon }^{\prime }\dint_{-\infty
}^{+\infty }dx\left( T_{+-}+T_{--}\right) =-\overline{q}\left( z^{-2}\left[ 
\overline{\epsilon },\overline{\epsilon }^{\prime }\right] \right)  \notag
\end{equation}%
and we have that 
\begin{equation}
\left\{ \overline{q}(z^{-1}\overline{\epsilon }),\overline{q}(z^{-1}%
\overline{\epsilon }^{\prime })\right\} _{2}(\mathcal{P})=-\overline{q}%
\left( z^{-2}\left[ \overline{\epsilon },\overline{\epsilon }^{\prime }%
\right] \right) \text{.}  \label{-- supers}
\end{equation}

Now we compute the mixed bracket of $q(z\epsilon )$ with $\overline{q}(z^{-1}%
\overline{\epsilon }),$ which is the novel part here. We need to find 
\begin{eqnarray*}
\left\{ \left( j_{+}(z\epsilon )\right) ,\left( \overline{j}_{+}(z^{-1}%
\overline{\epsilon })\right) \right\} _{2}(\mathcal{L}_{+}) &=&-\left( D_{+},%
\left[ d_{q}\text{\ }j_{+}(z\epsilon ),d_{q}\text{\ }\overline{j}_{+}(z^{-1}%
\overline{\epsilon })\right] \right) -\left( \Lambda ,\left[ d_{\psi
}j_{+}(z\epsilon ),d_{\psi }\overline{j}_{+}(z^{-1}\overline{\epsilon })%
\right] \right) , \\
\left\{ \left( j_{-}(z\epsilon )\right) ,\left( \overline{j}_{-}(z^{-1}%
\overline{\epsilon })\right) \right\} _{2}(\mathcal{L}_{-}) &=&-\left(
\partial _{-},\left[ d_{\overline{q}}\text{\ }j_{-}(z\epsilon ),d_{\overline{%
q}}\ \overline{j}_{-}(z^{-1}\overline{\epsilon })\right] \right) -\left(
\gamma ^{-1}\Lambda \gamma ,\left[ d_{\overline{\psi }}j_{-}(z\epsilon ),d_{%
\overline{\psi }}\overline{j}_{-}(z^{-1}\overline{\epsilon })\right] \right)
.
\end{eqnarray*}%
The differentials (\ref{differentials})\ including the action of (\ref%
{infinite ambiguity}) are given by 
\begin{eqnarray*}
\text{ \ \ \ \ \ }d_{\psi }j_{+}(z\epsilon ) &=&\left[ y_{-2}+\theta
_{-2},\epsilon \right] +\frac{1}{2}\left[ y_{-1},\left[ y_{-1},\epsilon %
\right] \right] +\frac{1}{2}\left[ \theta _{-1},\left[ \theta _{-1},\epsilon %
\right] \right] +\left[ y_{-1},\left[ \theta _{-1},\epsilon \right] \right] ,
\\
d_{q}\text{\ }\overline{j}_{+}(z^{-1}\overline{\epsilon }) &=&-\gamma ^{-1}%
\left[ y_{+1}+\theta _{+1},\overline{\epsilon }\right] \gamma ,\text{ \ \ \
\ \ }d_{q}\text{\ }j_{+}(z\epsilon )=\left[ y_{-1}+\theta _{-1},\epsilon %
\right] ,\text{\ \ \ \ \ }d_{\psi }\overline{j}_{+}(z^{-1}\overline{\epsilon 
})=-\gamma ^{-1}\overline{\epsilon }\gamma
\end{eqnarray*}%
and%
\begin{eqnarray*}
d_{\overline{q}}\text{\ }j_{-}(z\epsilon ) &=&-\left[ y_{-1}+\theta
_{-1},\epsilon \right] ,\text{ \ \ \ \ \ }d_{\overline{q}}\text{\ }\overline{%
j}_{-}(z^{-1}\overline{\epsilon })=\gamma ^{-1}\left[ y_{+1}+\theta _{+1},%
\overline{\epsilon }\right] \gamma ,\text{ \ \ \ \ \ }d_{\overline{\psi }%
}j_{-}(z\epsilon )=-\epsilon , \\
\text{ \ \ \ \ \ }d_{\overline{\psi }}\overline{j}_{-}(z^{-1}\overline{%
\epsilon }) &=&\gamma ^{-1}\left( \left[ y_{+2}+\theta _{+2},\overline{%
\epsilon }\right] +\frac{1}{2}\left[ y_{+1},\left[ y_{+1},\overline{\epsilon 
}\right] \right] +\frac{1}{2}\left[ \theta _{+1},\left[ \theta _{+1},%
\overline{\epsilon }\right] \right] +\left[ y_{+1},\left[ \theta _{+1},%
\overline{\epsilon }\right] \right] \right) \gamma .
\end{eqnarray*}

Let us compute the easiest term first%
\begin{equation*}
\left( \Lambda ,\left[ d_{\psi }j_{+}(z\epsilon ),d_{\psi }\overline{j}%
_{+}(z^{-1}\overline{\epsilon })\right] \right) =-\left( q^{\parallel },%
\left[ \epsilon ,\gamma ^{-1}\overline{\epsilon }\gamma \right] \right)
-\left( \psi ,\left[ \left[ \theta _{-1},\epsilon \right] ,\gamma ^{-1}%
\overline{\epsilon }\gamma \right] \right) .
\end{equation*}%
Now we consider%
\begin{eqnarray*}
\left( D_{+},\left[ d_{q}\text{\ }j_{+}(z\epsilon ),d_{q}\text{\ }\overline{j%
}_{+}(z^{-1}\overline{\epsilon })\right] \right) &=&-\left( \left[
D_{+},d_{q}\text{\ }\overline{j}_{+}(z^{-1}\overline{\epsilon })\right]
,d_{q}\text{\ }j_{+}(z\epsilon )\right) \\
&=&\left( \psi ,\left[ \left[ y_{-1}+\theta _{-1},\epsilon \right] ,\gamma
^{-1}\overline{\epsilon }\gamma \right] \right) ,
\end{eqnarray*}%
where we have used%
\begin{equation*}
\left[ D_{+},d_{q}\text{\ }\overline{j}_{+}(z^{-1}\overline{\epsilon })%
\right] =\left[ \psi ,\gamma ^{-1}\overline{\epsilon }\gamma \right] ,\text{
\ \ \ \ \ }\partial _{+}\left( y_{+1}+\theta _{+1}\right) =-\gamma \psi
\gamma ^{-1}.
\end{equation*}%
Then, we have%
\begin{equation*}
\left\{ \left( j_{+}(z\epsilon )\right) ,\left( \overline{j}_{+}(z^{-1}%
\overline{\epsilon })\right) \right\} _{2}(\mathcal{L}_{+})=\dint%
\nolimits_{-\infty }^{+\infty }dx^{+}Str\left( q,\left[ \epsilon ,\gamma
^{-1}\overline{\epsilon }\gamma \right] \right) .
\end{equation*}

On the other hand we have%
\begin{eqnarray*}
\left( \gamma ^{-1}\Lambda \gamma ,\left[ d_{\overline{\psi }%
}j_{-}(z\epsilon ),d_{\overline{\psi }}\overline{j}_{-}(z^{-1}\overline{%
\epsilon })\right] \right) &=&-\left( \overline{q}^{\parallel },\left[
\gamma \epsilon \gamma ^{-1},\overline{\epsilon }\right] \right) -\left( 
\overline{\psi },\left[ \gamma \epsilon \gamma ^{-1},\left[ \theta _{+1},%
\overline{\epsilon }\right] \right] \right) , \\
\left( \partial _{-},\left[ d_{\overline{q}}\text{\ }j_{-}(z\epsilon ),d_{%
\overline{q}}\ \overline{j}_{-}(z^{-1}\overline{\epsilon })\right] \right)
&=&\left( \overline{\psi },\left[ \gamma \epsilon \gamma ^{-1},\left[
y_{+1}+\theta _{+1},\overline{\epsilon }\right] \right] \right) ,
\end{eqnarray*}%
where we have used 
\begin{equation*}
\left[ \partial _{-},d_{\overline{q}}\text{\ }j_{-}(z\epsilon )\right] =%
\left[ \gamma ^{-1}\overline{\psi }\gamma ,\epsilon \right] \text{, \ \ \ \
\ }\partial _{-}\left( y_{-1}+\theta _{-1}\right) =-\gamma ^{-1}\overline{%
\psi }\gamma .
\end{equation*}%
Thus, we get%
\begin{equation*}
\left\{ \left( j_{-}(z\epsilon )\right) ,\left( \overline{j}_{-}(z^{-1}%
\overline{\epsilon })\right) \right\} _{2}(\mathcal{L}_{-})=\dint%
\nolimits_{-\infty }^{+\infty }dx^{-}Str\left( \overline{q},\left[ \gamma
\epsilon \gamma ^{-1},\overline{\epsilon }\right] \right) .
\end{equation*}

Putting all together we have, at fixed time ($dx^{\pm }\rightarrow \pm dx$),
that 
\begin{eqnarray*}
\left\{ q(z\epsilon ),\overline{q}(z^{-1}\overline{\epsilon })\right\} _{2}(%
\mathcal{L}_{x}) &=&\left\{ \left( j_{+}(z\epsilon )\right) ,\left( 
\overline{j}_{+}(z^{-1}\overline{\epsilon })\right) \right\} _{2}(\mathcal{L}%
_{+})-\left\{ \left( j_{-}(z\epsilon )\right) ,\left( \overline{j}_{-}(z^{-1}%
\overline{\epsilon })\right) \right\} _{2}(\mathcal{L}_{-}) \\
&=&\dint\nolimits_{-\infty }^{+\infty }dx\partial _{x}Str\left( \gamma
\epsilon \gamma ^{-1}\overline{\epsilon }\right) =Z_{\epsilon ,\overline{%
\epsilon }},\text{ \ \ \ \ \ }Z_{\epsilon ,\overline{\epsilon }}\equiv
Str\left( \gamma \epsilon \gamma ^{-1}\overline{\epsilon }\right) \mid
_{-\infty }^{+\infty }
\end{eqnarray*}%
and we have shown that\footnote{%
If we compute $\left\{ q(z\epsilon ),\overline{q}(z^{-1}\overline{\epsilon }%
)\right\} _{2}\left( \mathcal{L}_{x}^{\prime }\right) ,$ we get the same
answer.}%
\begin{equation}
\left\{ q(z\epsilon ),\overline{q}(z^{-1}\overline{\epsilon })\right\} _{2}(%
\mathcal{P})=Z_{\epsilon ,\overline{\epsilon }}\text{.}  \label{mixed super}
\end{equation}

Let us consider $Z_{\epsilon ,\overline{\epsilon }}$ for some particular
models in order to have a better feeling about its meaning\footnote{%
Use the results of \cite{yo03} with $B\rightarrow \gamma ,$ $%
F_{1i}\rightarrow f_{i}^{(1)}$ and $F_{3j}\rightarrow f_{j}^{(3)}.$}. First
we take\footnote{%
The supraindexes $(1),(3)$ denote the $%
\mathbb{Z}
_{4}$ grading.} $\epsilon =\epsilon _{i}f_{i}^{(1)},\overline{\epsilon }=%
\overline{\epsilon }_{j}f_{j}^{(3)}$ and rewrite 
\begin{equation*}
Str\left( \gamma \epsilon \gamma ^{-1}\overline{\epsilon }\right) =\epsilon
_{i}\overline{\epsilon }_{j}K_{ij},\text{ \ \ \ \ \ }K_{ij}=Str\left( \gamma
f_{i}^{(1)}\gamma ^{-1},f_{j}^{(3)}\right) .
\end{equation*}%
For the Pohlmeyer reduced $AdS_{2}\times S^{2}$ $\sigma $-model we have%
\begin{equation*}
K=2\left( 
\begin{array}{cc}
\cos \varphi \cosh \phi & -\sin \varphi \sinh \phi \\ 
\sin \varphi \sinh \phi & \cos \varphi \cosh \phi%
\end{array}%
\right) .
\end{equation*}%
The $N=(1,1)$ sine-Gordon model is obtained by taking $\phi =0$ and by
supressing some fermions and SUSY parameters. Putting $\phi =0$ above, we
have $K_{SG}=2\mu I\cos \varphi ,$ where we have re-installed the mass
parameter $\mu .$ In this case the bracket (\ref{mixed super}) reduces to
the well known result \cite{Witten-Olive} (if we take $\epsilon _{1}=%
\overline{\epsilon }_{1}=0$ or $\epsilon _{2}=\overline{\epsilon }_{2}=0$)%
\begin{equation*}
\left\{ q(z\epsilon ),\overline{q}(z^{-1}\overline{\epsilon })\right\} _{2}(%
\mathcal{P})=\epsilon \cdot \overline{\epsilon }\left( 2\mu \cos \varphi
\mid _{-\infty }^{+\infty }\right) .
\end{equation*}

For the Pohlmeyer reduced $AdS_{3}\times S^{3}$ models we find%
\begin{equation*}
K=2\left( 
\begin{array}{cccc}
\cos \varphi \cosh \phi \sin \left( \theta -\chi \right) & \cos \varphi
\cosh \phi \cos \left( \theta -\chi \right) & -\sin \varphi \sinh \phi \cos
\left( t-t^{\prime }\right) & \sin \varphi \sinh \phi \sin \left(
t-t^{\prime }\right) \\ 
-\cos \varphi \cosh \phi \cos \left( \theta -\chi \right) & \cos \varphi
\cosh \phi \sin \left( \theta -\chi \right) & -\sin \varphi \sinh \phi \sin
\left( t-t^{\prime }\right) & -\sin \varphi \sinh \phi \cos \left(
t-t^{\prime }\right) \\ 
\sin \varphi \sinh \phi \cos \left( t-t^{\prime }\right) & \sin \varphi
\sinh \phi \sin \left( t-t^{\prime }\right) & \cos \varphi \cosh \phi \sin
\left( \theta -\chi \right) & -\cos \varphi \cosh \phi \cos \left( \theta
-\chi \right) \\ 
-\sin \varphi \sinh \phi \sin \left( t-t^{\prime }\right) & \sin \varphi
\sinh \phi \cos \left( t-t^{\prime }\right) & \cos \varphi \cosh \phi \cos
\left( \theta -\chi \right) & \cos \varphi \cosh \phi \sin \left( \theta
-\chi \right)%
\end{array}%
\right) .
\end{equation*}%
At spatial infinity $x\rightarrow \pm \infty $ we have that $\gamma
\rightarrow \gamma _{0}\in H\ $which is satisfied by the vacuum values $\phi
(\pm \infty )=0$ and $\varphi (\pm \infty )=\pi n_{\pm },$ $n_{\pm }\in 
\mathbb{Z}
.$ Then, we can set $K_{AdS_{3}\times S^{3}}=diag(K_{+},K_{-})$, where%
\footnote{%
The Pauli matrices are 
\begin{equation*}
\sigma ^{1}=\left( 
\begin{array}{cc}
0 & 1 \\ 
1 & 0%
\end{array}%
\right) ,\text{ \ }\sigma ^{2}=\left( 
\begin{array}{cc}
0 & -i \\ 
i & 0%
\end{array}%
\right) ,\text{ \ }\sigma ^{3}=\left( 
\begin{array}{cc}
1 & 0 \\ 
0 & -1%
\end{array}%
\right)
\end{equation*}%
} 
\begin{equation*}
K_{\pm }=2\mu \cos \varphi \left( I\sin \left( \theta -\chi \right) \pm
i\sigma ^{2}\cos \left( \theta -\chi \right) \right) .
\end{equation*}

Strictly speaking, the $N=(2,2)$ complex sine-Gordon model cannot be
obtained by truncation \cite{Grigo-Tseytlin II}, but for illustrative
purposes we ignore this and take the limit\footnote{%
This limit corresponds to $\gamma =diag(\gamma _{A},\gamma _{S})\rightarrow
diag(I_{2},\gamma _{S}),$ where $A$, $S$ refers to the $AdS$ and $S$ spaces.}
$\phi =\chi =t=0.$ In this case we have that $K_{\text{cSG}}=diag(\overline{K%
}_{+},\overline{K}_{-}),$ where $\overline{K}_{\pm }=K_{\pm }(\chi =0).$ If
we take $\epsilon _{1}$=$\epsilon _{2}=\overline{\epsilon }_{1}=\overline{%
\epsilon }_{2}=0$ or $\epsilon _{3}$=$\epsilon _{4}=\overline{\epsilon }_{3}=%
\overline{\epsilon }_{4}=0$ we use $\overline{K}_{-}$ or $\overline{K}_{+}$
respectively. For example, if we choose $\overline{K}_{+}$ and rotate it to $%
K^{\prime }=M^{\dagger }\overline{K}_{+}M$ we recover the well known result 
\cite{Napo-Sciuto}%
\begin{equation*}
\left\{ q(z\epsilon )^{\prime },\overline{q}(z^{-1}\overline{\epsilon }%
)^{\prime }\right\} _{2}=\epsilon _{i}\overline{\epsilon }_{j}K_{ij}^{\prime
}\mid _{-\infty }^{+\infty },\text{ \ \ \ \ \ }K^{\prime }=-2\mu i\cos
\varphi \left( iI\sin \theta +\sigma ^{3}\cos \theta \right) ,
\end{equation*}%
where $M=-\left( \sigma ^{2}-\sigma ^{3}\right) /\sqrt{2}.$ Thus, we
conclude that $Z_{\epsilon ,\overline{\epsilon }}$ is the general expression
for the\ central charge of the 2d supersymmetry algebra associated to the
Pohlmeyer reduced GS $\sigma $-models on $AdS_{n}\times S^{2},$ $n=2,3,5.$

Finally, we write the Poisson superalgebra associated to the action $%
\mathfrak{s}\circlearrowright \mathcal{P}$%
\begin{eqnarray}
\left\{ q(z\epsilon ),q(z\epsilon ^{\prime })\right\} _{2}(\mathcal{P})
&=&-q\left( z^{2}\left[ \epsilon ,\epsilon ^{\prime }\right] \right) ,\text{
\ \ \ \ }\left\{ \overline{q}(z^{-1}\overline{\epsilon }),\overline{q}(z^{-1}%
\overline{\epsilon }^{\prime })\right\} _{2}(\mathcal{P})=-\overline{q}%
\left( z^{-2}\left[ \overline{\epsilon },\overline{\epsilon }^{\prime }%
\right] \right) ,  \label{Poisson superalgebra} \\
\left\{ q(z\epsilon ),\overline{q}(z^{-1}\overline{\epsilon })\right\} _{2}(%
\mathcal{P}) &=&Z_{\epsilon ,\overline{\epsilon }}\text{.}  \notag
\end{eqnarray}

\begin{remark}
The kink central charge $Z_{\epsilon ,\overline{\epsilon }}$ and the $R_{\pm
}$ matrices in the Poisson brackets are related. By construction, the
soliton solutions satisfy the constraints $\mu _{l/r}=0$ and at quantum
level, the symmetry superalgebra $\mathfrak{s}$ gets q-deformed $\mathfrak{%
s\rightarrow s}_{q}.$ Then, it would be interesting to study the precise
relation among the saturation of the quantum Bogomolny bound, the soliton
masses, the deformation parameter $q$ and the quantum $R_{\pm }$ matrices.
\end{remark}

\subsection{The superalgebra using the symplectic form $\Omega _{SSSSG}.$}

Here we compute the Poisson superalgebra in a different way by using moments
and Hamiltonian vector fields as presented in section 6. In particular, we
want to verify the result for the mixed bracket (\ref{mixed super}).

Start with the mixed bracket and consider a contraction of equation (\ref%
{susy moment +1}) with $X_{\overline{\epsilon }}=\left( \delta _{\overline{%
\epsilon }}\gamma ,\delta _{\overline{\epsilon }}\psi ,\delta _{\overline{%
\epsilon }}\overline{\psi }\right) $ 
\begin{equation*}
\Omega _{WZNW}(X_{\epsilon },X_{\overline{\epsilon }})=\Omega
_{+}(X_{\epsilon },X_{\overline{\epsilon }})-\Omega _{-}(X_{\epsilon },X_{%
\overline{\epsilon }}),
\end{equation*}%
where we have to use (\ref{on-shell susy II})%
\begin{eqnarray*}
\delta _{\overline{\epsilon }}\gamma &=&\text{\ }\widehat{\overline{\epsilon 
}}\gamma ,\text{ \ \ \ \ \ }\widehat{\overline{\epsilon }}=\left[ \widetilde{%
\Lambda }\overline{\psi },\overline{\epsilon }\right] -\overline{w}^{\perp },%
\text{\ } \\
\delta _{\overline{\epsilon }}\overline{\psi } &=&\left[ \overline{q}%
^{\parallel },\overline{\epsilon }\right] -\left[ \overline{w}^{\perp },%
\overline{\psi }\right] ,\text{ \ \ \ \ \ }\delta _{\overline{\epsilon }%
}\psi =-\left[ \Lambda ,\gamma ^{-1}\overline{\epsilon }\gamma \right]
\end{eqnarray*}%
with $\delta \gamma (X_{\overline{\epsilon }})=\widehat{\overline{\epsilon }}%
\gamma ,$ $\delta \psi (X_{\overline{\epsilon }})=-\left[ \Lambda ,\gamma
^{-1}\overline{\epsilon }\gamma \right] ,$ $\delta \overline{\psi }(X_{%
\overline{\epsilon }})=\left[ \overline{q}^{\parallel },\overline{\epsilon }%
\right] -\left[ \overline{w}^{\perp },\overline{\psi }\right] .$ From (\ref%
{omega +}) we have%
\begin{eqnarray*}
\Omega _{+}(X_{\epsilon },X_{\overline{\epsilon }}) &=&-\left( \widehat{%
\epsilon },\delta q(X_{\overline{\epsilon }})\right) +\left( \epsilon ,\left[
q,\widetilde{\Lambda }\delta \psi (X_{\overline{\epsilon }})\right] \right)
+\left( w^{\perp },\left[ \psi ,\widetilde{\Lambda }\delta \psi (X_{%
\overline{\epsilon }})\right] \right) \\
&=&-\left( q^{\parallel },\left[ \epsilon ,\gamma ^{-1}\overline{\epsilon }%
\gamma \right] \right) -\left( \left[ \left[ \epsilon ,\widetilde{\Lambda }%
\psi \right] ,\psi \right] ,\gamma ^{-1}\overline{\epsilon }\gamma \right) \\
&=&-\left( q,\left[ \epsilon ,\gamma ^{-1}\overline{\epsilon }\gamma \right]
\right) +\left( \mu _{L},\left[ \epsilon ,\gamma ^{-1}\overline{\epsilon }%
\gamma \right] \right) ,
\end{eqnarray*}%
where we have used $\delta q(X_{\overline{\epsilon }})=D_{+}\left( \gamma
^{-1}\delta \gamma (X_{\overline{\epsilon }})\right) =D_{+}\left( \gamma
^{-1}\widehat{\overline{\epsilon }}\gamma \right) =\left[ \psi ,\gamma ^{-1}%
\overline{\epsilon }\gamma \right] .$ In a similar manner, we have%
\begin{eqnarray*}
\Omega _{-}(X_{\epsilon },X_{\overline{\epsilon }}) &=&\left( \epsilon
,\gamma ^{-1}\delta \overline{\psi }(X_{\overline{\epsilon }})\gamma +\left[
\gamma ^{-1}\overline{\psi }\gamma ,\gamma ^{-1}\delta \gamma (X_{\overline{%
\epsilon }})\right] \right) \\
&=&-\left( \overline{q}^{\parallel },\left[ \gamma \epsilon \gamma ^{-1},%
\overline{\epsilon }\right] \right) +\left( \gamma \epsilon \gamma ^{-1},%
\left[ \overline{\psi },\left[ \widetilde{\Lambda }\overline{\psi },%
\overline{\epsilon }\right] \right] \right) \\
&=&-\left( \overline{q},\left[ \gamma \epsilon \gamma ^{-1},\overline{%
\epsilon }\right] \right) +\left( \mu _{R},\left[ \gamma \epsilon \gamma
^{-1},\overline{\epsilon }\right] \right) .
\end{eqnarray*}

Putting all together and restricting to the symplectic quotient (\ref%
{reduced symplectic form}), we have%
\begin{equation*}
\Omega _{SSSSG}(X_{\epsilon },X_{\overline{\epsilon }})=-\dint\nolimits_{-%
\infty }^{+\infty }dx\partial _{x}Str\left( \gamma \epsilon \gamma ^{-1},%
\overline{\epsilon }\right) =-Z_{\epsilon ,\overline{\epsilon }}.
\end{equation*}

After a tedious but straightforward calculation we can show that%
\begin{eqnarray*}
\Omega _{+}(X_{\epsilon },X_{\epsilon ^{\prime }}) &=&2\epsilon \cdot
\epsilon ^{\prime }\dint\nolimits_{-\infty }^{+\infty }dx^{+}T_{++}+\left(
\mu _{R},2\epsilon \cdot \epsilon ^{\prime }q^{\perp }+\left[ w_{\epsilon
}^{\perp },w_{\epsilon ^{\prime }}^{\perp }\right] \right) , \\
-\Omega _{-}(X_{\epsilon },X_{\epsilon ^{\prime }}) &=&-2\epsilon \cdot
\epsilon ^{\prime }\dint\nolimits_{-\infty }^{+\infty }dx^{-}T_{-+}
\end{eqnarray*}%
and by restriction we get%
\begin{equation*}
\Omega _{SSSSG}(X_{\epsilon },X_{\epsilon ^{\prime }})=2\epsilon \cdot
\epsilon ^{\prime }\dint\nolimits_{-\infty }^{+\infty }dx\left(
T_{++}+T_{-+}\right) =q\left( z^{2}\left[ \epsilon ,\epsilon ^{\prime }%
\right] \right) .
\end{equation*}%
In a similar way we have%
\begin{equation*}
\Omega _{SSSSG}(X_{\overline{\epsilon }},X_{\overline{\epsilon }^{\prime
}})=2\overline{\epsilon }\cdot \overline{\epsilon }^{\prime
}\dint\nolimits_{-\infty }^{+\infty }dx\left( T_{+-}+T_{--}\right) =q\left(
z^{-2}\left[ \overline{\epsilon },\overline{\epsilon }^{\prime }\right]
\right) .
\end{equation*}

Note that the last expressions are independent of the non-local terms
present in the field variations (\ref{on-shell susy}),(\ref{on-shell susy II}%
). This is precisely what we found above in the computation done by using
differentials, i.e, the invariance of the brackets (\ref{++ super}) and (\ref%
{-- supers}) under the action of (\ref{infinite ambiguity}).

Comparing with (\ref{Poisson superalgebra}) we have that\footnote{%
Roughly, if we use (\ref{moment definition}) and (\ref{moments for s}) we
can write $\left\{ H_{a},H_{a^{\prime }}\right\} _{2}=H_{\left[ a,a^{\prime }%
\right] }+c,$ i.e, the action $\mathfrak{s\circlearrowright }\mathcal{P}$ is
not Poissonian.} 
\begin{equation*}
\left\{ q(a),q(a^{\prime })\right\} _{2}(\mathcal{P})=-\Omega
_{SSSSG}(X_{a},X_{a^{\prime }}),\text{ \ \ \ \ \ }a\in \widehat{\mathfrak{s}}%
.
\end{equation*}

\section{Concluding remarks.}

We have shown explicitly through simple arguments the existence 2d
supersymmetry on the reduced phase space of the GSs$\sigma $ models on the
target spaces $AdS_{n}\times S^{n},$ $n=2,3,5.$ However, several question
are still open and there are some interesting and important directions to be
followed in the future. For example, the implementation of the Pohlmeyer
reduction on the path integral, e.g. \cite{Iwashita}, and the study of the
supersymmetry in which perhaps, localization techniques can be used. The
role of the q-deformed supersymmetry. The supersymmetry associated to the
reduction of the GSs$\sigma $ model on $AdS_{4}\times CP^{3}$. A refine
study of the relations between boundary terms and integrability, etc. We
hope to address some of these questions in the near future.

\paragraph{Acknowledgements}

The author thanks Tim Hollowood and Luis Miramontes for a very fruitful and
pleasant collaboration at the early stages of the present work in relation
to the non-local Poisson structures. This work is supported by the Capes
PNPD 2416093 post-doc grant.

\end{document}